\documentclass[11pt,a4paper,preprintnumbers,nofootinbib, superscriptaddress]{revtex4}

\usepackage[T1]{fontenc}
\usepackage[utf8]{inputenc}
\usepackage{amsmath}
\usepackage{amssymb}
\usepackage{hyperref}
\usepackage[final]{graphicx}

\usepackage[english]{babel}

\usepackage{graphicx}
\usepackage[dvipsnames]{xcolor}
\usepackage{subfigure}

\usepackage{braket}

\usepackage{amsfonts}
\usepackage{amstext}
\usepackage{slashed}
\usepackage{microtype}

\usepackage{multirow}
\usepackage{booktabs}

\usepackage{placeins}
\usepackage{verbatim}

\newcommand{\dd}{d}

\def\GeV{{\rm GeV}}

\def\A{\mathcal{A}}

\def\L{\mathcal{L}}

\def\O{\mathcal{O}}

\def\ubar{\overline{u}}

\def\qbar{\overline{q}}

\def\Kbar{\overline{K}}

\begin{document}

\title{Probing QCD dynamics and the standard model with $D_{(s)} \to P^+ P^0 \gamma$ decays}

\author{Nico Adolph}
\email{nico.adolph@tu-dortmund.de}
\affiliation{Fakult\"at Physik, TU Dortmund, Otto-Hahn-Str.4, D-44221 Dortmund, Germany}

\author{Gudrun Hiller}
\email{gudrun.hiller@tu-dortmund.de}
\affiliation{Fakult\"at Physik, TU Dortmund, Otto-Hahn-Str.4, D-44221 Dortmund, Germany}
\preprint{DO-TH 21/06}

\begin{abstract}
We compute 10 radiative three-body decays of charged charmed mesons $D^+ \to P^+ P^0 \gamma$  and $D_s  \to P^+ P^0 \gamma$, $P=\pi, K$,
in leading order QCDF, HH$\chi$PT and the soft photon approximation.
We work out decay distributions and asymmetries in the standard model and with new physics
in the electromagnetic dipole operators.
The forward-backward asymmetry is suitable to probe the QCD frameworks, in particular
the $s$-channel dependent weak annihilation contributions in QCDF against the markedly different
resonance structure in  HH$\chi$PT. These studies can be performed with Cabibbo-favored  modes 
$D_s \to \pi^+ \pi^0 \gamma$, $D^+ \to \pi^+ \Kbar^0 \gamma$ and $D_s \to K^+ \Kbar^0 \gamma$  with ${\cal{O}}(10^{-4}-10^{-3})$-level branching ratio, 
which are standard model-like and induced by different hadronic dynamics. Understanding of the latter can therefore be improved in a data-driven way
and sharpens the interpretation of standard model tests.
Singly Cabibbo-suppressed modes such as
$ D^+ \to \pi^+ \pi^0 \gamma$,  $D_s \to \pi^+ K^0 \gamma$, $D_s \to K^+ \pi^0 \gamma$  with branching ratios within
$\sim 10^{-5}-10^{-4}$
are sensitive to new physics that can be signalled in the forward-backward asymmetry
and  in the CP-asymmetry of the rate, ideally in the Dalitz region but also in single differential distributions.
Results complement those with neutral $D^0 \to PP \gamma$ decays.
\end{abstract}

\maketitle

\section{Introduction}

Rare semileptonic and radiative charm decays are sensitive to the flavor sector in and  beyond the standard model \cite{Burdman:2001tf}.
They are accessible to precision study  at the LHC \cite{Cerri:2018ypt}, the Belle II  \cite{Kou:2018nap}, and BES III experiments \cite{Ablikim:2019hff},  and complement searches in beauty and kaon decays.
Disentangling resonance contributions from those of BSM physics is key, and particularly challenging in charm
as the charm quark mass is not as well separated from the QCD scale as the $b$-quark mass, and branching ratios are often dominated by resonances on and 
off peak. A well-known workaround
strategy is to use null tests, that is,  observables designed to  have an unmeasurably small or otherwise controlled SM background, due to protection by approximate symmetries of the SM 
\cite{Fajfer:2015mia, deBoer:2018buv, Bause:2019vpr}.
On the other hand,   the $D$-system is advantageous as many partner modes exist which are  linked by swapping light quarks, or flavor symmetry.
A combined analysis of these partner modes
allows therefore to gain insights into the resonance  dynamics and new physics
simultaneously, as shown for instance for radiative decays in  \cite{deBoer:2018zhz, Adolph:2018hde}.

In this regard, radiative three-body $D$-decays turn  out to be a useful laboratory to explore  QCD dynamics and to test the SM.
Here we consider, in total ten,
Cabibbo-favored (CF), singly Cabibbo-suppressed (SCS), and doubly Cabibbo-suppressed (DCS) decays of $D^+$ and $D_s$ mesons
\begin{equation} \label{eq:all}
  \begin{alignedat}{2}
    &\text{CF:} \quad && D_s \to \pi^+ \pi^0 \gamma\, , ~D_s \to K^+ \Kbar^0 \gamma \, , ~D^+ \to \pi^+ \Kbar^0 \gamma\, , ~(D^0 \to \pi^0 \Kbar^0 \gamma\, , ~D^0 \to \pi^+ K^- \gamma)\\
    &\text{SCS:} \quad && D^+ \to \pi^+ \pi^0 \gamma\, , ~D_s \to \pi^+ K^0 \gamma\, , ~ D_s \to K^+ \pi^0 \gamma\, ,\\
    &~ &&D^+ \to K^+ \Kbar^0 \gamma \, , (D^0 \to \pi^+ \pi^- \gamma\, , ~D^0 \to K^+ K^- \gamma)\\
    &\text{DCS:} \quad && D^+ \to \pi^+ K^0 \gamma\, , D^+ \to K^+ \pi^0 \gamma\, , ~ D_s \to K^+ K^0 \gamma\\ 
  \end{alignedat}
\end{equation}
The $D^0$ decays  given in parentheses are covered in \cite{Adolph:2020ema}.
The modes $D^+ \to K^+ K^0 \gamma$ and  $D_s \to \pi^+ \Kbar^0 \gamma$  are $|\Delta s|=2$ processes and are not induced by dimension 6 operators,
and not considered in this works.
Three-body radiative $D$ decays have been investigated previously in  \cite{Fajfer:2002bq, Fajfer:2002xf}.

As in  \cite{Adolph:2020ema} we work out decay amplitudes in  different QCD frameworks: leading order QCD factorization (QCDF), heavy hadron chiral perturbation theory
(HH$\chi$PT) and the soft photon approximation, each of which is expected to hold in specific regions of phase space.
We show that the forward-backward asymmetry efficiently disentangles different resonance contributions and therefore probes the QCD frameworks.
These tests can be performed  with SM-like modes, the CF and DCS ones from (\ref{eq:all}), and assist NP searches in the SCS ones.

This paper is organized as follows: In Sec.~\ref{sec:QCD} we give the kinematics and observables for $D \to PP \gamma$ decays, the weak effective low energy Lagrangian  and briefly review
QCD frameworks. We also present the weak annihilation contribution  in QCDF.
Analytical results in HH$\chi$PT are presented in Appendix~\ref{app:HQCHPT form factors}.
SM predictions are shown in Sec.~\ref{sec:compare}.
We work out predictions with BSM physics in Sec.~\ref{sec:BSM}.
In Sec.~\ref{sec:con} we summarize. In appendix \ref{app:HQCHPT form factors} we provide
HH$\chi$PT-amplitudes and Feynman diagrams for CF, SCS and DCS modes, and a list of differences we found with previous works \cite{Fajfer:2002bq}
on $D^+ \to \pi^+ \Kbar^0 \gamma$ in appendix \ref{sec:diff}.

\section{Generalities   \label{sec:QCD}}

\subsection{Kinematics and observables}

The double differential decay rate of the radiative three-body decay 
$D_{(s)}^+(P) \to P^+(p_1)P^0(p_2)\gamma(k, \epsilon^*)$ can be written as \cite{Adolph:2020ema} 
\begin{equation}
  \begin{split}
   \frac{\dd{}^2\Gamma}{\dd{}s \dd{}t} & = \frac{ |A_-|^2 + |A_+|^2}{128 (2\pi)^3 m_D^3}\\
   & \quad \times \big[m_1^2(t-m_2^2)(s-m_D^2)-m_2^4m_D^2 - st(s+t-m_D^2) + m_2^2(st+(s+t)m_D^2 - m_D^4)\big]\,,
  \end{split}
\end{equation}
where the parity-even ($\A_+$) and parity-odd ($\A_-$) contributions
are defined by the general Lorentz decomposition of the decay amplitude
\begin{align} \label{eq:mainA}
  \A(D_{(s)}^+ \rightarrow P^+ P^0 \gamma) = A_-(s,t)  \left[(p_1 \cdot k)(p_2 \cdot  \epsilon^{*}) - (p_2 \cdot  k)(p_1 \cdot  \epsilon^{*})\right] + A_+(s,t) \epsilon^{\mu\alpha\beta\gamma}
   \epsilon^{*}_\mu p_{1\alpha} p_{2\beta} k_\gamma \,.
\end{align}
The subscript 1 (2) of four-momenta and masses refer to the (un)charged 
final state $P^+$ ($P^0$), while the initial state $D_{(s)}^+$-meson is 
denoted by capital letters. Furthermore, $s=(p_1 + p_2)^2$, $t=(p_2 + k)^2$ and $u=(p_1+k)^2=m_D^2+m_1^2+m_2^2-s-t$.
We employ $m_D$ for both the $D^+$ and the $D_s$ mass and specify the spectator quark only when needed.
The polarization vector of the photon is denoted by $\epsilon$.

The three-body decay allows to define a forward-backward asymmetry \cite{Adolph:2020ema}
\begin{align} \label{eq:AFB}
  \begin{split}
    &A_{\rm FB}(s) = \frac{\int_{t_{\rm min}}^{t_{\rm 0}} dt \frac{\dd{}^2\Gamma}{\dd{}s \dd{}t} -\int_{t_{\rm 0}}^{t_{\rm max}} dt  \frac{\dd{}^2\Gamma}{\dd{}s \dd{}t}}{\int_{t_{\rm min}}^{t_{\rm 0}} dt \frac{\dd{}^2\Gamma}{\dd{}s \dd{}t} + \int_{t_{\rm 0}}^{t_{\rm max}} dt  \frac{\dd{}^2\Gamma}{\dd{}s \dd{}t}} \, ,\\
    &t_{\rm min}= \frac{(m_D^2 - m_{1}^2 + m_{2}^2)^2}{4s}  - \left(\sqrt{\frac{(s - m_{1}^2 + m_{2}^2)^2}{4s} - m_2^2} + \frac{m_D^2 - s}{2\sqrt{s}}\right)^2\, ,\\ 
    &t_{\rm max}= \frac{(m_D^2 - m_{1}^2 + m_{2}^2)^2}{4s}  - \left(\sqrt{\frac{(s - m_{1}^2 + m_{2}^2)^2}{4s} - m_2^2} - \frac{m_D^2 - s}{2\sqrt{s}}\right)^2 \, , \\
    &t_{\rm 0} = \frac{1}{2s}\left(-s^2 + s(m_D^2 + m_1^2 + m_2^2) + m_D^2(m_2^2 - m_1^2)\right)\, ,    
  \end{split}
\end{align}
where the first (second) term in the numerator corresponds to $0 \leq
\cos(\theta_{2\gamma}) \leq 1$ $(-1 \leq \cos(\theta_{2\gamma}) \leq
0)$. Here, $\theta_{2\gamma}$ is the angle between $P^0$ and the
photon in the $P^+ - P^0$ center-of-mass frame. If the forward-backward 
asymmetry would be defined in terms of $\cos(\theta_{1\gamma})$, one would
obtain an additional minus sign in the definition of $A_{\rm FB}$. The  $A_{\rm FB}$  turns out 
to be suitable to distinguish resonance contributions in $s$-channel 
(in $PP$), versus $t,u$-channel ones (in $P \gamma$) in the decay amplitudes.

The most promissing observable to test for BSM physics
is the single- or double-differential CP asymmetry defined as
respectively, by
\begin{align}  \label{eq:ACP}
 A_{\rm CP}(s) =\int \dd{} t    A_{\rm CP}(s,t) \, , ~~~
  A_{\rm CP}(s,t) =\frac{1}{\Gamma + \overline{\Gamma}}\left(\frac{\dd{}^2\Gamma}{\dd{}s\dd{}t} - \frac{\dd{}^2\overline{\Gamma}}{\dd{}s\dd{}t}\right) \, . 
\end{align}
Here, $\overline{\Gamma}$ refers to the decay rate of the
CP-conjugated mode.

\subsection{Weak effective Lagrangian}

The effective weak Lagrangian for $c \rightarrow u \gamma$ processes can be written as~\cite{deBoer:2017que}
\begin{align}
    \L_{\text{eff}} = \frac{4G_F}{\sqrt{2}}\left(\sum_{q, q'\in \{d, s\}}V_{cq}^* V_{uq'}\sum_{i=1}^2 C_i O_i^{(q, q')}  +\sum_{i=3}^6 C_i O_i + \sum_{i=7}^8 \left(C_i O_i + C_i^\prime O_i^\prime\right)\right)\, ,
\end{align}
where $G_F$ is Fermi's constant and $V_{ij}$ are elements of the
Cabibbo-Kobayashi-Maskawa (CKM) matrix. For the purpose of this work
only the operators
\begin{equation}
  \begin{alignedat}{2}
    &O_1^{(q, q')} = \left(\ubar_L \gamma_\mu T^a q_L'\right) \left(\qbar_L \gamma^\mu T^a c_L\right)\, , \quad &&O_2^{(q, q')} = \left(\ubar_L \gamma_\mu q_L'\right) \left(\qbar_L \gamma^\mu c_L\right)\, ,\\
    &O_7 = \frac{e m_c}{16\pi^2}  \left(\ubar_L \sigma^{\mu \nu} c_R\right)F_{\mu \nu}\, , &&O_7^\prime = \frac{e m_c}{16\pi^2}  \left(\ubar_R \sigma^{\mu \nu} c_L\right)F_{\mu \nu}\, 
  \end{alignedat}
\end{equation}
are relevant. Here, the subscripts $L(R)$ denote left-(right-)handed 
quark fields, $T^a$ are generators of $SU(3)$ normalized to $\text{Tr}\{T^aT^b\} =
\delta^{ab}/2$ and $F_{\mu \nu}$ is the photon field strength tensor, 
respectively. For $\mu_c \in \left[m_c/\sqrt{2}, \sqrt{2}m_c\right]$,
the leading order Wilson Coefficients of the four quark operators 
$O_{1,2}^{(q, q')}$ are given by \cite{deBoer:2017que}
\begin{align}
  C_1 \in [-1.28, -0.83]\, , \qquad C_2 \in [1.14, 1.06]\, .
\end{align}
Within the SM, $C^{(\prime)}_{3-8}$ can be neglected due to the 
efficient GIM cancellation. However, we also consider 
BSM contributions to $C_7^{(\prime)}$ in section \ref{sec:BSM}.

\subsection{QCD frameworks for $D \to PP \gamma$ decays \label{sec:frame}}

In the following we use QCDF, Low's Theorem and  HH$\chi$PT as QCD frameworks.
We refrain from a detailed introduction of the individual models at 
this point. Instead, we refer to the literature cited below, as well as
\cite{Adolph:2020ema} where we provide further information on kinematics,
the models and our notation.

The leading weak annihilation contribution ($\O(\alpha_s^0 (\Lambda_\text{QCD}/m_c)^1)$) 
can be determined with QCDF methods \cite{Beneke:2000ry, Bosch:2001gv, DescotesGenon:2002mw} as
\begin{align}
  \begin{split}
    &\A^{\text{WA}}_- = i \frac{G_F e}{\sqrt{2}} C_2  \frac{f_{D_{(s)}} Q_d}{\lambda_{D_{(s)}} (v\cdot k)} \sum_{q\in \{d, s\}} V_{cd(s)}^* V_{uq}  f^{P_1 P_2}_{(q)}(s) \, ,\\
    &\A^{\text{WA}}_+ = \frac{G_F e}{\sqrt{2}} C_2  \frac{f_{D_{(s)}} Q_d}{\lambda_{D_{(s)}} (v\cdot k)} \sum_{q\in \{d, s\}} V_{cd(s)}^* V_{uq}  f^{P_1 P_2}_{(q)}(s)  \, .  \label{eq:QCDF-Amplitude}
  \end{split}
\end{align}
Here, $Q_d$ is the electric charge of the down-type quarks 
and $\lambda_{D_{(s)}}$ is a non perturbative parameter of 
$\O(\Lambda_\text{QCD})$ which is poorly known. Furthermore, 
$v^\mu = P^\mu/m_D$ is the four-velocity of the $D_{(s)}^+$ 
meson. Note that unlike the decays of the neutral $D$ meson, 
the amplitudes \eqref{eq:QCDF-Amplitude} do not depend on the 
color-suppressed combination of the Wilson coefficients
\cite{deBoer:2017que,Adolph:2020ema}. 
Therefore, the scale uncertainty is significantly smaller. 
For $f^{P_1 P_2}_{(q)}$ we use the electromagnetic pion and 
kaon form factors \cite{Bruch:2004py}, form factors extracted 
from $\tau^- \to \nu_\tau K_S \pi^-$ decays \cite{Boito:2008fq} 
and isospin relations
\begin{align}  \label{eq:iso}
  \begin{split}
    &f^{\pi^+ \pi^0}_{(d)}(s) = - \sqrt{2} F^{\text{em}}(s)\, ,\\
    &f^{K^+ \Kbar^0}_{(d)}(s) = 2F^{(I=1)}_{K^+}(s)\, ,\\
    &f^{\pi^+K^0}_{(s)}(s)  = f^{\overline{K}\pi^-}_+(s)\, , \\
    &f^{K^+ \pi^0}_{(s)}(s)  = -\frac{1}{\sqrt{2}}f^{\overline{K}\pi^-}_+(s) \, .
  \end{split}
\end{align}
Detailed information on the form factors can also be found 
in appendix B1 of \cite{Adolph:2020ema}. Note that QCDF is only 
applicable for small invariant masses $s \lesssim 1.5 \GeV^2$ of 
the hadronic system in the final state. We remark that there are 
no weak annihilation contributions for $D^+ \to \pi^+ \Kbar^0 \gamma$
and $D_s \to K^+ K^0 \gamma$.

Low's theorem~\cite{Low:1958} relates the bremsstrahlungsamplitude
of the radiative three-body decays to the amplitudes 
of the hadronic two-body decays
\begin{align}
  \A_-^{\rm Low}=-\frac{e\A(D^+_{(s)} \to P^+_1 P^0_2)}{(p_1\cdot k)(P\cdot k)}\,, \qquad \A_+^{\rm Low}=0\, . \label{eq:Low_theorem}
\end{align}
This approach is valid for soft photons with an energy below 
$m_{P^+}^2/E_{P^+}$~\cite{DelDuca:1990} and thus describes
the opposite part of the phase space compared to QCDF. We 
don't rely on a theory prediction for $\A(D_{(s)}^+ \to P^+ P^0)$,
but extract the modulus from data on branching ratios~\cite{Tanabashi:2018oca}.
We obtain
\begin{equation} \label{eq:lowdata}
  \begin{alignedat}{2}
    &\left| \A(D^+ \to \pi^+ \pi^0)\right| = (2.74 \pm 0.04) \cdot 10^{-7}\, \text{GeV}\, , \quad &&\left| \A(D_s \to \pi^+ \pi^0)\right| < 2.11  \cdot 10^{-7}\, \text{GeV}\, ,\\
    &\left| \A(D^+ \to \pi^+ K^0)\right| < 4.70 \cdot 10^{-8}\, \text{GeV}\, , \quad &&\left| \A(D_s \to \pi^+ K^0)\right| = (5.75 \pm 0.12) \cdot 10^{-7}\, \text{GeV}\, ,\\
    &\left| \A(D^+ \to K^+ \pi^0)\right| = (1.16 \pm 0.06) \cdot 10^{-7}\, \text{GeV}\, , \quad &&\left| \A(D_s \to K^+ \pi^0)\right| = (2.9 \pm 0.5) \cdot 10^{-7}\, \text{GeV}\, ,\\
    &\left| \A(D^+ \to K^+ \Kbar^0)\right| = (6.62 \pm 0.10) \cdot 10^{-7}\, \text{GeV}\, ,\quad &&\left| \A(D_s \to K^+ \Kbar^0)\right| = (2.10 \pm 0.05) \cdot 10^{-6}\, \text{GeV}\, ,\\
    &\left| \A(D^+ \to \pi^+ \Kbar^0)\right| = (1.396 \pm 0.014) \cdot 10^{-6}\, \text{GeV}\qquad && \left|\A(D_s \to K^+ K^0)\right| < 5.36 \cdot 10^{-8}\, \text{GeV}\, .
  \end{alignedat}
\end{equation}
Note that there is only an upper limit on the $D_s \to \pi^+ \pi^0$ 
branching ratio. We remark that it turns out that HH$\chi$PT implies 
an amplitude about two orders of magnitude smaller. 
Moreover, in the isospin limit, the amplitude vanishes.
Furthermore, for the decays into neutral kaons, 
only the branching ratio of $D_s \to K^+ \Kbar^0$ is measured.
We extract the corresponding amplitudes for the remaining decays
from data for final states with $K_S$ and $K_L$. This procedure 
leads to huge uncertainties for the DCS decays. Therefore, we only 
give upper limits for these amplitudes. All upper limits refer to 
a 90\% confidence level. As it is sizable we explicitly take  the uncertainty of the 
$D_s \to K^+ \pi^0$ amplitude in our numerical analysis into account, but neglect the uncertainties in (\ref{eq:lowdata}) 
from the other decay channels.

Furthermore, we employ HH$\chi$PT \cite{Wise:1992,Burdman:1992,Yan:1992} extended 
by light vector resonances \cite{Casalbuoni:1993}. The parity-even 
and parity-odd amplitudes are given in terms of 
form factors $A_i^{(q, q')}, B_i^{(q, q')}, D_i^{(q, q')}$ and $E_i^{(q, q')} $
\begin{align} \label{eq:HHA}
  \begin{split}
    &A_-^{\rm HH\chi PT} = \frac{G_F e}{\sqrt{2}} \sum_{q, q'\in \{d, s\}}V_{cq}^*V_{uq'}  \left[(C_2 - \frac{1}{6}C_1) \sum_{i} A_i^{(q, q')} + \frac{1}{2}C_1 \sum_{i} E_i^{(q, q')}\right]\, ,\\
    &A_+^{\rm HH\chi PT} = \frac{G_F e}{\sqrt{2}} \sum_{q, q'\in \{d, s\}}V_{cq}^*V_{uq'}  \left[(C_2 - \frac{1}{6}C_1) \sum_{i} B_i^{(q, q')} + \frac{1}{2}C_1 \sum_{i} D_i^{(q, q')}\right]\, .
  \end{split}
\end{align}
In Appendix~\ref{app:HQCHPT form factors} the corresponding 
diagrams are shown in Fig.~\ref{fig:Diagramme_Tensorstrom} to 
\ref{fig:Diagramme_Ds_KplusK} and the non-zero
contributions to the form factors are listed. 
We consider the masses of the light pseudoscalars only in phase space
and in their propagators. Otherwise we neglect them in the form factors.
To enforce Low's theorem, we remove the bremsstrahlung contributions
$A_{1,2,3}$ and $E_{1,2}$ in (\ref{eq:HHA}) and add (\ref{eq:Low_theorem}) while using the strong and weak phases predicted by HH$\chi$PT.
For the diagrams $A_{6,1}$, the contributions of 
longitudinal polarization of the $D^+_{(s)} \to V^+ V^0$ subdiagram have 
to be removed in order to obtain a gauge invariant amplitude \cite{Golowich:1994zr}.

\section{SM predictions \label{sec:compare}}

In this section we compare SM predictions for the branching ratios (Section \ref{sec:BR}) and forward-backward asymmetries (Section \ref{sec:AFB}) of the QCD frameworks given in Sec.~\ref{sec:frame}.
SM CP asymmetries are discussed in Section \ref{sec:SMCP}.

\subsection{Branching ratios \label{sec:BR}}

\begin{figure}
  \centering
  \includegraphics[width=0.9\linewidth]{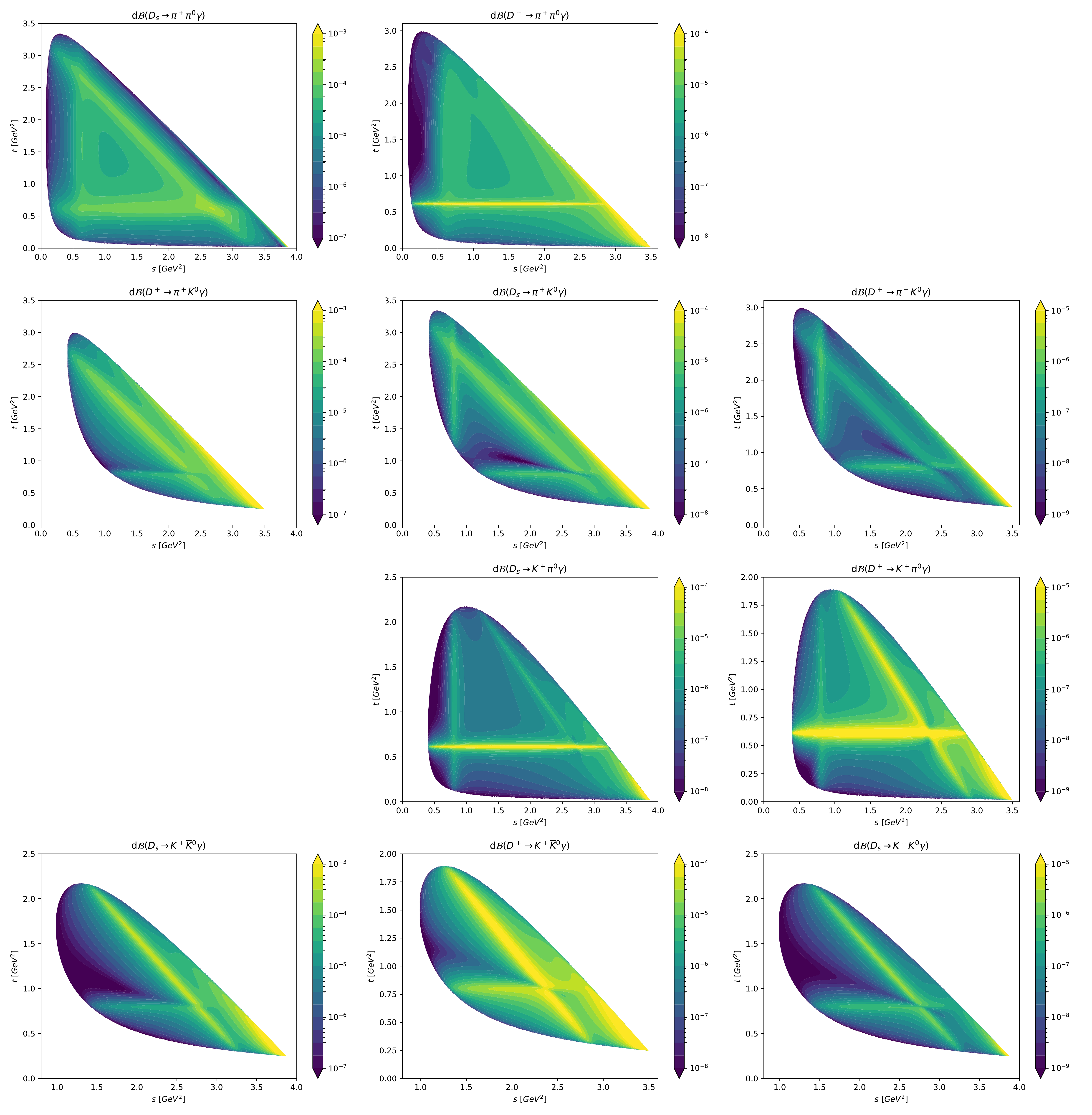}
  \caption{Dalitz plots based on the SM HH$\chi$PT predictions for $\mu_c=m_c$ and the mean value of the $D_{(s)}\to P^+ P^0$ amplitude.
  The first, second and third columns show the CF, SCS and DCS decay modes, respectively. Each row shows the same final state or the final state with $K^0 \leftrightarrow \Kbar^0$.}
  \label{fig:SM_d2BR}
\end{figure}

The Dalitz plots in Fig.~\ref{fig:SM_d2BR} illustrate the SM
predictions based on HH$\chi$PT. The bands in $s$, $t$ and $u$, which are generated by the 
intermediate $\rho$, $\omega$ and $K^\star$ resonances, are
particularly distinctive. Furthermore, the significant 
bremsstrahlung effects can be seen for large $s$.
For the decays $D_{(s)}^+ \to K^+ \Kbar^0 \gamma$
$D^+ \to \pi^+ \Kbar^0 \gamma$ and $D_s \to K^+ K^0 \gamma$, a preference 
of the $u$-channel resonance over the $t$-channel resonance is 
observed. This is due to constructive interference of the 
$u$-channel contributions from the form factors $B$ and $D$, 
while the $t$-channel contributions interfere destructively.

\begin{figure}
  \centering
  \includegraphics[width=0.9\linewidth]{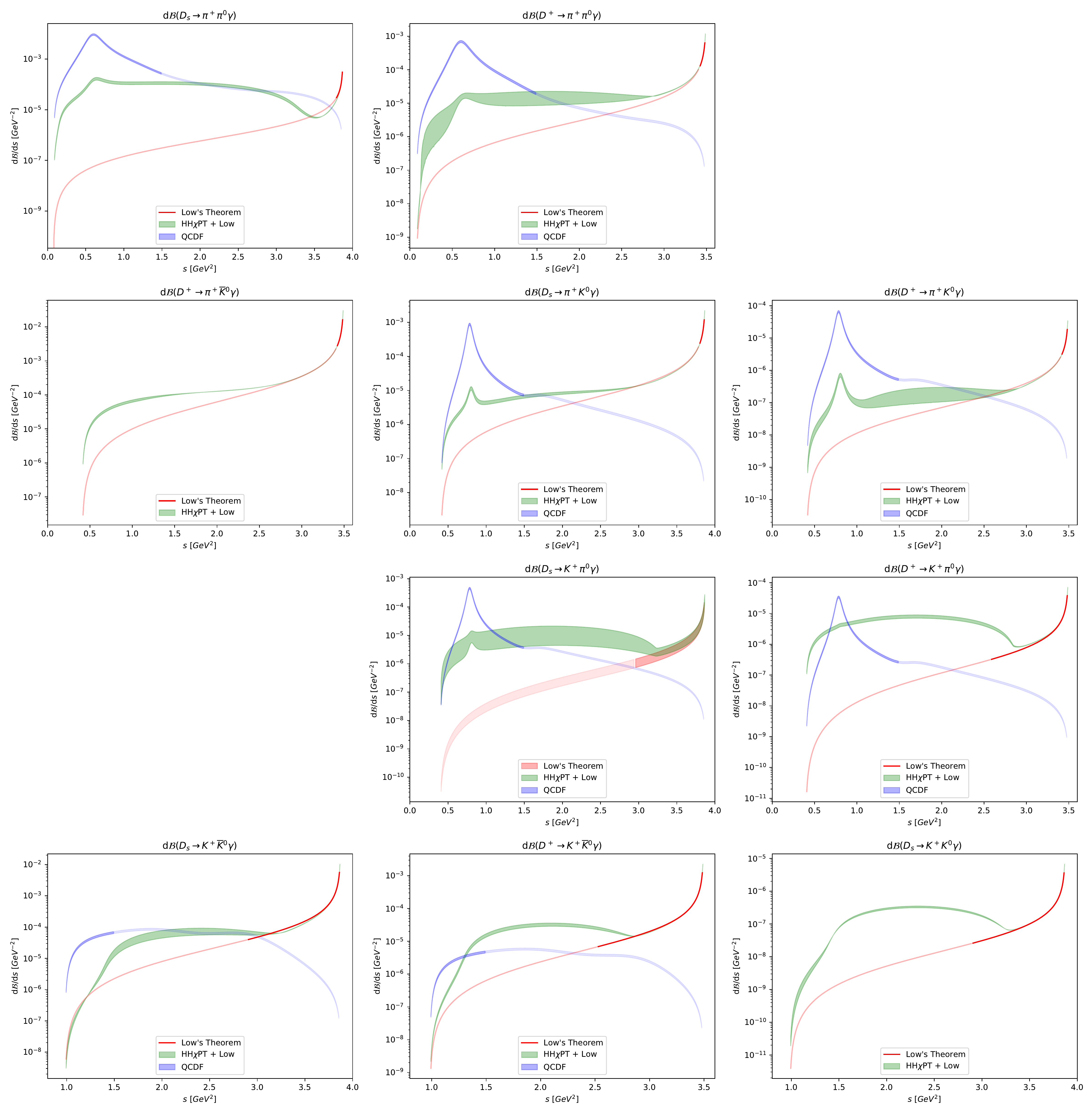}
  \caption{Single differential branching ratios based on Low's Theorem (red), HH$\chi$PT (green) and QCDF (blue) within the SM.
  The darker shaded areas and lines correspond to the model's region of applicability.
  QCDF predictions are obtained for $\lambda_{D_{(s)}}=0.1\, \GeV$ and scale with $(0.1\, \GeV/\lambda_{D_{(s)}})^2$.
  The first, second and third columns show the CF, SCS and DCS decay modes, respectively. Each row shows the same final state or the final state with $K^0 \leftrightarrow \Kbar^0$.}
  \label{fig:SM_dBr_Vergleich}
\end{figure}

Figure~\ref{fig:SM_dBr_Vergleich} shows the differential branching ratios
of all considered decay channels as a function of $s$. For each decay, 
the predictions of QCDF (blue), HH$\chi$PT (green) and Low's Theorem (red) 
are illustrated. The width of the bands arise from the $\mu_c$ dependence 
of the Wilson Coefficients and the uncertainty on the 
$D_s \rightarrow K^+ \pi^0$ amplitude.

The distinct resonance in the QCDF results is a consequence of the
$P^+-P^0$ form factors dominating the shape of the distributions.
For the final state $K^+ K \gamma$ no resonance peak 
can be identified, since the lowest $\rho$ resonance is outside 
of the phase space. In comparison, the $s$-channel resonances 
in the HH$\chi$PT predictions only result in a distinct peak for the final 
state $\pi^+ K^0 \gamma$. The other decays are mostly dominated by the 
$t$- and $u$-channel resonances in the range of small and intermediate 
values of $s$. In particular, the contributions of the $\omega$ resonances
are large due to the significant $\omega \pi \gamma$ coupling.
For large $s$, the bremsstrahlung is dominating and the results are 
approaching to those of Low's theorem due to the replacement of the 
model's own bremsstrahlung contribution.

In Table~\ref{tbl:branching_ratios_CF} and \ref{tbl:branching_ratios_DCS}
we give the branching ratios for the SM-like decay modes. The QCDF 
results only include the phase space up to $s \leq 1.5\GeV$. For 
comparison, we show the HH$\chi$PT branching ratios for this cut as well. 
Furthermore, we employ the phase space cut $E_\gamma \geq 0.1\GeV$ to 
avoid the soft photon pole, where $E_\gamma=(m_D^2-s)/(2 m_D)$ is the photon 
energy in the $D_{(s)}^+$ meson's rest frame. QCDF leads to larger values 
for the decays $D_{s} \to \pi^+ \pi^0 \gamma$, $D^+_{s} \to K^+ \Kbar^0 \gamma$
and $D^+ \to \pi^+ K^0 \gamma$ as a result of the large $s$-channel 
contributions and the small value of $\lambda_{D_{(s)}}$. We
note that the branching ratio of $D^+_{s} \to K^+ \Kbar^0 \gamma$
is about two orders of magnitude smaller than the one of $D_{s} \to \pi^+ \pi^0 \gamma$,
although both decays are CF, since the lightest $\rho$ is not inside the
phase space of $D^+_{s} \to K^+ \Kbar^0 \gamma$.
Due to the missing $t$- and $u$-channel resonances in QCDF
 and the large $\omega \pi \gamma$ coupling, 
HH$\chi$PT predicts larger branching ratios for 
$D^+ \to K^+ \pi^0 \gamma$. SM branching ratios for the  SCS modes 
are given in Table~\ref{tbl:branching_ratios_BSM}.

\begingroup
\renewcommand{\arraystretch}{1.4}
\begin{table}
  \centering
  \begin{tabular}{l|c|c|c}
                                                          & $D_s \to \pi^+ \pi^0 \gamma$ & $D^+ \to \pi^+ \Kbar^0 \gamma$ & $D_s \to K^+ \Kbar^0 \gamma$ \\
    \hline
    $\text{QCDF}\big|^\text{SM}_{s \leq 1.5~\GeV^2} $            & $(2.8 - 3.2)\cdot 10^{-3}$ & $-$ & $(1.8 - 2.1)\cdot 10^{-5}$ \\
    $\text{HH$\chi$PT}\big|^\text{SM}_{s \leq 1.5~\GeV^2} $      & $(1.1 - 1.5)\cdot 10^{-4}$ & $(6.0 - 6.7)\cdot 10^{-5}$ & $(2.5 - 3.2)\cdot 10^{-6}$ \\
    $\text{HH$\chi$PT}\big|^\text{SM}_{E_\gamma \geq 0.1~\GeV} $ & $(2.4 - 3.0)\cdot 10^{-4}$ & $(3.5 - 3.6)\cdot 10^{-4}$ & $(1.5 - 2.0)\cdot 10^{-4}$ \\
  \end{tabular}
  \caption{Branching ratios for the CF decays. 
  The branching ratios are given in the region of applicability of QCDF $s \lesssim 1.5\, \GeV^2$ for QCDF and HH$\chi$PT to enable a comparison of both models. Additionally, HH$\chi$PT predictions are given for $E_\gamma \geq 0.1\, \GeV$, see text for details. The QCDF branching ratios are obtained for $\lambda_{D_{(s)}}=0.1\, \GeV$ and are $\propto (0.1\, \GeV/\lambda_{D_{(s)}})^2$.}
  \label{tbl:branching_ratios_CF}
\end{table}
\endgroup

\begingroup
\renewcommand{\arraystretch}{1.4}
\begin{table}
  \centering
  \begin{tabular}{l|c|c|c}
                                                              & $ D^+ \to \pi^+ K^0 \gamma$ & $D^+ \to K^+ \pi^0 \gamma$ & $D_s \to K^+ K^0 \gamma$ \\
    \hline
    $\text{QCDF}\big|^\text{SM}_{s \leq 1.5~\GeV^2} $            & $(7.2 - 8.3)\cdot 10^{-6}$ & $(3.7 - 4.3)\cdot 10^{-6}$ & $-$ \\
    $\text{HH$\chi$PT}\big|^\text{SM}_{s \leq 1.5~\GeV^2} $      & $(1.2 - 2.0)\cdot 10^{-7}$ & $(4.5 - 5.8)\cdot 10^{-6}$ & $(7 - 8) \cdot 10^{-9}$ \\
    $\text{HH$\chi$PT}\big|^\text{SM}_{E_\gamma \geq 0.1~\GeV} $ & $(3.9 - 6.6)\cdot 10^{-7}$ & $(1.2 - 1.5)\cdot 10^{-5}$ & $(4.3 - 4.7)\cdot 10^{-7}$ \\
  \end{tabular}
  \caption{As in Table~\ref{tbl:branching_ratios_CF} but for the DCS decay modes.}
  \label{tbl:branching_ratios_DCS}
\end{table}
\endgroup

\subsection{Forward-Backward Asymmetry\label{sec:AFB}}

\begin{figure}
  \centering
  \includegraphics[width=0.9\linewidth]{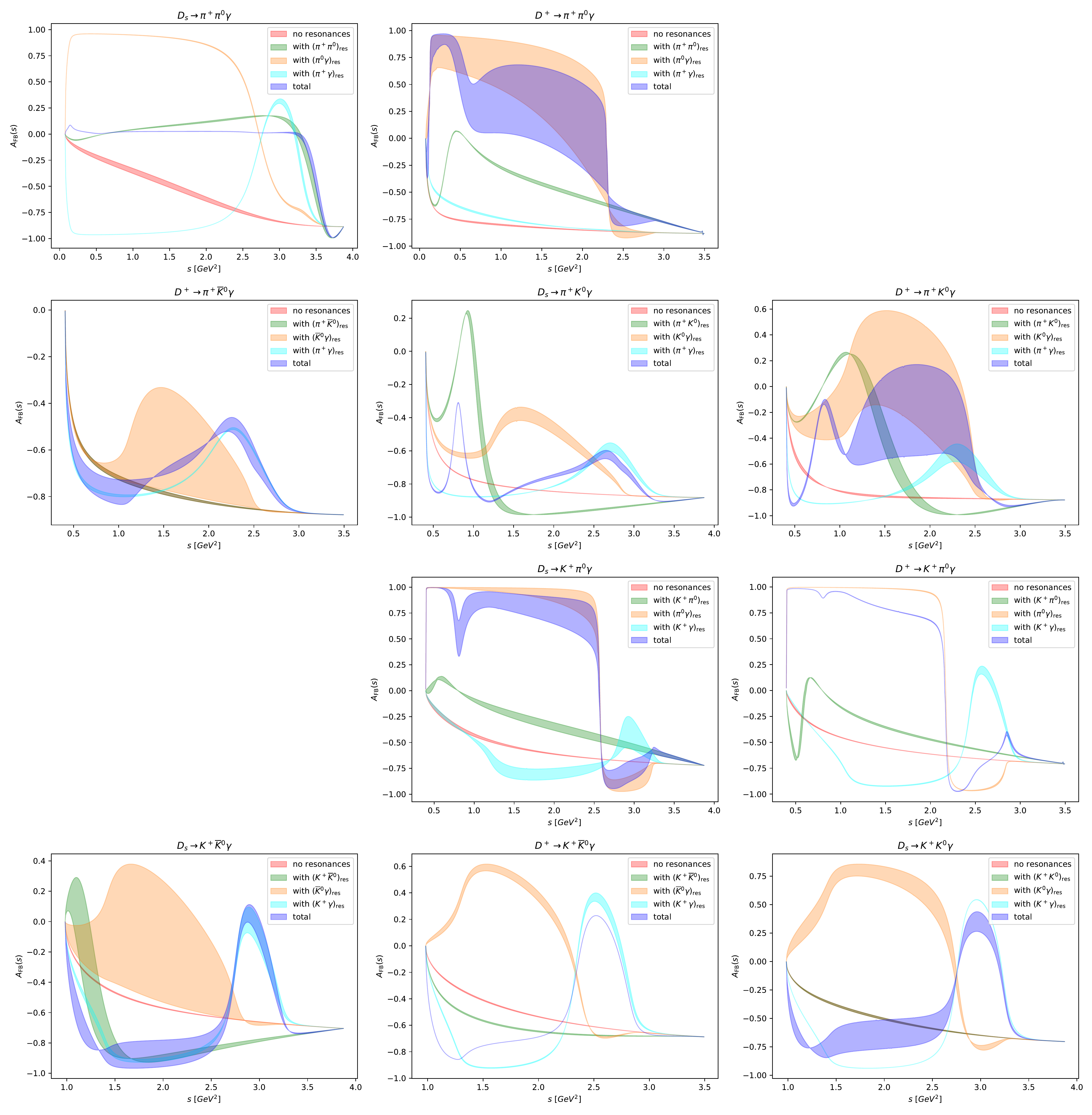}
  \caption{The SM forward backward asymmetry $A_{\rm FB}(s) $ (\ref{eq:AFB}) as a function of $s$ based on HH$\chi$PT. Purely non-resonant contributions are shown by the red bands. The green, orange and light blue bands contain additional contributions of a specific resonance channel. The dark blue bands represent the complete FB asymmetries.
  $A^{\rm SM}_{\rm FB}(s)$ vanishes for the leading order QCDF contribution.
  The first, second and third columns show the CF, SCS and DCS decay modes, respectively. Each row shows the same final state or the final state with $K^0 \leftrightarrow \Kbar^0$.}
  \label{fig:A_FB_SM}
\end{figure}

Angular observables are also suitable for testing QCD models. Here we study $A_{\rm FB}(s)$ as defined in (\ref{eq:AFB}).
Within the SM, QCDF predicts $A_{\rm FB}(s)=0$ since the amplitudes depend only on $s$ (and not on $t,u$). The SM forward-backward asymmetry 
based on HH$\chi$PT is shown in Fig.~\ref{fig:A_FB_SM}. We illustrate 
$A_{\rm FB}(s)$ without contributions from resonances (red), only with 
individual resonances and the complete results (dark blue). The forward-backward 
asymmetry is dominated by resonances in the region of small and 
intermediate $s$. The effect of the $\omega$ is particularly large as 
can be seen for the decays $D^+ \to \pi^+ \pi^0 \gamma$ and 
$D^+_{(s)} \to K^+ \pi^0 \gamma$. The decay $D_s \to \pi^+ \pi^0 \gamma$
has no contribution from the $\omega$ despite the $\pi^0$ in the final 
state. The effects of the $\rho^+$ and $\rho^0$ cancel almost exactly.
Deviations from $A_{\rm FB}(s)=0$ originate from the small difference 
in the $\rho^0 \pi^0 \gamma$ and $\rho^+ \pi^+ \gamma$ coupling and 
from the non-resonant contributions. Since the $s$-channel resonances 
only generate dependencies on $s$, they can significantly reduce 
the asymmetry at the $(P^+ P^0)_\text{res}$ peak for some decay 
channels. In some cases, however, the interference term can even increase 
the asymmetry. There are no $s$-channel resonances for $D^+ \to \pi^+ \Kbar^0 \gamma$
and $D_s \to K^+ K^0 \gamma$. The corresponding plots are therefore 
identical to the non-resonant case. 
For the decays $D^+ \to \pi^+ \Kbar^0 \gamma$, $D_s \to K^+ \Kbar^0 \gamma$,
$D^+ \to \pi^+ \pi^0 \gamma$ and $D^+ \to \pi^+ K^0 \gamma$ significant
scale uncertainties arise for the $t$-channel contribution due to the 
interference of the form factors $B$ and $D$. Since the $t$-channel resonances
provide the only large contribution in the forward region for 
small and intermediate $s$, this uncertainty is clearly visible in the plots.
In the high s region, the forward-backward asymmetry 
is dominated by the bremsstrahlung. Complementary to the 
$D^0 \to P^+ P^- \gamma$ decays \cite{Adolph:2020ema}, 
the bremsstrahlung of the $D^+_{(s)}$ decays always results in a 
large asymmetry, since one of the final state mesons is uncharged 
and thus cannot emit a photon.

\subsection{SM CP Asymmetries  \label{sec:SMCP}}

\begin{figure}
  \centering
  \includegraphics[width=0.9\linewidth]{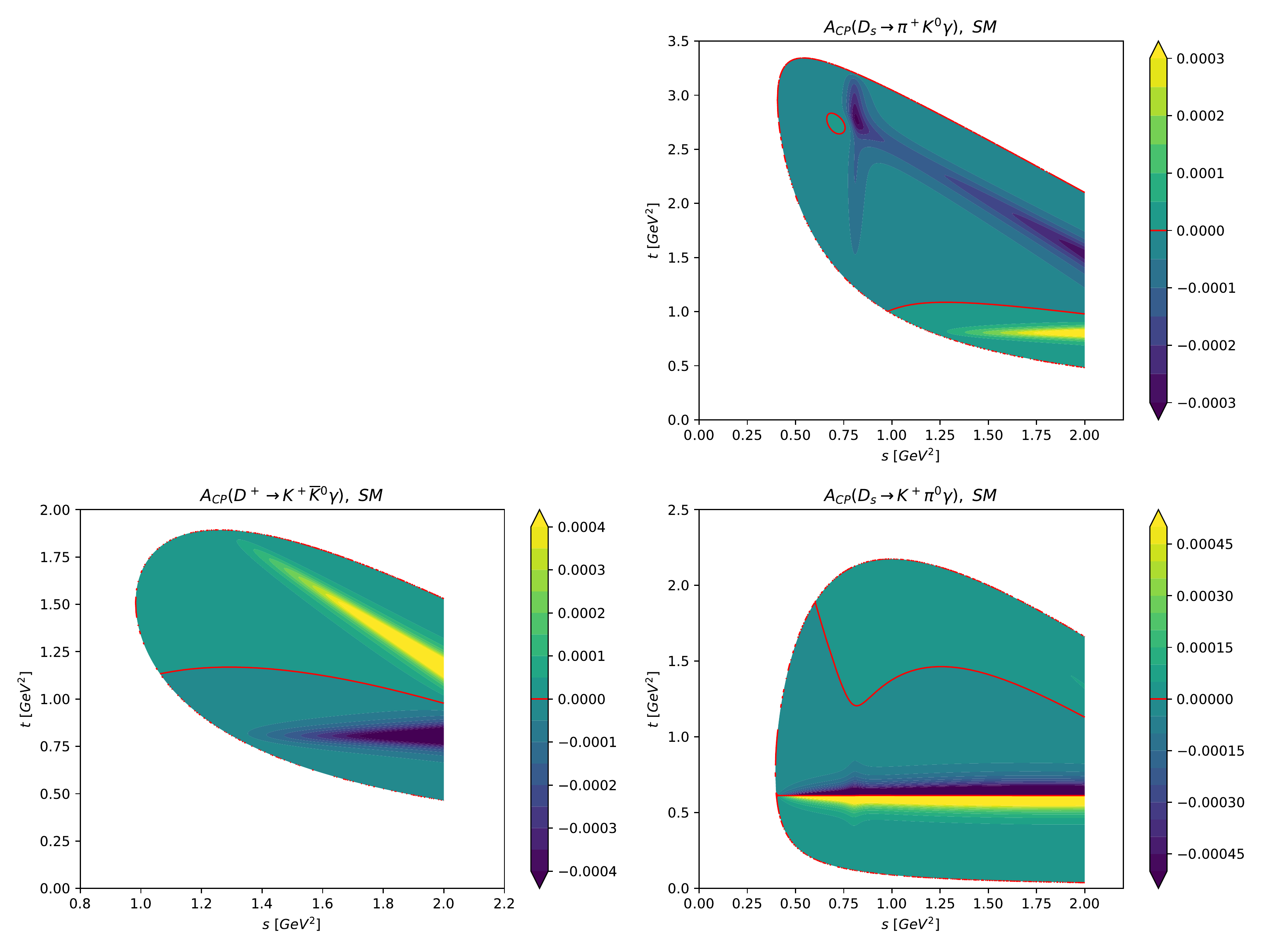}
  \caption{Dalitz plot of  $A_{\rm CP}(s,t)$ for $D_s \to \pi^+ K^0 \gamma$ (upper right), $D^+ \to K^+ \Kbar^0 \gamma$ (lower left) and $D_s \to K^+ \pi^0 \gamma$ (lower right) based on HH$\chi$PT within the SM. We employed a cut $s \leq 2\, \GeV^2$ to avoid large bremsstrahlung contributions in the normalization.
  $A^{\rm SM}_{\rm CP}(s,t)$ vanishes for $D^+ \to \pi^+ \pi^0 \gamma$ and is therefore not shown as the CP asymmetries of the CF and DCS modes.}
  \label{fig:d2A_CP_SM}
\end{figure}

Within the SM, QCDF at leading order implies a vanishing  CP asymmetry 
since the WA amplitude contains only one weak phase. On the other hand, 
HH$\chi$PT predicts non-zero CP asymmetries for the SCS modes $D^+ \to K^+ \Kbar^0 \gamma$,
$D_s \to K^+ \pi^0 \gamma$ and $D_s \to \pi^+ K^0 \gamma$ 
as shown in the Dalitz plots in Fig.~\ref{fig:d2A_CP_SM}. For the 
first two decays, it can be seen that there are large cancellations
when integrating over $t$. This leads to rather small single-differential 
CP-asymmetries of $|A_{\text{CP}}(s)| \lesssim 1.8 \cdot 10^{-5}$ and 
$|A_{\text{CP}}(s)| \lesssim 4.5 \cdot 10^{-5}$, respectively. For 
$D_s \to \pi^+ K^0 \gamma$, there are also cancellations between the 
$t$- and $u$-channel resonance. However, significantly larger values of 
$|A_{\text{CP}}(s)| \lesssim 2.5 \cdot 10^{-4}$ can be seen at the $(P^+P^0)_{\text{res}}$
peak. The SM CP asymmetry for $D^+ \to \pi^+ \pi^0 \gamma$ vanishes in the isospin
limit.
The CP asymmtries of the CF and DCS modes vanish as there is only one weak (CKM) phase involved.

\section{BSM  signatures \label{sec:BSM}}

We study the impact of BSM physics for the SCS decays. We consider contributions of the electromagnetic dipole operators
\begin{align} 
  \begin{split} 
    &\A_-^{\rm BSM} = i \frac{G_F e}{\sqrt{2}} \frac{m_c }{4\pi^2 }(C_7 + C_7^\prime)\frac{(b^\prime - a^\prime)}{v\cdot k} \, ,\\
    &\A_+^{\rm BSM} = \frac{G_F e}{\sqrt{2}} \frac{m_c m_D}{2\pi^2 } (C_7 - C_7^\prime) h^\prime\, . \label{eq:BSM-Amplituden}
  \end{split}
\end{align}
Corresponding contributions within the SM can be neglected due to the strongly suppressed Wilson 
coefficients. On the other hand, $C_7^{(\prime)}$ can reach $\O(0.1)$ values through BSM 
physics. Model-independent analyses of $D \to \rho^0 \gamma$ and
$D \to \pi \ell \ell$ decays yield the constraints 
\cite{deBoer:2018buv, Abdesselam:2016yvr, Bause:2019vpr}
\begin{align} \label{eq:max}
  |C_7|, |C_7^\prime| \lesssim 0.3\, .
\end{align}
The tensor current form factors $a^\prime$, $b^\prime$, $c^\prime$
and $h^\prime$ are defined in appendix \ref{app:HQCHPT form factors}.

In the following we discuss the NP impact 
on branching ratios (Sec.~\ref{sec:bsm-br}), the forward-backward asymmetry
(Sec.~\ref{sec:bsm-afb})
and the CP-asymmetry (Sec.~\ref{sec:CP}).

\subsection{BSM effects in the branching ratios \label{sec:bsm-br}}

\begin{figure}
  \centering
  \includegraphics[width=0.9\linewidth]{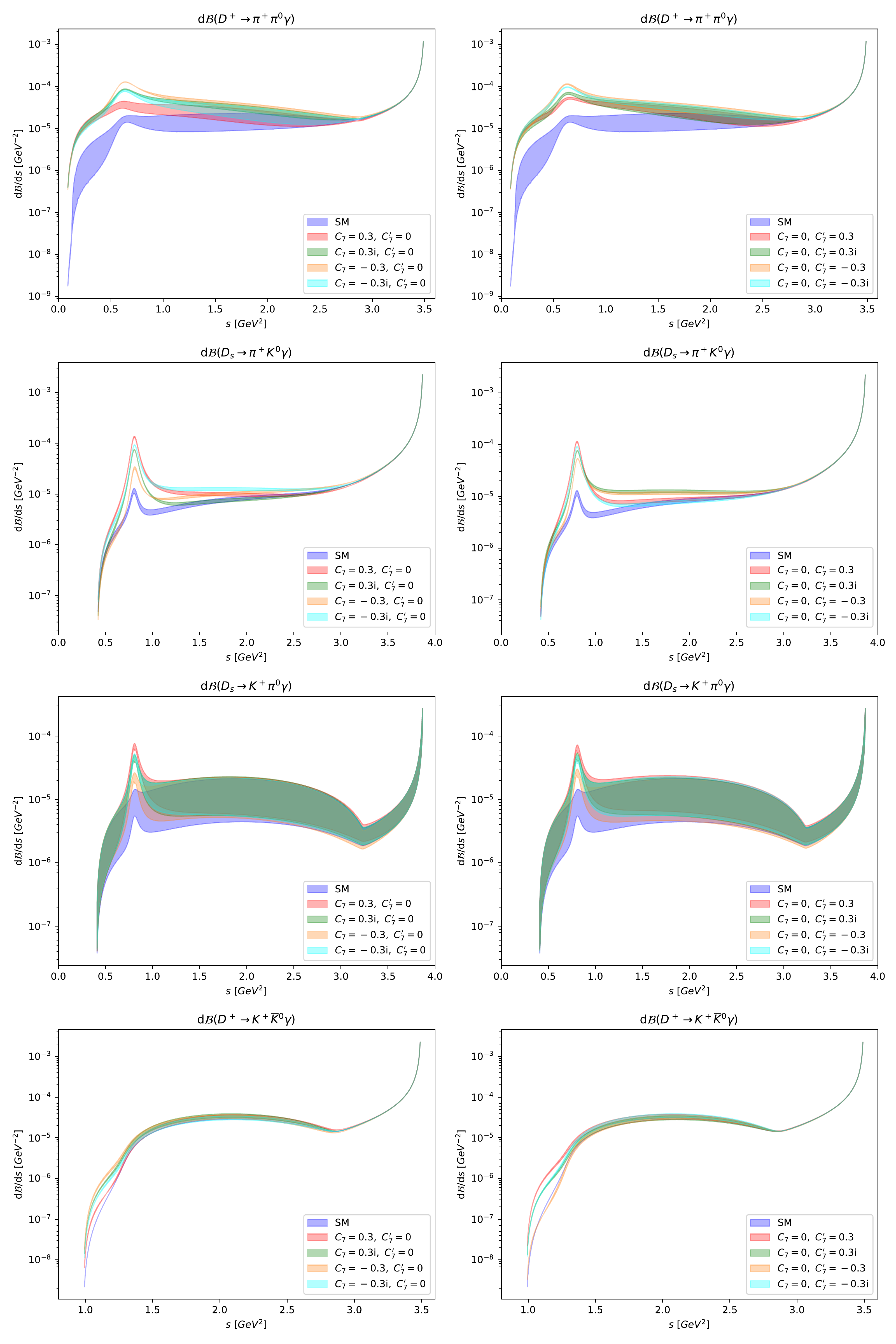}
  \caption{Differential branching ratios for several BSM scenarios based on HH$\chi$PT. We exhaust the limit of one BSM coefficient \eqref{eq:max} and set the other one to zero. We overlay distributions obtained for   CP-phases set to $0, \pm \pi/2$ and  $\pi$.}
   \label{fig:d1BR_HQCHIPT_BSM}
\end{figure}

In Fig.~\ref{fig:d1BR_HQCHIPT_BSM} 
we show a comparison of SM predictions (blue) and different BSM 
scenarios of the FCNC modes based on HH$\chi$PT.
We set one of the BSM coefficients $C_7^{(\prime)}$ to zero and 
exhaust the limit \eqref{eq:max} of the other one. The CP-phases 
are set to $0, \pm \pi/2, \pi$. The differential branching ratios 
can be increased by one order of magnitude at the $s$-channel peak.
As noted previously, the decay $D_s \to K^+ \Kbar^0 \gamma$  is an exception. Since 
the $\rho^+$ resonance is outside of the phase space, the BSM contributions
are negligibly small in the entire phase space. 
For QCDF, the largest deviation between
SM and BSM scenarios arise above  the $s$-channel peak. 
The deviations are below one order of magnitude for $\lambda_{D_{(s)}}=0.1 \GeV$. At the $s$-channel peak, the 
deviations are significantly smaller. Especially for $C_7^\prime = 0$,
the SM predictions and BSM scenarios hardly differ from each other.
Since the deviations between SM and BSM are not very large and
sizable $t$- and $u$-channel contributions are to be expected beyond 
the $s$-channel peak, which are not taken into account in leading order 
QCDF, we do not show plots of differential branching ratios within QCDF.
In Table~\ref{tbl:branching_ratios_BSM} we give (integrated) branching ratios  within QCDF and HH$\chi$PT
for the FCNC modes. We employ the same phase space cuts as in 
Table~\ref{tbl:branching_ratios_CF}. 
We conclude that the (differential) branching ratios of SCS decays are affected by BSM physics, 
however, are not sufficiently clean to unambiguously signal NP.

\begingroup
\renewcommand{\arraystretch}{1.4}
\begin{table}
  \centering
  \begin{tabular}{l|c|c|c|c}
                                                               & $D^+ \to \pi^+ \pi^0 \gamma$ & $ D_s \to \pi^+ K^0 \gamma$ & $D_s \to K^+ \pi^0 \gamma$ & $D^+ \to K^+ \Kbar^0 \gamma$ \\
    \hline
    $\text{QCDF}\big|^\text{SM}_{s \leq 1.5~\GeV^2} $             & $(2.1 - 2.4)\cdot 10^{-4}$ & $(0.9 - 1.1)\cdot 10^{-4}$ & $(5.0 - 5.8)\cdot 10^{-5}$ & $(1.3 - 1.5)\cdot 10^{-6}$ \\
    $\text{HH$\chi$PT}\big|^\text{SM}_{s \leq 1.5~\GeV^2} $       & $(1.0 - 2.2)\cdot 10^{-5}$ & $(4.4 - 5.5)\cdot 10^{-6}$ & $(0.3 - 1.4)\cdot 10^{-5}$ & $(1.8 - 2.3)\cdot 10^{-6}$ \\
    $\text{HH$\chi$PT}\big|^\text{SM}_{E_\gamma \geq 0.1~\GeV} $  & $(3.0 - 5.4)\cdot 10^{-5}$ & $(2.8 - 3.1)\cdot 10^{-5}$ & $(1.0 - 4.3)\cdot 10^{-5}$ & $(3.8 - 4.6)\cdot 10^{-5}$ \\
    \hline
    $\text{QCDF}\big|^\text{BSM}_{s \leq 1.5~\GeV^2} $            & $(1.2 - 4.0)\cdot 10^{-4}$ & $(0.5 - 1.8)\cdot 10^{-4}$ & $(2.7 - 9.5)\cdot 10^{-5}$ & $(0.5 - 3.1)\cdot 10^{-6}$ \\
    $\text{HH$\chi$PT}\big|^\text{BSM}_{s \leq 1.5~\GeV^2} $      & $(2.7 - 7.7)\cdot 10^{-5}$ & $(0.9 - 2.3)\cdot 10^{-5}$ & $(0.5 - 2.3)\cdot 10^{-5}$ & $(1.6 - 3.1)\cdot 10^{-6}$ \\
    $\text{HH$\chi$PT}\big|^\text{BSM}_{E_\gamma \geq 0.1~\GeV} $ & $(0.5 - 1.3)\cdot 10^{-4}$ & $(3.5 - 5.0)\cdot 10^{-5}$ & $(1.3 - 5.5)\cdot 10^{-5}$ & $(3.7 - 5.0)\cdot 10^{-5}$
  \end{tabular}
  \caption{Branching ratios for the BSM sensitive SCS decays in the SM (top entries) and with BSM physics (lower entries). For the BSM branching ratios, we employed the same 
  general scenarios as for Fig.~\ref{fig:d1BR_HQCHIPT_BSM}.
  The branching ratios are given in the region of applicability of QCDF $s \lesssim 1.5\, \GeV^2$ for QCDF and HH$\chi$PT to enable a comparison of both models. Additionally, HH$\chi$PT predictions are given for $E_\gamma \geq 0.1\, \GeV$, see text for details. The QCDF branching ratios are obtained for $\lambda_{D_{(s)}}=0.1\, \GeV$ and are $\propto (0.1\, \GeV/\lambda_{D_{(s)}})^2$ in the SM.}
  \label{tbl:branching_ratios_BSM}
\end{table}
\endgroup

\subsection{BSM effects in $A_{\rm FB}$ \label{sec:bsm-afb} }

\begin{figure}
  \centering
  \includegraphics[width=0.9\linewidth]{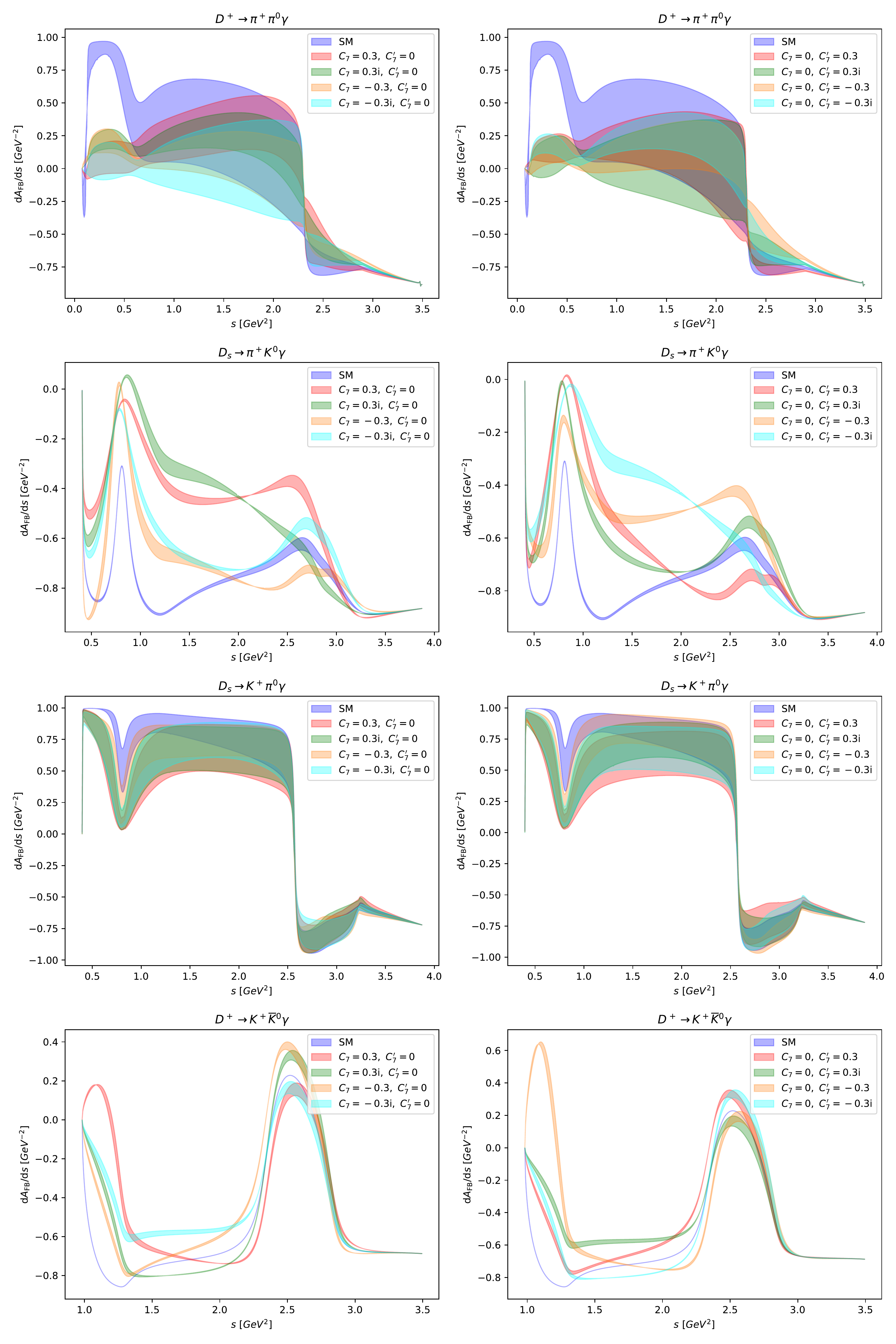}
  \caption{A comparison of SM and BSM FB asymmetries based on HH$\chi$PT. We show the same BSM scenarios as for the branching ratios in Fig.~\ref{fig:d1BR_HQCHIPT_BSM}}
  \label{fig:A_FB_HHCHIPT}
\end{figure}

We investigate $A_{\rm FB}$ based 
on HH$\chi$PT in Fig.~\ref{fig:A_FB_HHCHIPT}  in  the same BSM scenarios as those considered in Sec.~\ref{sec:bsm-br}.
Since the dipole operators $O_7^{(\prime)}$ induce a $t$-dependence
only by non-resonant contributions, there are only small forward-backward
asymmetries for SM QCDF contributions. For $\lambda_{D_{(s)}}=0.1 \, \GeV$,
$A_{\rm FB}(s)$ reach values of $\O(10^{-3})-\O(10^{-2})$. For $\lambda_{D_{(s)}}=0.3 \, \GeV$,
the SM contribution is significantly reduced, leading to larger asymmetries.
These are typically in the range of $\O(10^{-2})$, but can occasionally 
reach larger values of $|A_{\rm FB}(s)| \lesssim 0.15$.
The minor $t$-dependence of the BSM contributions can lead to a significant suppression of
the substantial SM asymmetries predicted by HH$\chi$PT.
These are considerably reduced, especially in the region of the 
$s$-channel resonance. This becomes particularly evident in case of 
$D^+ \to \pi^+ \pi^0 \gamma$.

Similarly to the decays of neutral $D$ mesons, sizable effects of 
the dipole operators can be seen for differential branching ratios 
and forward backward asymmetries \cite{Adolph:2020ema}. However, 
it is difficult to claim sensitivity to NP due to the uncertainties
of the leading order calculation and the intrinsic uncertainty of 
the Breit-Wigner contributions. Nevertheless, these observables 
are suitable for testing the various QCD models with SM-like decays 
and for understanding the decay mechanisms. Due to the small 
CP-violating phases in the charm sector of the SM, 
CP-asymmetries, discussed in the next section, have the best 
sensitivity to NP.

\subsection{BSM CP violation  \label{sec:CP}}

The most promissing observable to test  BSM physics
is the single- or double-differential CP asymmetry defined in (\ref{eq:ACP}).
Note that we perform a cut $s \leq 2\, \GeV^2$
for HH$\chi$PT to avoid  large bremsstrahlung contributions  in 
the normalization. For QCDF, we include the contributions with $s \leq 1.5\,\GeV^2$.

\begin{figure}
  \centering
  \includegraphics[width=0.9\linewidth]{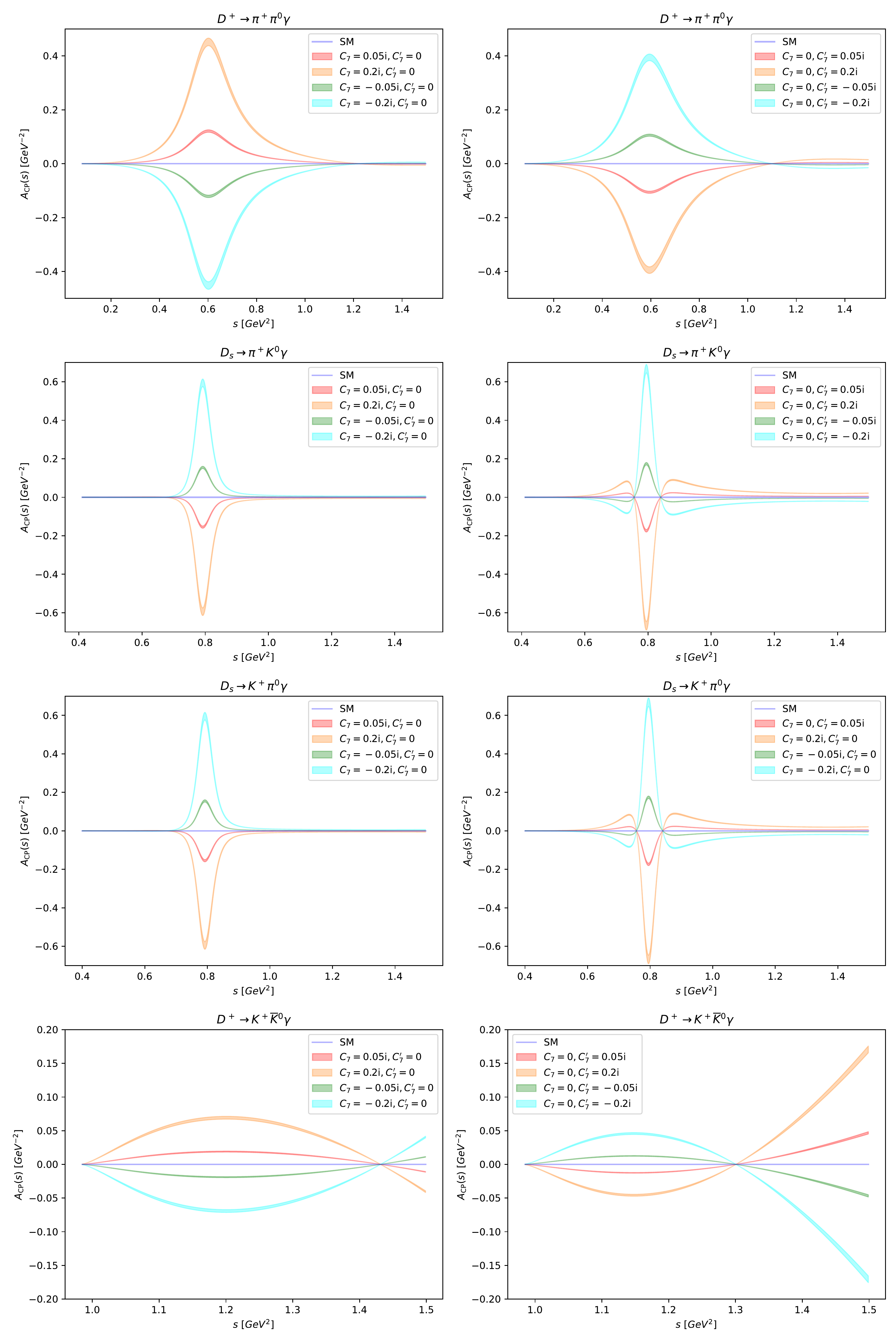}
  \caption{The CP asymmetry as a function of $s$, based on SM QCDF and (\ref{eq:BSM-Amplituden}). 
  Plots in the left column correspond to $C_7=\pm   0.05\, i$ or $\pm  0.2\, i$ and $C_7^{\prime}=0$.
  Plots in the right column correspond to $C_7^\prime=\pm  0.05\, i$ or $\pm  0.2\, i$ and $C_7=0$.
   We performed a cut $s \leq 1.5\, \GeV^2$  to remain within the region where QCDF applies.}
  \label{fig:dA_CP_QCDF}
\end{figure}

Considering possible BSM contributions in the electromagnetic dipole 
operator, all FCNC decay modes can exhibit sizable CP-asymmetries.
In Fig.~\ref{fig:dA_CP_QCDF} we show $A_{\text{CP}}(s)$ 
for different BSM scenarios. We set one of the BSM coefficients to
zero and the other one to $0.05i $ or $0.2i$. 
To maximize $A_{\text{CP}}(s)$,
we choose $C_7^{(\prime)}$ to be purely imaginary.
Since the QCDF amplitude does not contain effects of the $t$- and 
$u$-channel resonances, significant strong phases and thus also 
CP asymmetry only arise at the $(P^+P^0)_{\text{res}}$ peak. 
Therefore, we do not show Dalitz plots for QCDF. Within these BSM scenarios, 
$A_{\text{CP}}(s)$ can reach $\mathcal{O}(0.01)$ values for 
$D^+ \to K^+ \Kbar^0 \gamma$ and $\mathcal{O}(0.1)$ values otherwise.
The CP asymmetries for $D_s \to \pi^+ K^0 \gamma$ and 
$D_s \to K^+ \pi^0 \gamma$ are almost identical, since the amplitudes 
differ basically only by an isospin factor of $-1/\sqrt{2}$. Minor
deviations are caused by different momenta in the heavy meson propagators
of the tensor form factors and the different shape of the phase space.

\begin{figure}
  \centering
  \includegraphics[width=0.9\linewidth]{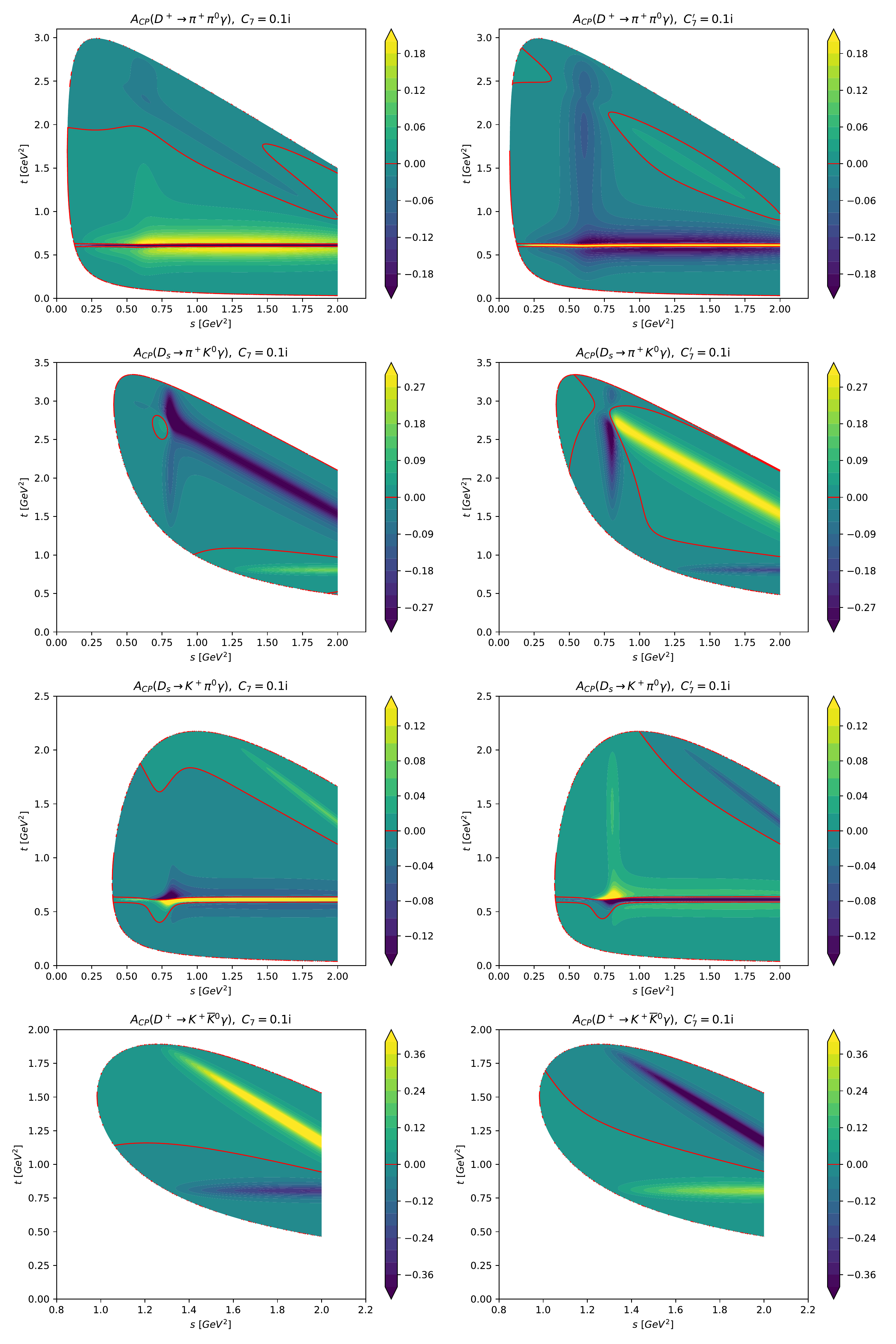}
  \caption{Dalitz plot of  $A_{\rm CP}(s,t)$ for $D^+ \to \pi^+ \pi^0 \gamma$ (first row), $D_s \to \pi^+ K^0 \gamma$ (second row), $D_s \to K^+ \pi^0 \gamma$ (third row) and $D^+ \to K^+ \Kbar^0 \gamma$ (fourth row) based on HH$\chi$PT. 
    Plots to the left (right)  correspond to $C_7= 0.1\, i$ and $C_7^{\prime}=0$ ($C_7^\prime= 0.1\, i$ and $C_7=0$).
  We employed a cut $s \leq 2\, \GeV^2$ to avoid large bremsstrahlung contributions in the normalization.}
  \label{fig:d2A_CP}
\end{figure}

\begin{figure}
  \centering
  \includegraphics[width=0.9\linewidth]{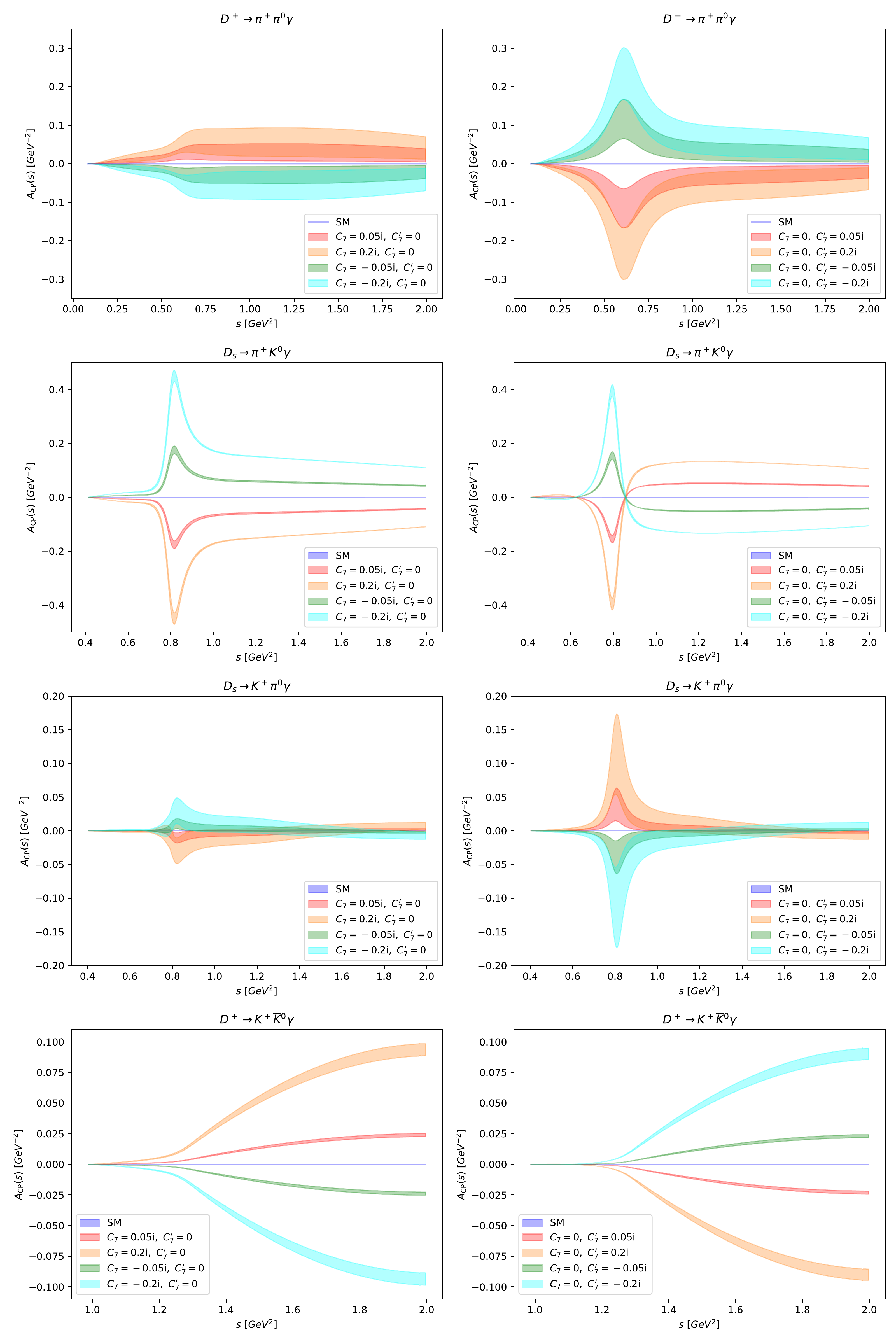}
  \caption{As in Fig.~\ref{fig:dA_CP_QCDF} but for HH$\chi$PT and with cut $s \leq 2\, \GeV^2$ to avoid large bremsstrahlung contributions in the normalization.}
  \label{fig:dA_CP_HHchiPT}
\end{figure}

In Fig.~\ref{fig:d2A_CP} we show Dalitz plots for the double 
differential CP asymmetry for HH$\chi$PT where we set one of 
the coefficients $C_7^{(\prime)}$ to $0.1i$. It can be seen 
that the values for $A_{\text{CP}}(s,t)$ are increased by a factor 
of $\sim 10^3$ compared to the SM. Furthermore, the single differential 
CP asymmetries are shown in Fig.~\ref{fig:dA_CP_HHchiPT} for the same 
BSM scenarios as for QCDF. For $D^+ \to \pi^+ \pi^0 \gamma$ and 
$D_s \to K^+ \pi^0 \gamma$, the parity even and parity odd 
amplitudes contribute to the $s$-channel peak to the same extend.
The relative sign between $C_7$ and $C_7^\prime$ in 
\eqref{eq:BSM-Amplituden} results in a cancellation
for $C_7$ and a constructive increase for $C_7^\prime$, respectively.
This characteristic feature can also be observed in 
$D^0 \to \pi^+ \pi^- \gamma$ \cite{Adolph:2020ema}. In contrast,
for $D_s \to \pi^+ K^0 \gamma$ the $s$-channel peak is dominated 
by $\A_-$. Complementary, the tails, generated by the $t$- and 
$u$-channel resonance, change sign in the two BSM scenarios. 
Therefore, it is dominated by $\A_+$. Since the $t$- and 
$u$-channel resonances are different vector mesons, the 
cancellation is not as effective as for $D^0 \to P^+ P^- \gamma$ decays
\cite{Adolph:2020ema}.

We stress that QCD renormalization-group evolution connects the 
electromagnetic and chromomagnetic dipole operators. Therefore, 
data on $\Delta A_\text{CP}= A_\text{CP}(D^0 \to K^+ K^-) - A_\text{CP}(D^0 \to \pi^+ \pi^-)$
can constrain the phase of the photon dipoles  as $|\text{Im}(C_7^{(\prime)})| \lesssim 2 \cdot 10^{-3}$
in several BSM models, see \cite{Adolph:2020ema} for further details.
Since $A_\text{CP}$ scales approximately
linear with $\text{Im}(C_7^{(\prime)})$ for $|C_7^{(\prime)}| \lesssim 0.1$ this would lead to a suppression factor of $50$ relative to the asymmetries shown in 
Fig.~\ref{fig:d2A_CP}.
However, resulting CP-asymmetries reach few permille level and are more than one order of magnitude larger than the ones in the the SM.

For larger photon dipole coefficients $|C_7^{(\prime)}| \gtrsim 0.1$, the BSM contributions lead to enlarged branching ratios for 
$D^+ \to \pi^+ \pi^0 \gamma$, $D_s \to \pi^+ K^0 \gamma$ and $D_s \to K^+ \pi^0 \gamma$, as shown in Fig.~\ref{fig:d1BR_HQCHIPT_BSM}.
This corresponds to  an increase of the normalization of the CP asymmetry (\ref{eq:ACP}).
Thus, for sizable $|C_7^{(\prime)}| \gtrsim 0.1$, the CP asymmetry no longer scales linearly with $\text{Im}(C_7^{(\prime)})$, which can also be seen for $A_{\text{CP}}(s)$ in Fig.~\ref{fig:dA_CP_HHchiPT}.

\FloatBarrier
 
\section{Summary \label{sec:con}}

We analyzed ten radiative three-body decays of charged, charmed mesons $D^+_{(s)} \to P^+ P^0 \gamma$  in the standard model and beyond.
This work complements earlier works on neutral meson decays \cite{Adolph:2020ema}.
The decay amplitudes and distributions are computed in QCDF, HH$\chi$PT, and in the region of high $PP$-invariant mass, using the soft photon approximation.
The DCS and CF modes (\ref{eq:all}) are SM-like and probe the QCD dynamics. Branching ratios are shown in Fig.~\ref{fig:SM_dBr_Vergleich} and compared in 
Table~\ref{tbl:branching_ratios_CF} and \ref{tbl:branching_ratios_DCS}.
As in \cite{Adolph:2020ema} the forward-backward asymmetry (\ref{eq:AFB}) is an observable that efficiently differentiates between predictions of lowest order QCDF, 
weak annihilation-type contributions with $s$-channel dependence only,
and HH$\chi$PT, subject to more complex resonance structures.
An understanding of the dominant decay dynamics can therefore be achieved from experimental study, and increases the sensitivity of the NP searches with
the SCS modes.

The SCS modes are sensitive to $|\Delta c|=|\Delta u|=1$ effects from BSM physics encoded in electromagnetic dipole couplings $C_7$ and $C_7^\prime$.
Branching ratios are in the $\sim 10^{-5}-10^{-4}$ range, except for $D^+ \to K^+ \bar K^0 \gamma$ in QCDF, which is about one  order of magnitude smaller due to smaller phase space, see Table.~\ref{tbl:branching_ratios_BSM}.
Not unexpected, we find that NP effects cannot be cleanly separated from the SM background in the  branching ratio nor its distribution.
On the other hand, $A_{\rm FB}$ allows for qualitatively different distributions, and to signal NP, see Fig.~\ref{fig:A_FB_HHCHIPT}.
The most clear-cut signals of NP are possible in CP-asymmetries, ideally in the Dalitz region, as in Fig.~\ref{fig:d2A_CP}, but also in single differential distributions,
see Fig.~\ref{fig:dA_CP_HHchiPT}.
The CP asymmetries can be sizable around resonance peaks, and reach ${\cal{O}}(0.1)$.

We conclude pointing out opportunities. The
best decay channels for 
\begin{itemize}
 \item testing QCD frameworks: The CF mode $D^+ \to \pi^+ \bar K^0 \gamma$ because it has no leading order QCDF contribution. The same is true also the DCS mode $D_s \to K^+ K^0 \gamma$.
  \item testing the QCD frameworks with $A_\text{FB}$, see Fig.~\ref{fig:A_FB_SM}: $D^0 \to \pi^0 \Kbar^0 \gamma$ (CF), $D_s \to \pi^+ \pi^0 \gamma$ (CF) and $D^+ \to K^+ \pi^0 \gamma$ (DCS), which feature small uncertainties and a very distinctive shape which is reasonably well understood.
  \item testing the SM with $A_\text{FB}$, see Fig.~\ref{fig:A_FB_HHCHIPT}: $D^0 \to \pi^+ \pi^- \gamma$, $D^+ \to \pi^+ \pi^0 \gamma$ and $D_s \to \pi^+ K^0 \gamma$ because differences between SM and BSM asymmetries in the other decays are small.
  \item testing the SM with $A_\text{CP}$: Here one should distinguish between the single and double differential CP-asymmetries. Dalitz plots (Fig.~\ref{fig:d2A_CP}) are suitable for all SCS decay channels. For $A_\text{CP}(s)$ (Fig.~\ref{fig:dA_CP_QCDF}, \ref{fig:dA_CP_HHchiPT}), the decays $D^+ \to \pi^+ \pi^0 \gamma$ and $D_s \to \pi^+ K^0 \gamma$ can be emphasized, because they are sensitive to both BSM coefficients $C_7$ and $C_7^\prime$ and exhibit less cancellations between the $t$ and $u$ channel resonances. $D^0 \to K^+ K^- \gamma$ is also a good option in the region of the $\Phi$ resonance. $D^0 \to \pi^+ \pi^- \gamma$, $D^+ \to K^+ \Kbar^0 \gamma$ and $D_s \to K^+ \pi^0 \gamma$ only have a good sensitivity in one of the BSM coefficients or smaller asymmetries have to be expected due to cancellations.
\end{itemize}

%\section*{Acknowledgements}

\appendix

\section{HH$\chi$PT form factors}
\label{app:HQCHPT form factors}

The $D_{(s)}^+ \to P^+ P^0$ matrix elements of the tensor currents can be parameterized as
\begin{align}
  \braket{P^+(p_1)P^0(p_2)| \ubar \sigma^{\mu \nu} k_\mu (1\pm \gamma_5) c| D^+_{(s)}(P)} = m_D \left[a^\prime p_1^\nu + b^\prime p_2^\nu + c^\prime P^\mu \mp 2 i  h^\prime \epsilon^{\nu\alpha\beta\gamma} p_{1\alpha} p_{2\beta} k_\gamma\right]\, . \label{eq:Tensor_Formfaktoren}
\end{align}
The form factors $a^\prime, b^\prime, c^\prime, h^\prime$ depend on $s$ and $t$ and satisfy
\begin{align}
a' p_1 \cdot k + b' p_2 \cdot k + c' P \cdot k=0 \, . 
\end{align}
The numerical values for the parameters in the form factors are given in appendix A of \cite{Adolph:2020ema}.

\begin{figure}
  \centering
  \includegraphics[width=0.8\linewidth]{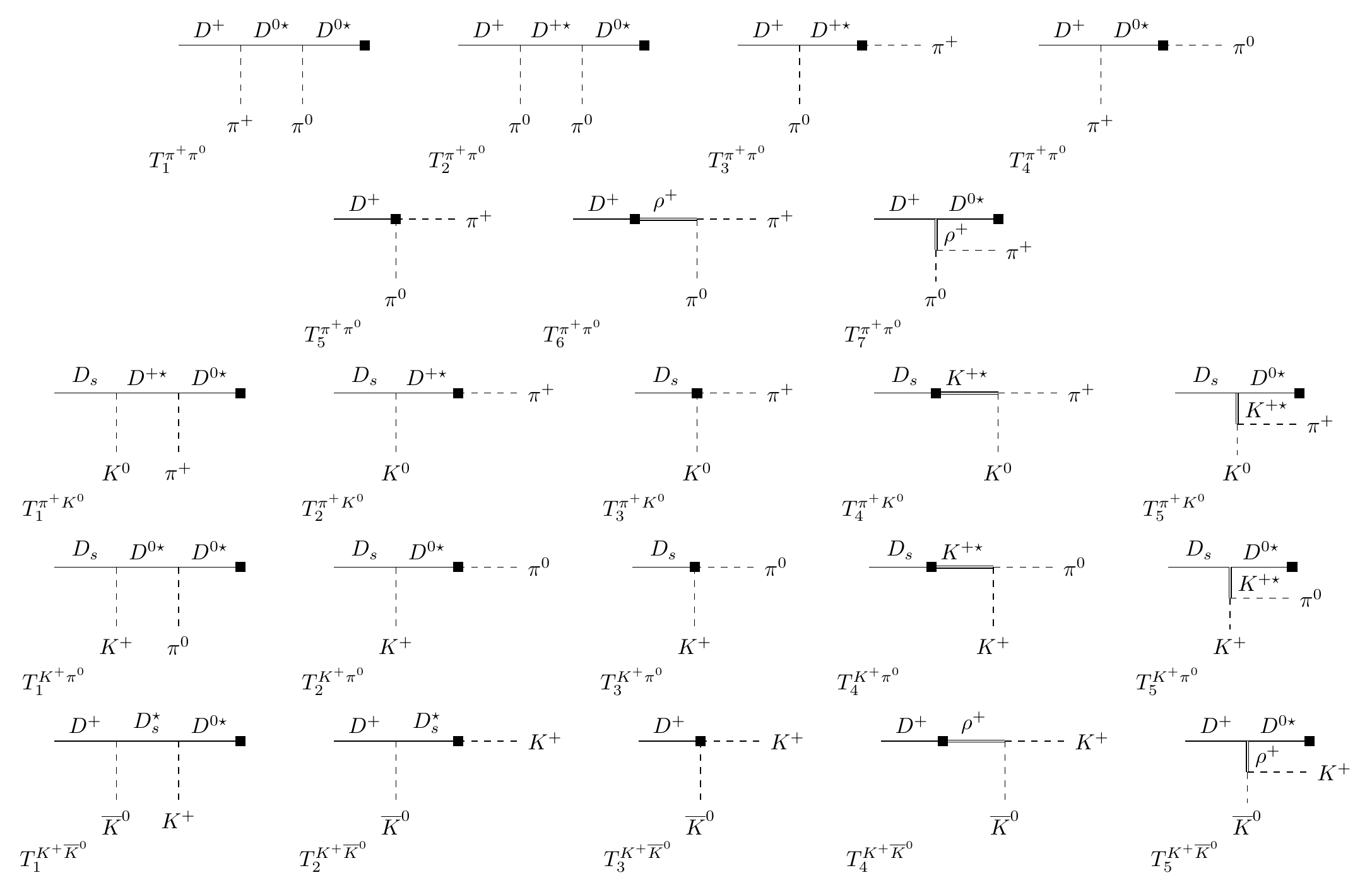}
  \caption{Feynman diagrams which contribute to the tensor form factors.}
  \label{fig:Diagramme_Tensorstrom}
\end{figure}

\subsection{Cabibbo-favored decay modes}

\noindent\underline{$D_s \to \pi^+ \pi^0 \gamma$}
\begin{figure}
  \centering
  \subfigure[Contributions to the parity-even form factors $A$ and $E$. Additionally, for each of the diagrams $A_{1,2}$, $A_{1,3}$ and $A_{2,3}$ there is another one where the photon is coupled via a vector meson.]{\includegraphics[width=0.45\linewidth]{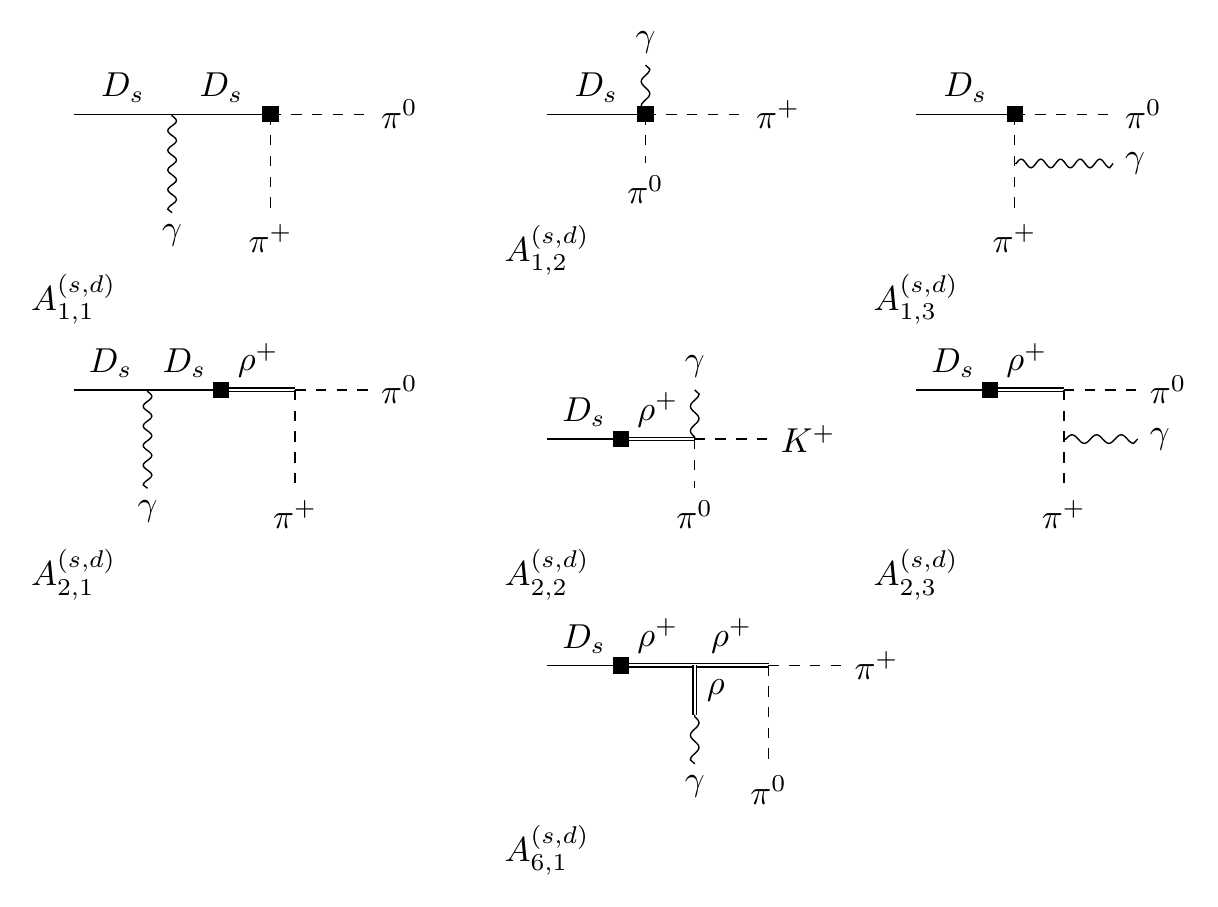}}
  \hfill
  \subfigure[Contributions to the parity-odd form factors $B$ and $D$.]{\includegraphics[width=0.45\linewidth]{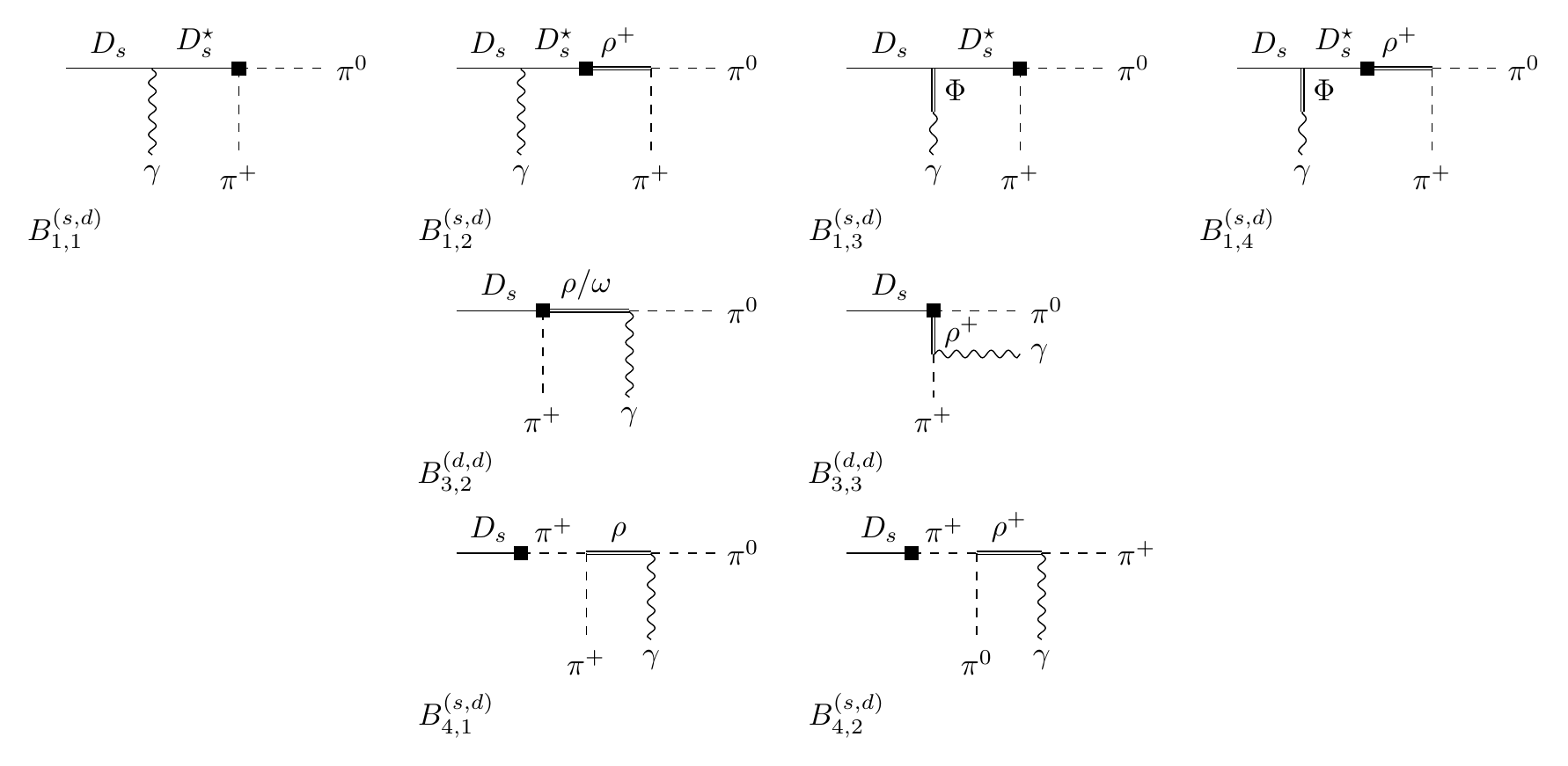}}
  \caption{Feynman diagrams contributing to the decay $D_s \to \pi^+ \pi^0 \gamma$ within the SM.}
  \label{fig:Diagramme_Ds_pipluspi}
\end{figure}
\begin{align}
    A_{1+2}^{(s,d)} &=  i \sqrt{2}f_{D_s}\frac{v \cdot p_2 - v \cdot p_1 - v \cdot k}{(v \cdot k)(p_1 \cdot k)}\\
    A_6^{(s,d)} &= i 2\sqrt{2} f_{D_s} \frac{p_1 \cdot p_2}{m_{D_s}(v \cdot k)} BW_{\rho^+}(p_1+p_2)
\end{align}
\begin{align}
  \begin{split}
    B_1^{(s,d)} &= \frac{f_{D_s}}{v \cdot k + \Delta} \left[1 - m_{\rho}^2 BW_{\rho^+}(p_1 +p_2)\right]\left[2\sqrt{2}\lambda^\prime - g_v\lambda\frac{2 g_\Phi}{3m_\Phi^2}\right]
  \end{split}\\
  \begin{split}
    B_3^{(s,d)} &= -\frac{\sqrt{2}f_{D_s} g_\rho}{f_\pi}\left(g_{\rho \pi \gamma}BW_\rho(p_2 + k) + g_{\rho^\pm \pi^\pm \gamma} BW_{\rho^+}(p_1 + k)\right)
  \end{split}\\
  B_4^{(s,d)} &= \frac{\sqrt{2}m_{D_s}^2 f_{D_s} f_\pi m_\rho^2}{g_\rho (m_{D_s}^2 - m_\pi^2)}\left(g_{\rho \pi \gamma}BW_\rho(p_2+k) + g_{\rho^\pm \pi^\pm \gamma}BW_{\rho^+}(p_1+k)\right)
\end{align}

\noindent\underline{$D_s \to K^+ \Kbar^0 \gamma$}

\begin{figure}
  \centering
  \subfigure[Contributions to the parity-even form factors $A$ and $E$. Additionally, for each of the diagrams $A_{1,2}$, $A_{1,3}$, $A_{2,3}$, $E_{1,2}$ und $E_{2,3}$ there is another one where the photon is coupled via a vector meson.]{\includegraphics[width=0.45\linewidth]{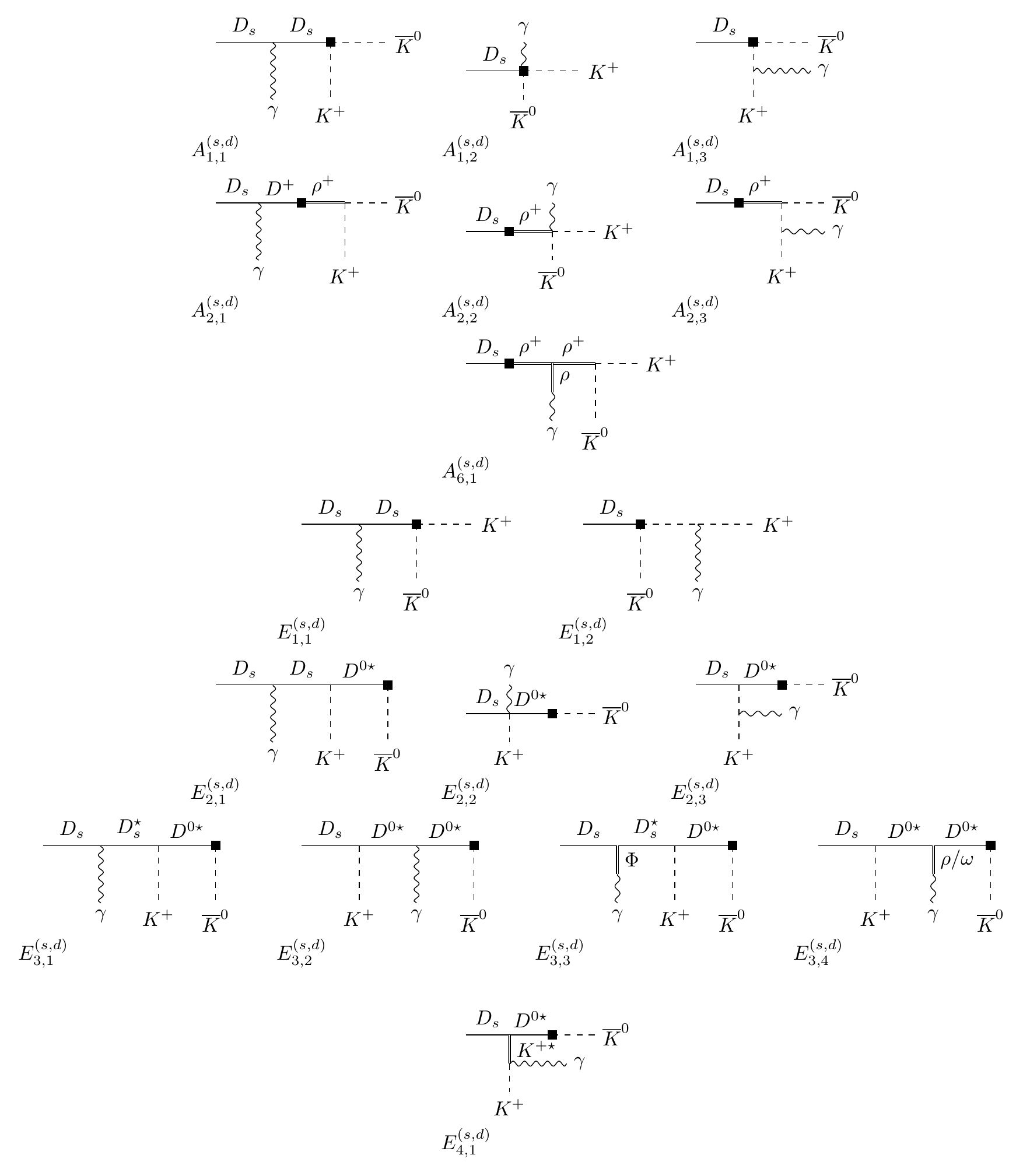}}
  \hfill
  \subfigure[Contributions to the parity-odd form factors $B$ and $D$.]{\includegraphics[width=0.45\linewidth]{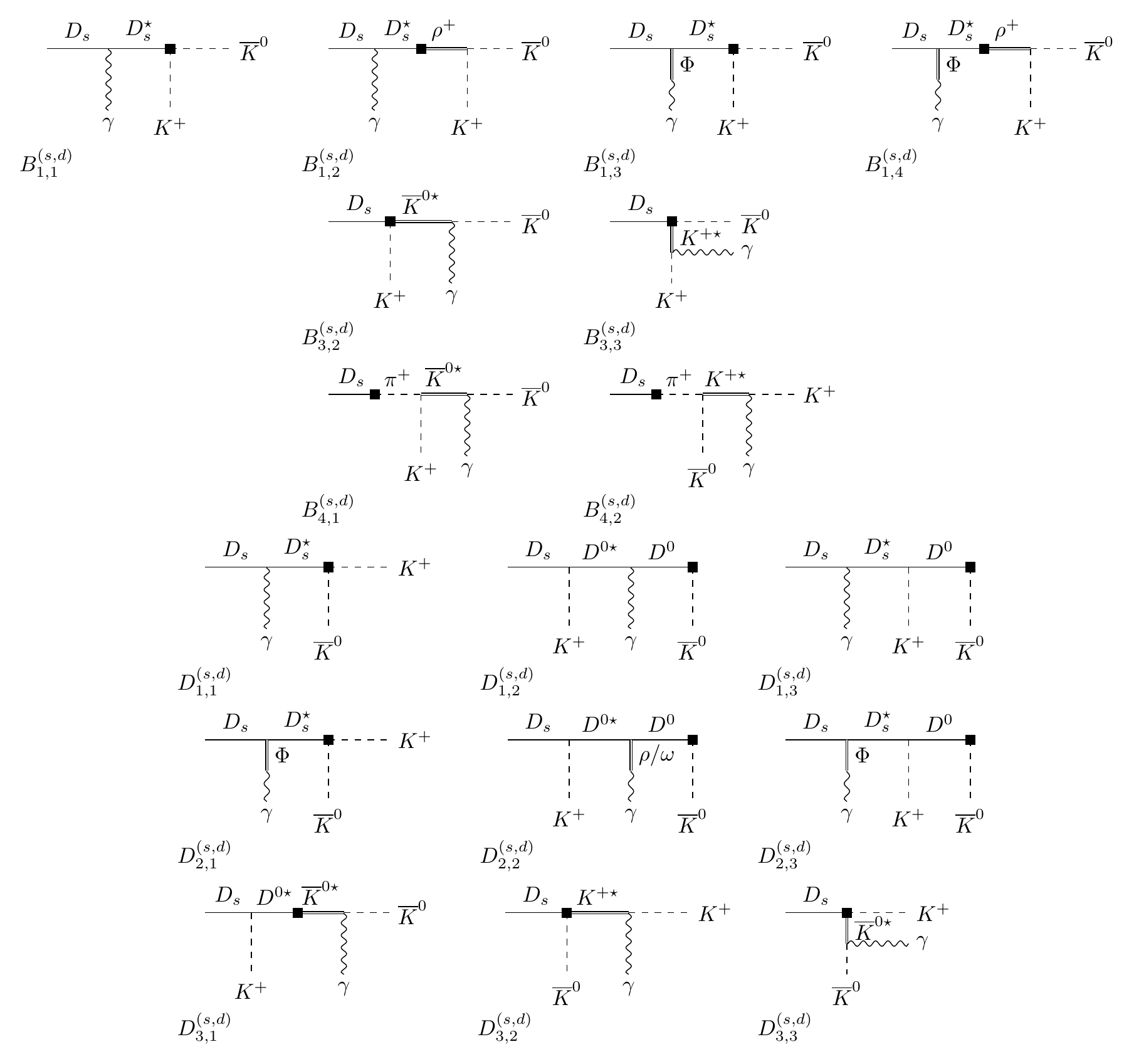}}
  \caption{Feynman diagrams contributing to the decay $D_s \to K^+ \Kbar^0 \gamma$ within the SM.}
  \label{fig:Diagramme_Ds_KplusKbar}
\end{figure}

\begin{align}
    A_{1+2}^{(s,d)} &= -i f_{D_s}\frac{v \cdot p_2 - v \cdot p_1 - v \cdot k}{(v \cdot k)(p_1 \cdot k)}\\
    A_6^{(s,d)} &= -i 2 f_{D_s} \frac{p_1 \cdot p_2}{m_{D_s}(v \cdot k)} BW_{\rho^+}(p_1+p_2)
\end{align}

\begin{align}
  E_1^{(s,d)} &= -i f_{D_s}\frac{v \cdot p_2}{(v \cdot k)(p_1 \cdot k)}\\
  E_2^{(s,d)} &= -i \sqrt{\frac{m_D}{m_{D_s}}}f_D g \frac{p_1 \cdot p_2 - (v \cdot p_1)(v \cdot p_2) + (v\cdot k)(m_{D_s} - v \cdot p_2)}{(v \cdot k + v \cdot p_1 + \Delta)(v \cdot k)(p_1 \cdot k)}\\
  E_3^{(s,d)} &= -i \sqrt{\frac{m_D}{m_{D_s}}}\frac{f_D g (v \cdot k)}{v \cdot k + v \cdot p_1 + \Delta} \left[\frac{2\lambda^\prime + \frac{1}{\sqrt{2}}g_v\lambda\left(\frac{g_\omega}{3m_\omega^2} + \frac{g_\rho}{m_\rho^2}\right)}{v \cdot p_1 + \Delta} - \frac{2\lambda^\prime - g_v\lambda\frac{\sqrt{2}g_\Phi}{3m_\Phi^2}}{v \cdot k + \Delta} \right]
\end{align}

\begin{align}
  \begin{split}
    B_1^{(s,d)} &= -\frac{f_{D_s}}{v \cdot k + \Delta} \left[1 - m_{\rho}^2 BW_{\rho^+}(p_1 +p_2)\right]\left[2\lambda^\prime - g_v\lambda\frac{\sqrt{2}g_\Phi}{3m_\Phi^2} \right]
  \end{split}\\
  \begin{split}
    B_3^{(s,d)} &= \frac{f_{D_s} g_{K^\star}}{f_K} \left(g_{K^\star K \gamma}BW_{K^\star}(p_2 + k) + g_{K^{\pm\star} K^\pm \gamma} BW_{K^{+\star}}(p_1 + k)\right)
  \end{split}\\
  B_4^{(s,d)} &= -\frac{m_{D_s}^2 f_{D_s} f_\pi}{m_{D_s}^2 - m_\pi^2}\frac{m_{K^\star}^2}{g_{K^\star}} \left( g_{K^{\pm\star} K^\pm \gamma}BW_{K^{+\star}}(p_1 + k) + g_{K^\star K \gamma} BW_{K^\star}(p_2 + k)\right)
\end{align}
\begin{align}
  D_1^{(s,d)} &= -2 \lambda^\prime \left[\frac{f_{D_s}}{v \cdot k + \Delta} + \frac{\sqrt{m_D}f_{D}g (v \cdot p_2)}{\sqrt{m_{D_s}}(v \cdot k + v \cdot p_1)}\left(\frac{1}{v \cdot k + \Delta} + \frac{1}{v \cdot p_1 + \Delta}\right)\right]\\
  D_2^{(s,d)} &= -g_v \lambda \left[\frac{-f_{D_s}\frac{\sqrt{2}g_\Phi}{3m_\Phi^2}}{v \cdot k + \Delta} + \frac{\sqrt{m_D}f_{D}g (v \cdot p_2)}{\sqrt{m_{D_s}}(v \cdot k + v \cdot p_1)}\left(\frac{-\frac{\sqrt{2}g_\Phi}{3m_\Phi^2}}{v \cdot k + \Delta} + \frac{\frac{1}{\sqrt{2}}\left(\frac{g_\omega}{3m_\omega^2} + \frac{g_\rho}{m_\rho^2}\right)}{v \cdot p_1 + \Delta}\right)\right]\\
  \begin{split}
    D_3^{(s,d)} &= \frac{g_{K^\star} g_{K^\star K \gamma}}{f_K} \left(f_{D_s} + \sqrt{\frac{m_D}{m_{D_s}}} f_D g \frac{m_{D_s} - v \cdot p_1}{v \cdot p_1 + \Delta}\right) BW_{K^\star}(p_2 + k)\\
    &- \frac{2 f_K  \left(m_{D_s} \alpha_1 - \alpha_2 v \cdot p_2\right)}{\sqrt{m_{D_s}}} \frac{m_{K^\star}^2}{g_{K^\star}} g_{K^{\pm\star} K^\pm \gamma} BW_{K^{+\star}}(p_1 + k)
  \end{split}
\end{align}

\noindent\underline{$D^+ \to \pi^+ \Kbar^0 \gamma$}

\begin{figure}
  \centering
  \subfigure[Contributions to the parity-even form factors $A$ and $E$. For each of the diagrams $A_{1,2}$, $A_{1,3}$, $A_{3,3}$, $A_{3,4}$, $E_{1,2}$ und $E_{2,3}$ there is another one where the photon is coupled via a vector meson.]{\includegraphics[width=0.45\linewidth]{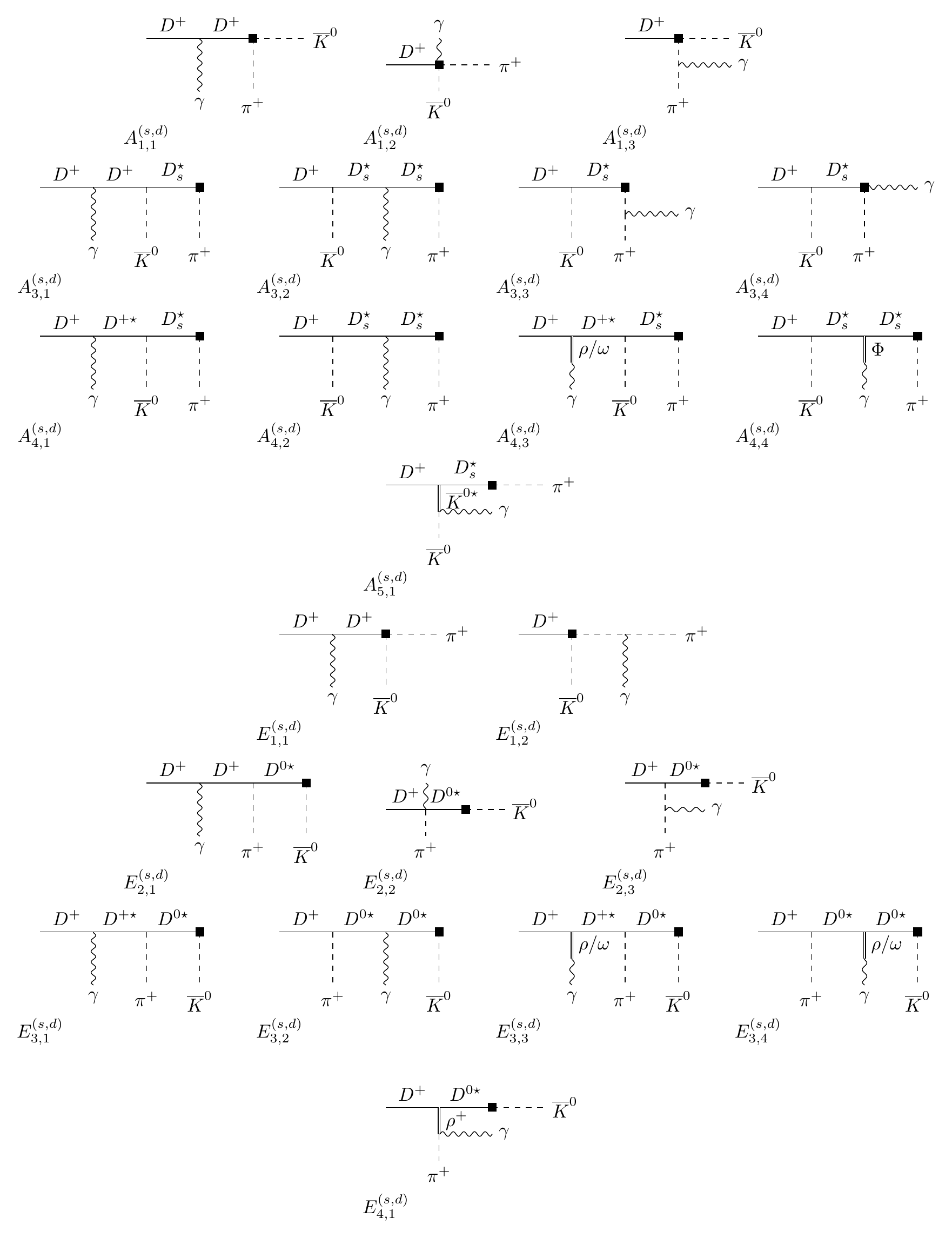}}
  \hfill
  \subfigure[Contributions to the parity-odd form factors $B$ and $D$.]{\includegraphics[width=0.45\linewidth]{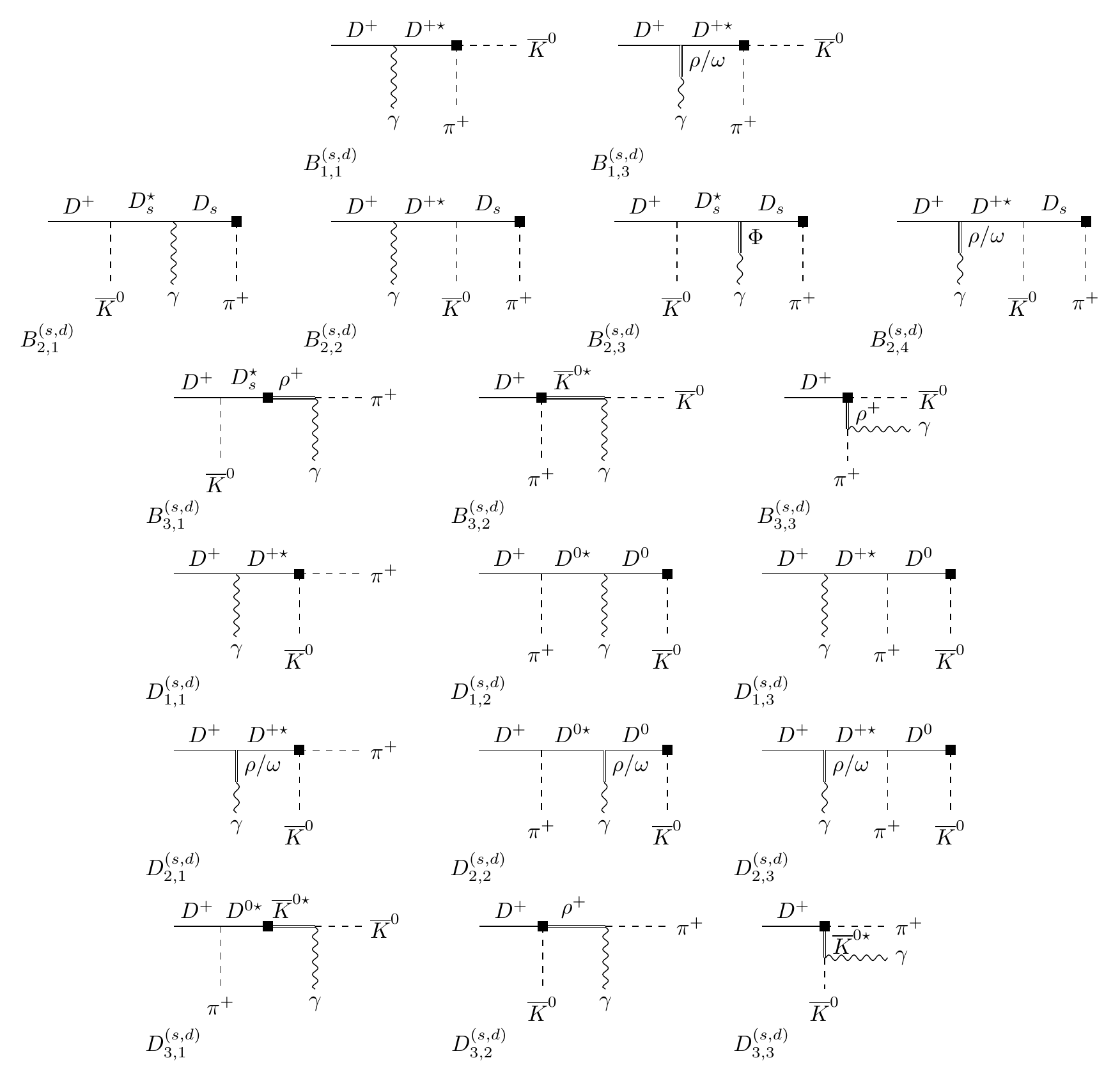}}
  \caption{Feynman diagrams contributing to the decay $D^+ \to \pi^+ \Kbar^0 \gamma$ within the SM.}
  \label{fig:Diagramme_Dplus_piplusKbar}
\end{figure}

\begin{align}
    A_1^{(s,d)} &= -i \frac{f_D f_\pi}{f_K}\frac{v \cdot p_1 + v \cdot k}{(v \cdot k)(p_1 \cdot k)}\\
    A_3^{(s,d)} &= -i \sqrt{\frac{m_{D_s}}{m_D}}\frac{f_{D_s} f_\pi g}{f_K} \frac{p_1 \cdot p_2 - (v \cdot p_1)(v \cdot p_2) + (v\cdot k)(m_D - v \cdot p_2)}{(v \cdot p_2 + \Delta)(v \cdot k)(p_1 \cdot k)}\\
    A_4^{(s,d)} &= -i \sqrt{\frac{m_{D_s}}{m_D}} \frac{f_{D_s} f_\pi g (v \cdot k)}{f_K(v \cdot k + v \cdot p_2 + \Delta)} \left[\frac{2\lambda^\prime + \frac{1}{\sqrt{2}}g_v\lambda\left(\frac{g_\omega}{3m_\omega^2} - \frac{g_\rho}{m_\rho^2}\right)}{v \cdot k + \Delta} - \frac{2\lambda^\prime - \frac{\sqrt{2}}{3}g_v\lambda \frac{g_\Phi}{m_\Phi^2}}{v \cdot p_2 + \Delta} \right]
\end{align}

\begin{align}
  E_1^{(s,d)} &= - i \frac{f_D f_K}{f_\pi}\frac{v \cdot p_2}{(v \cdot k)(p_1 \cdot k)}\\
  E_2^{(s,d)} &= - i \frac{f_D f_K g}{f_\pi} \frac{p_1 \cdot p_2 - (v \cdot p_1)(v \cdot p_2) + (v\cdot k)(m_D - v \cdot p_2)}{(v \cdot k + v \cdot p_1 + \Delta)(v \cdot k)(p_1 \cdot k)}\\
  E_3^{(s,d)} &= -i \frac{f_D f_K g (v \cdot k)}{f_\pi (v \cdot k + v \cdot p_1 + \Delta)} \left[\frac{2\lambda^\prime + \frac{1}{\sqrt{2}}g_v\lambda\left(\frac{g_\omega}{3m_\omega^2} + \frac{g_\rho}{m_\rho^2}\right)}{v \cdot p_1 + \Delta} - \frac{2\lambda^\prime + \frac{1}{\sqrt{2}}g_v\lambda\left(\frac{g_\omega}{3m_\omega^2} - \frac{g_\rho}{m_\rho^2}\right)}{v \cdot k + \Delta} \right]
\end{align}

\begin{align}
  B_1^{(s,d)} &= \frac{f_D f_\pi}{f_K(v \cdot k + \Delta)} \left[2\lambda^\prime + \frac{1}{\sqrt{2}}g_v\lambda\left(\frac{g_\omega}{3m_\omega^2} - \frac{g_\rho}{m_\rho^2}\right)\right]\\
  B_2^{(s,d)} &= \sqrt{\frac{m_{D_s}}{m_D}} \frac{f_{D_s} f_\pi g (v \cdot p_1)}{f_K(v \cdot k + v \cdot p_2)}\left[\frac{2\lambda^\prime - \frac{\sqrt{2}}{3}g_v\lambda \frac{g_\Phi}{m_\Phi^2}}{v \cdot p_2 + \Delta} + \frac{2\lambda^\prime + \frac{1}{\sqrt{2}}g_v\lambda\left(\frac{g_\omega}{3m_\omega^2} - \frac{g_\rho}{m_\rho^2}\right)}{v \cdot k + \Delta}\right]\\
  \begin{split}
    B_3^{(s,d)} &= -\frac{g_{\rho} g_{\rho^\pm \pi^\pm \gamma}}{f_K}\left(f_D + \sqrt{\frac{m_{D_s}}{m_D}} f_{D_s}g\frac{m_D - v \cdot p_2}{v \cdot p_2 + \Delta}\right)BW_{\rho^+}(p_1 + k) \\
    &+ \frac{2f_\pi (m_D \alpha_1 - \alpha_2 v \cdot p_1)}{\sqrt{m_D}}\frac{m_{K^\star}^2}{g_{K^\star}}g_{{K^\star} K \gamma} BW_{K^\star}(p_2 + k)
  \end{split}
\end{align}
\begin{align}
  D_1^{(s,d)} &= -2 \frac{f_D f_K}{f_\pi} \lambda^\prime \left[\frac{1}{v \cdot k + \Delta} + \frac{g (v \cdot p_2)}{v \cdot k + v \cdot p_1}\left(\frac{1}{v \cdot k + \Delta} + \frac{1}{v \cdot p_1 + \Delta}\right)\right]\\
  D_2^{(s,d)} &= -\frac{f_D f_K g_v \lambda}{\sqrt{2} f_\pi} \left[\frac{\frac{g_\omega}{3m_\omega^2} - \frac{g_\rho}{m_\rho^2}}{v \cdot k + \Delta} + \frac{g (v \cdot p_2)}{v \cdot k + v \cdot p_1}\left(\frac{\frac{g_\omega}{3m_\omega^2} - \frac{g_\rho}{m_\rho^2}}{v \cdot k + \Delta} + \frac{\frac{g_\omega}{3m_\omega^2} + \frac{g_\rho}{m_\rho^2}}{v \cdot p_1 + \Delta}\right)\right]\\
  \begin{split}
    D_3^{(s,d)} &= \frac{f_D g_{K^\star} g_{{K^\star} K \gamma}}{f_\pi}\left(1 + g \frac{m_D - v \cdot p_1}{v \cdot p_1 + \Delta}\right) BW_{K^\star}(p_2 + k) \\
    &- \frac{2 f_K   \left(m_D \alpha_1 - \alpha_2 v \cdot p_2\right)}{\sqrt{m_D}}\frac{m_\rho^2}{g_\rho} g_{\rho^\pm \pi^\pm \gamma} BW_{\rho^+}(p_1 + k)
  \end{split}
\end{align}

\subsection{Singly Cabibbo-suppressed decay modes}

\noindent\underline{$D^+ \to \pi^+ \pi^0 \gamma$}

\begin{figure}
  \centering
  \subfigure[Contributions to the parity-even form factors $A$ and $E$. Note that the diagrams $A_1$ have two different factorizations. Additionally, for each of the diagrams $A_{1,2}$, $A_{1,3}$, $A_{2,3}$, $A_{3,3}$, $A_{3,4}$, $E_{1,2}$ und $E_{2,3}$ there is another one where the photon is coupled via a vector meson.]{\includegraphics[width=0.45\linewidth]{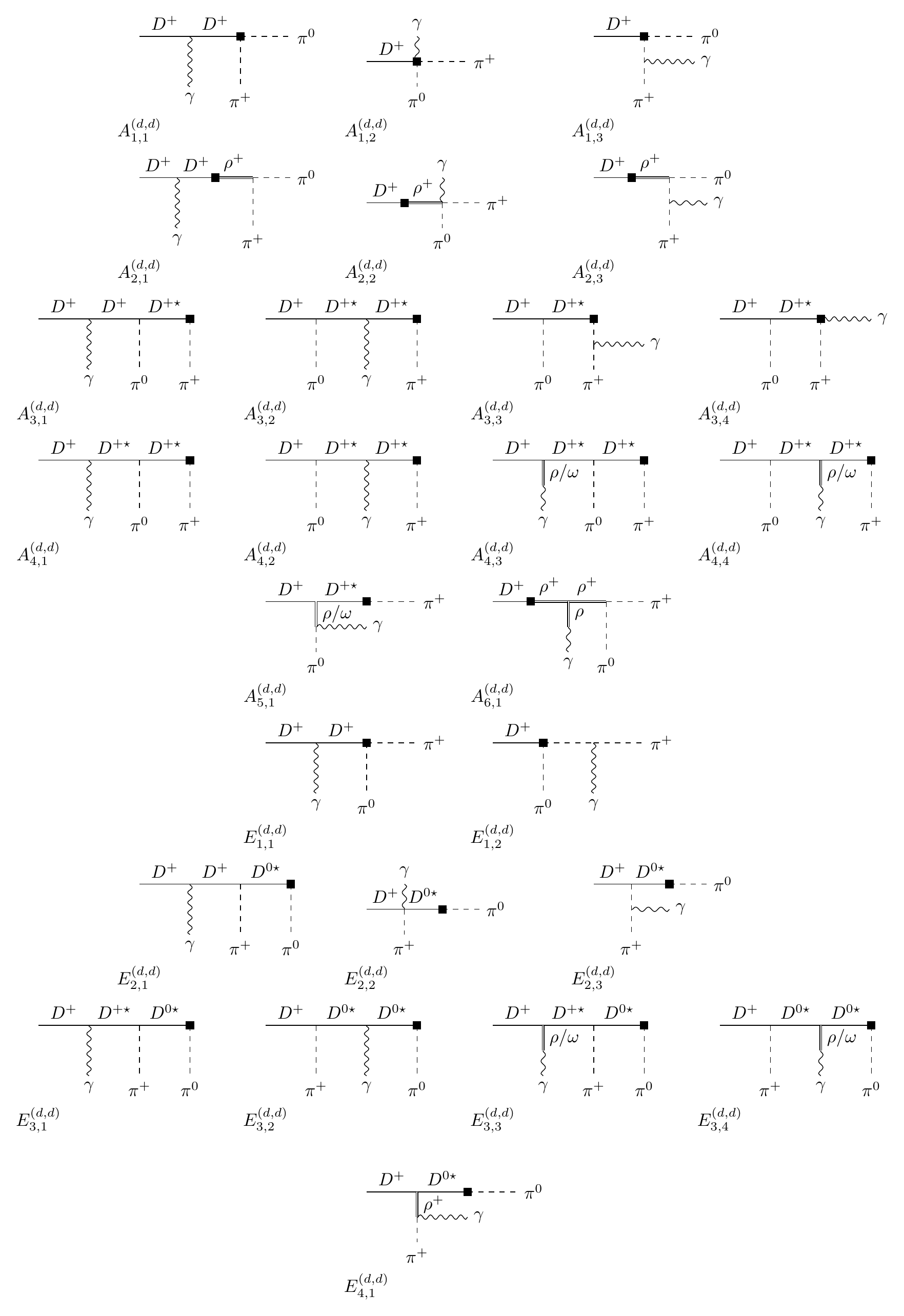}}
  \hfill
  \subfigure[Contributions to the parity-odd form factors $B$ and $D$. Note that the diagrams $B_{1,1/3}$ and $B_{3,2/3}$ have two different factorizations.]{\includegraphics[width=0.45\linewidth]{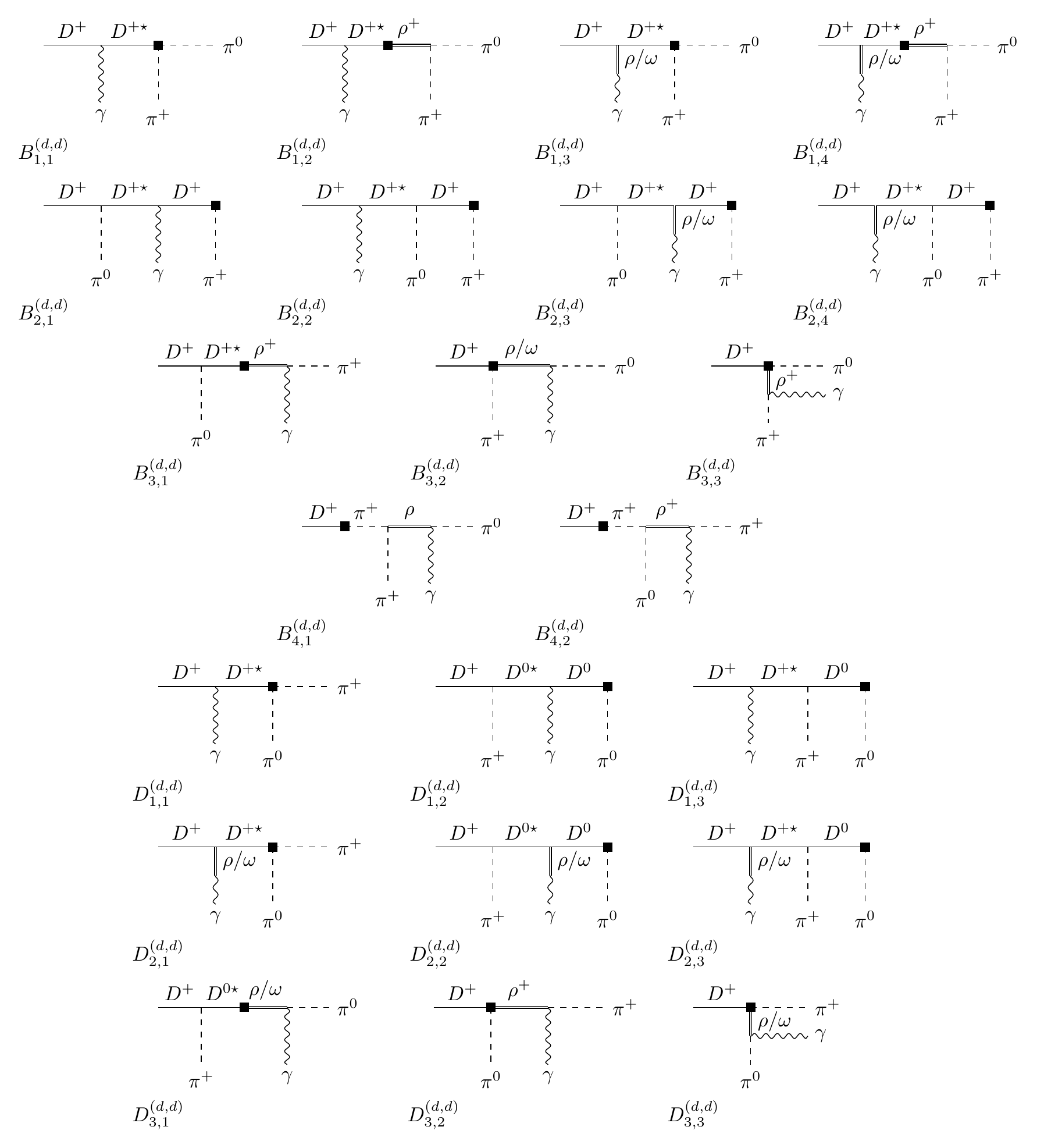}}
  \caption{Feynman diagrams contributing to the decay $D^+ \to \pi^+ \pi^0 \gamma$ within the SM.}
  \label{fig:Diagramme_Dplus_pipluspi}
\end{figure}

\begin{align}
    A_{1+2}^{(d,d)} &= i \sqrt{2}f_D\frac{v \cdot p_2 - v \cdot p_1 - v \cdot k}{(v \cdot k)(p_1 \cdot k)} + i \frac{f_D}{\sqrt{2}}\frac{v \cdot p_1 + v \cdot k}{(v \cdot k)(p_1 \cdot k)}\\
    A_3^{(d,d)} &= i \frac{f_D g}{\sqrt{2}} \frac{p_1 \cdot p_2 - (v \cdot p_1)(v \cdot p_2) + (v\cdot k)(m_D - v \cdot p_2)}{(v \cdot p_2 + \Delta)(v \cdot k)(p_1 \cdot k)}\\
    A_4^{(d,d)} &= i \frac{f_D g (v \cdot k)}{v \cdot k + v \cdot p_2 + \Delta} \left[\frac{\sqrt{2}\lambda^\prime + \frac{1}{2}g_v\lambda\left(\frac{g_\omega}{3m_\omega^2} - \frac{g_\rho}{m_\rho^2}\right)}{v \cdot k + \Delta} - \frac{\sqrt{2}\lambda^\prime + \frac{1}{2}g_v\lambda\left(\frac{g_\omega}{3m_\omega^2} - \frac{g_\rho}{m_\rho^2}\right)}{v \cdot p_2 + \Delta} \right]\\
    A_6^{(d,d)} &= i 2\sqrt{2} f_D \frac{p_1 \cdot p_2}{m_D(v \cdot k)} BW_{\rho^+}(p_1+p_2)
\end{align}

\begin{align}
  E_1^{(d,d)} &=i \frac{f_D}{\sqrt{2}}\frac{v \cdot p_2}{(v \cdot k)(p_1 \cdot k)}\\
  E_2^{(d,d)} &= i \frac{f_D g}{\sqrt{2}} \frac{p_1 \cdot p_2 - (v \cdot p_1)(v \cdot p_2) + (v\cdot k)(m_D - v \cdot p_2)}{(v \cdot k + v \cdot p_1 + \Delta)(v \cdot k)(p_1 \cdot k)}\\
  E_3^{(d,d)} &= i \frac{f_D g (v \cdot k)}{v \cdot k + v \cdot p_1 + \Delta} \left[\frac{\sqrt{2}\lambda^\prime + \frac{1}{2}g_v\lambda\left(\frac{g_\omega}{3m_\omega^2} + \frac{g_\rho}{m_\rho^2}\right)}{v \cdot p_1 + \Delta} - \frac{\sqrt{2}\lambda^\prime + \frac{1}{2}g_v\lambda\left(\frac{g_\omega}{3m_\omega^2} - \frac{g_\rho}{m_\rho^2}\right)}{v \cdot k + \Delta} \right]
\end{align}

\begin{align}
    B_1^{(d,d)}&= \frac{f_D}{v \cdot k + \Delta} \left[\frac{1}{2} - m_{\rho}^2 BW_{\rho^+}(p_1 +p_2)\right]\left[2\sqrt{2}\lambda^\prime + g_v\lambda\left(\frac{g_\omega}{3m_\omega^2} - \frac{g_\rho}{m_\rho^2}\right)\right]\\
    B_2^{(d,d)} &= -\frac{f_D g (v \cdot p_1)}{v \cdot k + v \cdot p_2}\left[\frac{\sqrt{2}\lambda^\prime + \frac{1}{2}g_v\lambda\left(\frac{g_\omega}{3m_\omega^2} - \frac{g_\rho}{m_\rho^2}\right)}{v \cdot p_2 + \Delta} + \frac{\sqrt{2}\lambda^\prime + \frac{1}{2}g_v\lambda\left(\frac{g_\omega}{3m_\omega^2} - \frac{g_\rho}{m_\rho^2}\right)}{v \cdot k + \Delta}\right]\\
  \begin{split}
    B_3^{(d,d)} &= \frac{f_D g_{\rho} g_{\rho^\pm \pi^\pm \gamma}}{\sqrt{2}f_\pi}\left(1 + g\frac{m_D - v \cdot p_2}{v \cdot p_2 + \Delta}\right)BW_{\rho^+}(p_1 + k)\\
    &+ \frac{\sqrt{2}f_\pi (m_D \alpha_1 - \alpha_2 v \cdot p_1)}{\sqrt{m_D}}\left(\frac{m_\omega^2}{g_\omega}g_{\omega \pi \gamma} BW_\omega(p_2 + k) - \frac{m_\rho^2}{g_\rho}g_{\rho \pi \gamma} BW_\rho(p_2 + k)\right)\\
    &-\frac{\sqrt{2}f_D g_\rho}{f_\pi}\left(g_{\rho \pi \gamma}BW_\rho(p_2 + k) + g_{\rho^\pm \pi^\pm \gamma} BW_{\rho^+}(p_1 + k)\right)
  \end{split}\\
  B_4^{(d,d)} &= \frac{\sqrt{2}m_D^2 f_D f_\pi m_\rho^2}{g_\rho (m_D^2 - m_\pi^2)}\left(g_{\rho \pi \gamma}BW_\rho(p_2+k) + g_{\rho^\pm \pi^\pm \gamma}BW_{\rho^+}(p_1+k)\right)
\end{align}
\begin{align}
  D_1^{(d,d)} &= \sqrt{2}f_D \lambda^\prime \left[\frac{1}{v \cdot k + \Delta} + \frac{g (v \cdot p_2)}{v \cdot k + v \cdot p_1}\left(\frac{1}{v \cdot k + \Delta} + \frac{1}{v \cdot p_1 + \Delta}\right)\right]\\
  D_2^{(d,d)} &= \frac{f_D g_v \lambda}{2} \left[\frac{\frac{g_\omega}{3m_\omega^2} - \frac{g_\rho}{m_\rho^2}}{v \cdot k + \Delta} + \frac{g (v \cdot p_2)}{v \cdot k + v \cdot p_1}\left(\frac{\frac{g_\omega}{3m_\omega^2} - \frac{g_\rho}{m_\rho^2}}{v \cdot k + \Delta} + \frac{\frac{g_\omega}{3m_\omega^2} + \frac{g_\rho}{m_\rho^2}}{v \cdot p_1 + \Delta}\right)\right]\\
  \begin{split}
    D_3^{(d,d)} &= \frac{f_D}{\sqrt{2}f_\pi}\left(1 + g \frac{m_D - v \cdot p_1}{v \cdot p_1 + \Delta}\right)\left(g_\omega g_{\omega \pi \gamma} BW_\omega(p_2 + k) - g_\rho g_{\rho \pi \gamma} BW_\rho(p_2 + k)\right)\\
    &+ \frac{\sqrt{2} f_\pi \left(m_D \alpha_1 - \alpha_2 v \cdot p_2\right)}{\sqrt{m_D}} \frac{m_\rho^2}{g_\rho} g_{\rho^\pm \pi^\pm \gamma}BW_{\rho^+}(p_1 + k)
  \end{split}
\end{align}
\begin{align}
  \begin{split}
    a^\prime &= \frac{f_D g (v \cdot k)}{\sqrt{2} f_\pi^2 (v \cdot p_1 + \Delta)} + \frac{f_D g^2 \left(p_2 \cdot k - (v \cdot k)(v \cdot p_2)\right)}{\sqrt{2}f_\pi^2 (v \cdot p_1 + v \cdot p_2 + \Delta)} \left[\frac{1}{v \cdot p_1 + \Delta} + \frac{1}{v \cdot p_2 + \Delta}\right]\\
    &- \frac{\sqrt{2}\alpha_1 (v \cdot k)}{f_\pi^2 \sqrt{m_D}} \left(1 + \frac{f_\pi^2m_\rho^4}{g_\rho^2} BW_{\rho^+}(p_1 + p_2)\right)\\
    &+\frac{2 f_D \lambda g_v m_\rho^2 \left(p_2 \cdot k - (v \cdot k)(v \cdot p_2)\right)}{g_\rho(v \cdot p_1 + v \cdot p_2 + \Delta)}BW_{\rho^+}(p_1 + p_2)
  \end{split}
\end{align}
\begin{align}
  \begin{split}
    b^\prime &= -\frac{f_D g (v \cdot k)}{\sqrt{2} f_\pi^2 (v \cdot p_2 + \Delta)} - \frac{f_D g^2 \left(p_1 \cdot k - (v \cdot k)(v \cdot p_1)\right)}{\sqrt{2}f_\pi^2 (v \cdot p_1 + v \cdot p_2 + \Delta)} \left[\frac{1}{v \cdot p_1 + \Delta} + \frac{1}{v \cdot p_2 + \Delta}\right]\\
    &+ \frac{\sqrt{2}\alpha_1 (v \cdot k)}{f_\pi^2 \sqrt{m_D}}\left(1 + \frac{f_\pi^2m_\rho^4}{g_\rho^2} BW_{\rho^+}(p_1 + p_2)\right)\\
    &-\frac{2f_D \lambda g_v m_\rho^2 \left(p_1 \cdot k - (v \cdot k)(v \cdot p_1)\right)}{g_\rho(v \cdot p_1 + v \cdot p_2 + \Delta)}BW_{\rho^+}(p_1 + p_2)
  \end{split}
\end{align}
\begin{align}
  \begin{split}
    c^\prime &= \frac{f_D g}{\sqrt{2} m_D f_\pi^2}\left(\frac{p_2 \cdot k}{v \cdot p_2 + \Delta} - \frac{p_1 \cdot k}{v \cdot p_1 + \Delta}\right) \\
    &- \frac{f_D g^2 \left((p_2 \cdot k)(v \cdot p_1) - (p_1 \cdot k)(v \cdot p_2)\right)}{\sqrt{2}m_D f_\pi^2 (v \cdot p_1 + v \cdot p_2 + \Delta)}\left(\frac{1}{v \cdot p_1 + \Delta} + \frac{1}{v \cdot p_2 + \Delta}\right)\\
    &+ \frac{\sqrt{2}\alpha_1 (p_1 \cdot k - p_2 \cdot k)}{\sqrt{m_D^3}f_\pi^2}\left(1 + \frac{f_\pi^2m_\rho^4}{g_\rho^2} BW_{\rho^+}(p_1 + p_2)\right)\\
    &-\frac{2 f_D \lambda g_v m_\rho^2 \left((p_2 \cdot k)(v \cdot p_1) - (p_1 \cdot k)(v \cdot p_2)\right)}{g_\rho m_D (v \cdot p_1 + v \cdot p_2 + \Delta)}BW_{\rho^+}(p_1 + p_2)
  \end{split}
\end{align}
\begin{align}
  \begin{split}
    h^\prime &=-\frac{f_D g }{2\sqrt{2} m_D f_\pi^2 }\left(\frac{1}{v \cdot p_1 + \Delta} + \frac{1}{v \cdot p_2 + \Delta}\right)\left(1 + g \frac{v \cdot k}{v \cdot p_1 + v \cdot p_2 + \Delta}\right)\\
    &- \frac{\sqrt{2}\alpha_1}{\sqrt{m_D^3}f_\pi^2}\left(1 + \frac{f_\pi^2m_\rho^4}{g_\rho^2} BW_{\rho^+}(p_1 + p_2)\right)\\
    &-\frac{f_D \lambda g_v m_\rho^2 (v \cdot k)}{g_\rho m_D (v \cdot p_1 + v \cdot p_2 + \Delta)}BW_{\rho^+}(p_1 + p_2)
  \end{split}
\end{align}

\noindent\underline{$D_s \to \pi^+ K^0 \gamma$}

\begin{figure}
  \centering
  \subfigure[Contributions to the parity-even form factors $A$ and $E$. Note that the diagrams $A_1$ have two different factorizations. Additionally, for each of the diagrams $A_{1,2}$, $A_{1,3}$, $A_{2,3}$, $A_{3,3}$ and $A_{3,4}$ there is another one where the photon is coupled via a vector meson.]{\includegraphics[width=0.45\linewidth]{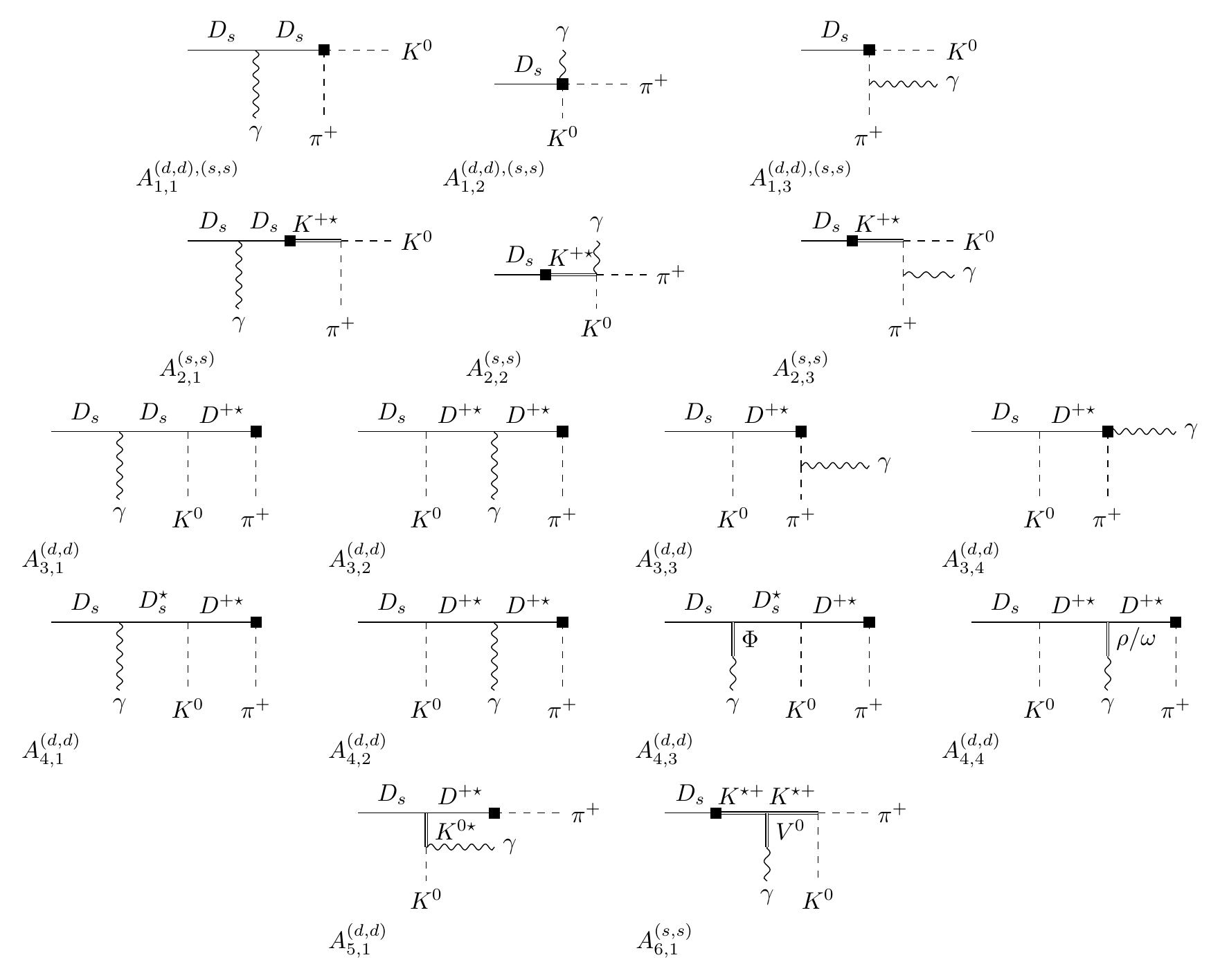}}
  \hfill
  \subfigure[Contributions to the parity-odd form factors $B$ and $D$. Note that the diagrams $B_{1,1/3}$ and $B_{3,2/3}$ have two different factorizations.]{\includegraphics[width=0.45\linewidth]{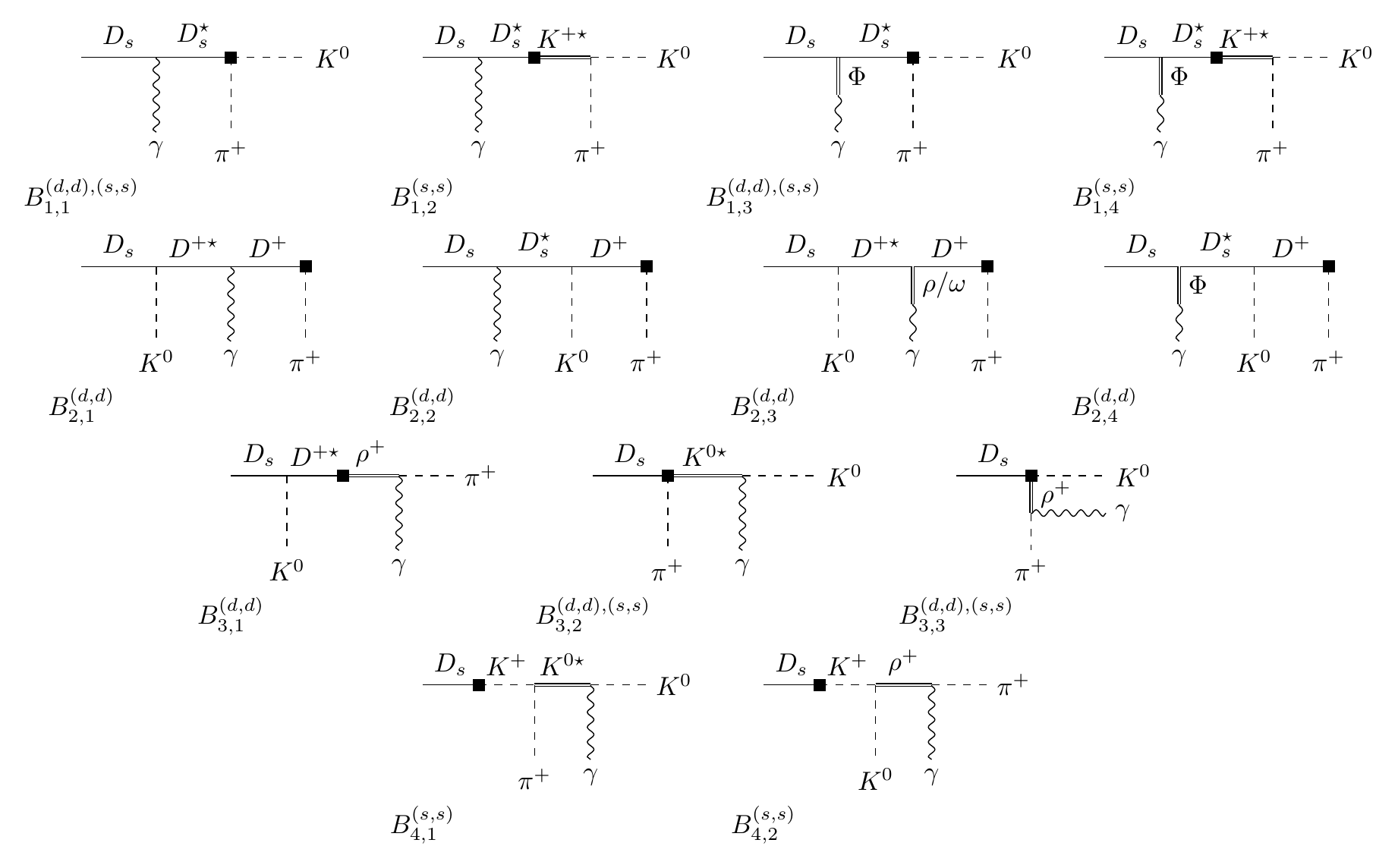}}
  \caption{Feynman diagrams contributing to the decay $D_s \to \pi^+ K^0 \gamma$ within the SM.}
  \label{fig:Diagramme_Ds_piplusK}
\end{figure}

\begin{align}
  \begin{split}
    A_1^{(d,d)} &= -i \frac{f_{D_s} f_\pi}{f_K}\frac{v \cdot p_1 + v \cdot k}{(v \cdot k)(p_1 \cdot k)}\\
    A_{1+2}^{(s,s)} &= -i f_{D_s}\frac{v \cdot p_2 - v \cdot p_1 - v \cdot k}{(v \cdot k)(p_1 \cdot k)}
  \end{split} \\
    A_3^{(d,d)} &= -i \frac{f_{D_s} f_\pi g}{f_K} \frac{p_1 \cdot p_2 - (v \cdot p_1)(v \cdot p_2) + (v\cdot k)(M - v \cdot p_2)}{(v \cdot p_2 + \Delta)(v \cdot k)(p_1 \cdot k)}\\
    A_4^{(d,d)} &= -i \frac{\sqrt{m_D}f_D f_\pi g (v \cdot k)}{\sqrt{m_{D_s}}f_K(v \cdot k + v \cdot p_2 + \Delta)} \left[\frac{2\lambda^\prime - \frac{\sqrt{2}}{3}g_v\lambda\frac{g_\Phi}{m_\Phi^2}}{v \cdot k + \Delta} - \frac{2\lambda^\prime + \frac{1}{\sqrt{2}}g_v\lambda\left(\frac{g_\omega}{3m_\omega^2} - \frac{g_\rho}{m_\rho^2}\right)}{v \cdot p_2 + \Delta} \right]\\
    A_6^{(s,s)} &= -i 2 f_{D_s} \frac{p_1 \cdot p_2}{m_{D_s}(v \cdot k)} BW_{K^{\star+}}(p_1+p_2)
\end{align}

\begin{align}
  \begin{split}
    B_1^{(d,d)} &= \frac{f_{D_s}f_\pi}{f_K(v \cdot k + \Delta)} \left[2\lambda^\prime - \frac{\sqrt{2}}{3}g_v\lambda\frac{g_\Phi}{m_\Phi^2}\right]\\
    B_1^{(s,s)} &= -\frac{f_{D_s}}{v \cdot k + \Delta} \left[1 - m_{K^\star}^2 BW_{K^{+\star}}(p_1 +p_2)\right]\left[2\lambda^\prime - \frac{\sqrt{2}}{3} g_v\lambda\frac{g_\Phi}{m_\Phi}\right]
  \end{split}\\
    B_2^{(d,d)} &= \frac{\sqrt{m_D}f_D f_\pi g (v \cdot p_1)}{\sqrt{m_{D_s}}f_K(v \cdot k + v \cdot p_2)}\left[\frac{2\lambda^\prime + \frac{1}{\sqrt{2}}g_v\lambda\left(\frac{g_\omega}{3m_\omega^2} - \frac{g_\rho}{m_\rho^2}\right)}{v \cdot p_2 + \Delta} + \frac{2\lambda^\prime - \frac{\sqrt{2}}{3}g_v\lambda\frac{g_\Phi}{m_\Phi^2}}{v \cdot k + \Delta}\right]\\
  \begin{split}
    B_3^{(d,d)} &= -\frac{ g_{\rho} g_{\rho^\pm \pi^\pm \gamma}}{f_K}\left(f_{D_s} + gf_D\sqrt{\frac{m_D}{m_{D_s}}}\frac{m_{D_s} - v \cdot p_2}{v \cdot p_2 + \Delta}\right)BW_{\rho^+}(p_1 + k) \\
    &+ \frac{2f_\pi (m_{D_s} \alpha_1 - \alpha_2 v \cdot p_1)}{\sqrt{m_{D_s}}}\frac{m_{K^\star}^2}{g_{K^\star}} g_{K^\star K \gamma} BW_{K^\star}(p_2 + k) )\\
    B_3^{(s,s)} &= \frac{f_{D_s} g_{K^\star}}{f_\pi}g_{K^\star K \gamma}BW_{K^\star}(p_2 + k) + \frac{f_{D_s} g_{\rho}}{f_K}g_{\rho^\pm \pi^\pm \gamma} BW_{\rho^+}(p_1 + k)
  \end{split}\\
  B_4^{(s,s)} &= -\frac{m_{D_s}^2 f_{D_s} f_K}{ (m_{D_s}^2 - m_K^2)}\left(\frac{m_{K^\star}^2}{g_{K^\star}}g_{K^\star K \gamma}BW_{K^\star}(p_2+k) + \frac{m_\rho^2}{g_\rho}g_{\rho^\pm \pi^\pm \gamma}BW_{\rho^+}(p_1+k)\right)
\end{align}

\begin{align}
  \begin{split}
    a^\prime &= - \frac{\sqrt{m_D} f_D g^2 \left(p_2 \cdot k - (v \cdot k)(v \cdot p_2)\right)}{f_\pi f_K \sqrt{m_{D_s}} (v \cdot p_1 + v \cdot p_2 + \Delta)(v \cdot p_2 + \Delta)}\\
    &+ \frac{\alpha_1 (v \cdot k)}{f_\pi f_K \sqrt{m_{D_s}}} \left(1 + \frac{f_\pi f_K m_{K^\star}^4}{g_{K^\star}^2} BW_{K^{+\star}}(p_1 + p_2)\right)\\
    &-\frac{\sqrt{2m_D} f_D \lambda g_v m_{K^\star}^2 \left(p_2 \cdot k - (v \cdot k)(v \cdot p_2)\right)}{g_{K^\star} \sqrt{m_{D_s}}(v \cdot p_1 + v \cdot p_2 + \Delta)}BW_{K^{+\star}}(p_1 + p_2)
  \end{split}
\end{align}
\begin{align}
  \begin{split}
    b^\prime &= \frac{\sqrt{m_D} f_D g }{f_\pi f_K \sqrt{m_{D_s}} (v \cdot p_2 + \Delta)} \left[v \cdot k + g\frac{p_1 \cdot k - (v \cdot k)(v \cdot p_1)}{v \cdot p_1 + v \cdot p_2 + \Delta}\right] \\
    &- \frac{\alpha_1 (v \cdot k)}{f_\pi f_K\sqrt{m_{D_s}}}\left(1 + \frac{f_\pi f_K m_{K^\star}^4}{g_{K^\star}^2} BW_{K^{+\star}}(p_1 + p_2)\right)\\
    &+ \frac{\sqrt{2m_D} f_D \lambda g_v m_{K^\star}^2 \left(p_1 \cdot k - (v \cdot k)(v \cdot p_1)\right)}{g_{K^\star} \sqrt{m_{D_s}}(v \cdot p_1 + v \cdot p_2 + \Delta)}BW_{K^{+\star}}(p_1 + p_2)
  \end{split}
\end{align}
\begin{align}
  \begin{split}
    c^\prime &= -\frac{\sqrt{m_D}f_D g}{\sqrt{m_{D_s}^3}f_\pi f_K(v \cdot p_2 + \Delta)} \left[p_2 \cdot k - g\frac{(p_2 \cdot k)(v \cdot p_1) - (p_1 \cdot k)(v \cdot p_2)}{v \cdot p_1 + v \cdot p_2 + \Delta}\right] \\
    &- \frac{\alpha_1 (p_1 \cdot k - p_2 \cdot k)}{\sqrt{m_{D_s}^3}f_\pi f_K}\left(1 + \frac{f_\pi f_K m_{K^\star}^4}{g_{K^\star}^2} BW_{K^{+\star}}(p_1 + p_2)\right)\\
    &+ \frac{\sqrt{2m_D } f_D \lambda g_v m_{K^\star}^2 \left((p_2 \cdot k)(v \cdot p_1) - (p_1 \cdot k)(v \cdot p_2)\right)}{\sqrt{m_{D_s}^3}g_{K^\star}(v \cdot p_1 + v \cdot p_2 + \Delta)}BW_{K^{+\star}}(p_1 + p_2)
  \end{split}
\end{align}
\begin{align}
  \begin{split}
    h^\prime &=\frac{\sqrt{m_D}f_D g }{2 \sqrt{m_{D_s}^3} f_\pi f_K (v \cdot p_2 + \Delta)}\left(1 + g \frac{v \cdot k}{v \cdot p_1 + v \cdot p_2 + \Delta}\right)\\
    &+ \frac{\alpha_1}{\sqrt{m_{D_s}^3}f_\pi f_K}\left(1 + \frac{f_\pi f_K m_{K^\star}^4}{g_{K^\star}^2} BW_{K^{+\star}}(p_1 + p_2)\right)\\
    &+ \frac{\sqrt{m_D } f_D \lambda g_v m_{K^\star}^2 \left(v \cdot k\right)}{\sqrt{2 m_{D_s}^3}g_{K^\star}(v \cdot p_1 + v \cdot p_2 + \Delta)}BW_{K^{+\star}}(p_1 + p_2)
  \end{split}
\end{align}

\noindent\underline{$D_s \to K^+ \pi^0 \gamma$}

\begin{figure}
  \centering
  \subfigure[Contributions to the parity-even form factors $A$ and $E$. Additionally, for each of the diagrams $A_{1,2}$, $A_{1,3}$, $A_{2,3}$, $E_{1,2}$ und $E_{2,3}$ there is another one where the photon is coupled via a vector meson.]{\includegraphics[width=0.45\linewidth]{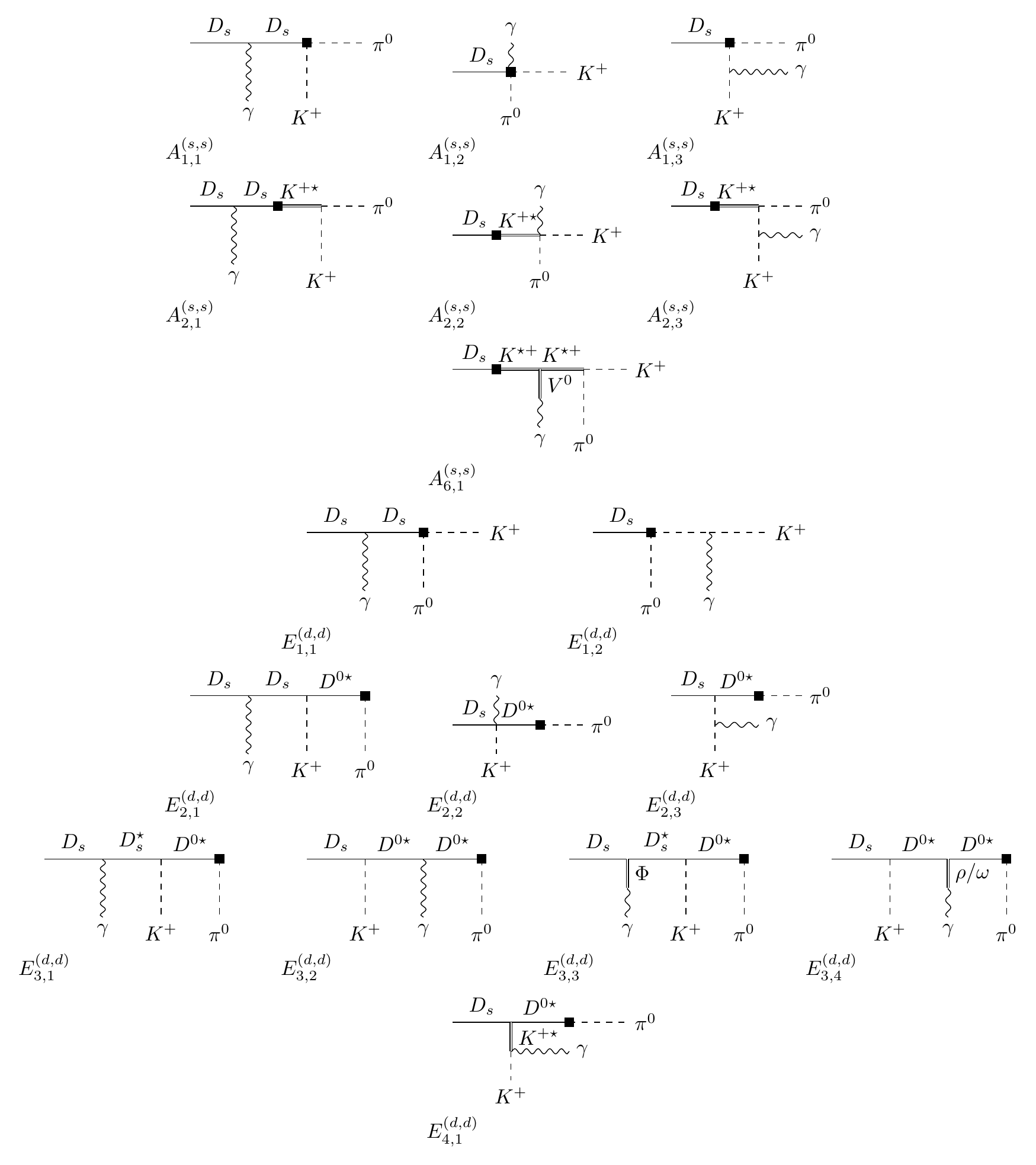}}
  \hfill
  \subfigure[Contributions to the parity-odd form factors $B$ and $D$.]{\includegraphics[width=0.45\linewidth]{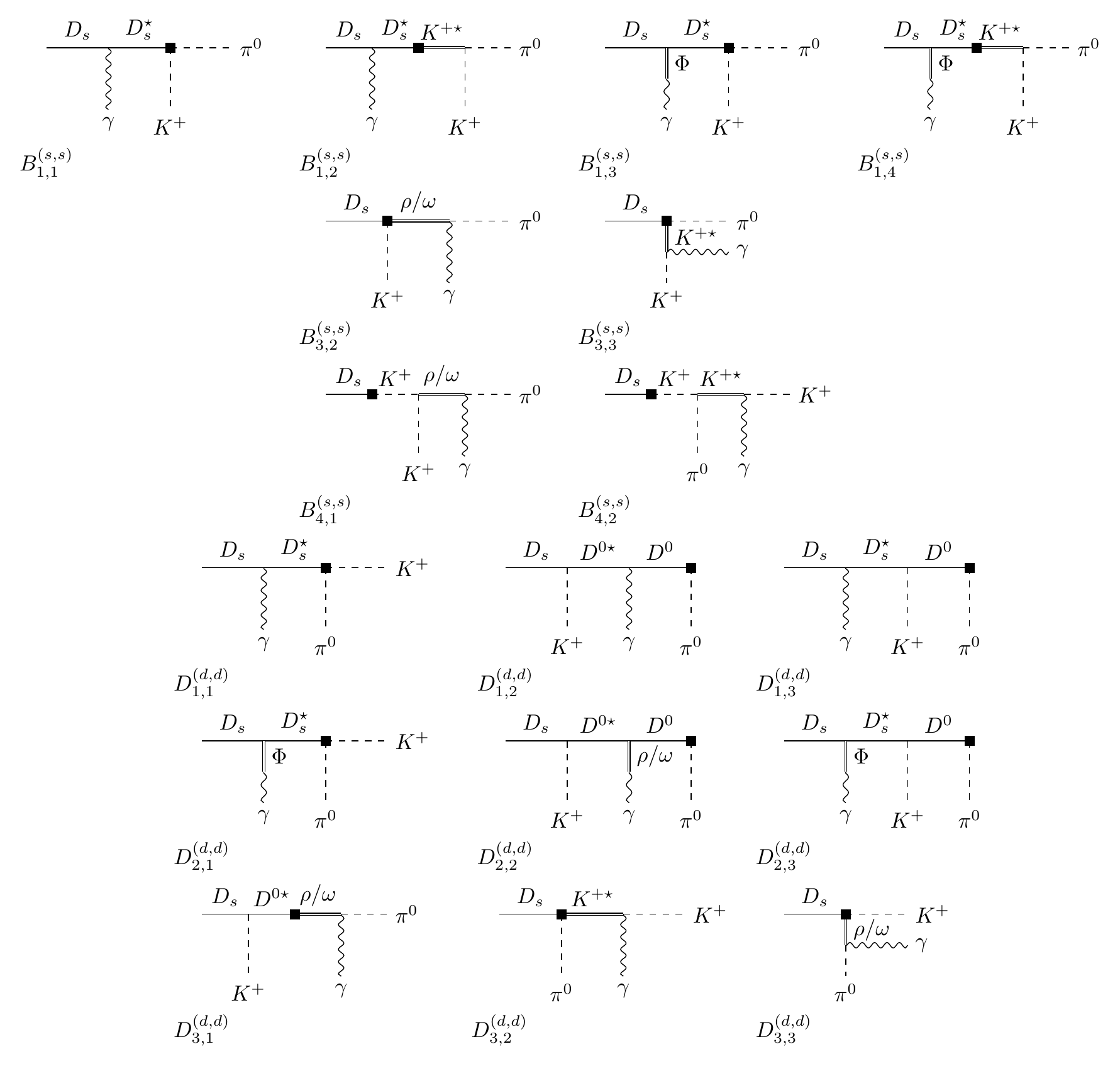}}
  \caption{Feynman diagrams contributing to the decay $D_s \to K^+ \pi^0 \gamma$ within the SM.}
  \label{fig:Diagramme_Ds_Kpluspi}
\end{figure}

\begin{align}
    A_{1+2}^{(s,s)} &= i \frac{f_{D_s}}{\sqrt{2}}\frac{v \cdot p_2 - v \cdot p_1 - v \cdot k}{(v \cdot k)(p_1 \cdot k)}\\
    A_6^{(s,s)} &= i \sqrt{2} f_{D_s} \frac{p_1 \cdot p_2}{m_{D_s}(v \cdot k)} BW_{K^{\star+}}(p_1+p_2)
\end{align}

\begin{align}
  E_1^{(d,d)} &= i \frac{f_{D_s} f_\pi}{\sqrt{2}f_K}\frac{v \cdot p_2}{(v \cdot k)(p_1 \cdot k)}\\
  E_2^{(d,d)} &= i \sqrt{\frac{m_D}{m_{D_s}}}\frac{f_D f_\pi g}{\sqrt{2}f_K} \frac{p_1 \cdot p_2 - (v \cdot p_1)(v \cdot p_2) + (v\cdot k)(m_{D_s} - v \cdot p_2)}{(v \cdot k + v \cdot p_1 + \Delta)(v \cdot k)(p_1 \cdot k)}\\
  E_3^{(d,d)} &= i \sqrt{\frac{m_D}{m_{D_s}}}\frac{f_D f_\pi g (v \cdot k)}{f_K(v \cdot k + v \cdot p_1 + \Delta)} \left[\frac{\sqrt{2}\lambda^\prime + \frac{1}{2}g_v\lambda\left(\frac{g_\omega}{3m_\omega^2} + \frac{g_\rho}{m_\rho^2}\right)}{v \cdot p_1 + \Delta} - \frac{\sqrt{2}\lambda^\prime - g_v\lambda\frac{g_\Phi}{3m_\Phi^2}}{v \cdot k + \Delta} \right]
\end{align}

\begin{align}
  \begin{split}
    B_1^{(s,s)} &= \frac{f_{D_s}}{v \cdot k + \Delta} \left[1 - m_{K^\star}^2 BW_{K^{+\star}}(p_1 +p_2)\right]\left[\sqrt{2}\lambda^\prime - g_v\lambda\frac{g_\Phi}{3m_\Phi^2} \right]
  \end{split}\\
  \begin{split}
    B_3^{(s,s)} &= -\frac{f_{D_s} g_{\rho}}{\sqrt{2}f_K} g_{\rho \pi \gamma}BW_{\rho}(p_2 + k) - \frac{f_{D_s} g_\omega}{\sqrt{2}f_K} g_{\omega \pi \gamma} BW_{\omega}(p_2 + k) - \frac{f_{D_s} g_{K^\star}}{\sqrt{2}f_\pi} g_{K^{\pm\star} K^\pm \gamma} BW_{K^{+\star}}(p_1 + k)
  \end{split}\\
  B_4^{(s,s)} &= \frac{m_{D_s}^2 f_{D_s} f_K}{\sqrt{2}(m_{D_s}^2 - m_K^2)}\left( \frac{2m_{K^\star}^2}{g_{K^\star}}g_{K^{\pm\star} K^\pm \gamma}BW_{K^{+\star}}(p_1 + k) + \frac{m_{\rho}^2}{g_{\rho}}g_{\rho \pi \gamma} BW_{\rho}(p_2 + k) + \frac{m_{\omega}^2}{g_{\omega}}g_{\omega \pi \gamma} BW_{\omega}(p_2 + k)\right)
\end{align}
\begin{align}
  D_1^{(d,d)} &= \sqrt{2}\frac{ f_\pi}{f_K} \lambda^\prime \left[\frac{f_{D_s}}{v \cdot k + \Delta} + \frac{\sqrt{m_D}f_{D}g (v \cdot p_2)}{\sqrt{m_{D_s}}(v \cdot k + v \cdot p_1)}\left(\frac{1}{v \cdot k + \Delta} + \frac{1}{v \cdot p_1 + \Delta}\right)\right]\\
  D_2^{(d,d)} &= \frac{f_\pi g_v \lambda}{f_K} \left[\frac{-f_{D_s}\frac{g_\Phi}{3m_\Phi^2}}{v \cdot k + \Delta} + \frac{\sqrt{m_D}f_{D}g (v \cdot p_2)}{\sqrt{m_{D_s}}(v \cdot k + v \cdot p_1)}\left(\frac{-\frac{g_\Phi}{3m_\Phi^2}}{v \cdot k + \Delta} + \frac{\frac{1}{2}\left(\frac{g_\omega}{3m_\omega^2} + \frac{g_\rho}{m_\rho^2}\right)}{v \cdot p_1 + \Delta}\right)\right]\\
  \begin{split}
    D_3^{(d,d)} &= \frac{1}{\sqrt{2}f_K}\left(f_{D_s} + \sqrt{\frac{m_D}{m_{D_s}}} f_D g \frac{m_{D_s} - v \cdot p_1}{v \cdot p_1 + \Delta}\right)\left(g_{\omega} g_{{\omega} \pi \gamma} BW_{\omega}(p_2 + k) - g_{\rho} g_{{\rho} \pi \gamma} BW_{\rho}(p_2 + k)\right)\\
    &+ \frac{\sqrt{2} f_\pi  \left(m_{D_s} \alpha_1 - \alpha_2 v \cdot p_2\right)}{\sqrt{m_{D_s}}} \frac{m_{K^\star}^2}{g_{K^\star}} g_{K^{\pm\star} K^\pm \gamma} BW_{K^{+\star}}(p_1 + k)
  \end{split}
\end{align}

\begin{align}
  \begin{split}
    a^\prime &= \frac{\sqrt{m_D} f_D g}{\sqrt{2}f_\pi f_K \sqrt{m_{D_s}} (v \cdot p_1 + \Delta)} \left[v \cdot k + g\frac{p_2 \cdot k - (v \cdot k)(v \cdot p_2)}{v \cdot p_1 + v \cdot p_2 + \Delta}\right]\\
    &- \frac{\alpha_1 (v \cdot k)}{\sqrt{2}f_\pi f_K\sqrt{m_{D_s}}} \left(1 + \frac{f_\pi f_K m_{K^\star}^4}{g_{K^\star}^2} BW_{K^{+\star}}(p_1 + p_2)\right)\\
    & +\frac{\sqrt{m_D } f_D \lambda g_v m_{K^\star}^2 \left(p_2 \cdot k - (v \cdot k)(v \cdot p_2)\right)}{g_{K^\star} \sqrt{m_{D_s}}(v \cdot p_1 + v \cdot p_2 + \Delta)}BW_{K^{+\star}}(p_1 + p_2)
  \end{split}
\end{align}
\begin{align}
  \begin{split}
    b^\prime &= -\frac{\sqrt{m_D} f_D g^2 \left(p_1 \cdot k - (v \cdot k)(v \cdot p_1)\right)}{\sqrt{2}f_\pi f_K \sqrt{m_{D_s}} (v \cdot p_1 + \Delta)(v \cdot p_1 + v \cdot p_2 + \Delta)} \\
    &+ \frac{\alpha_1 (v \cdot k)}{\sqrt{2}f_\pi f_K\sqrt{m_{D_s}}}\left(1 + \frac{f_\pi f_K m_{K^\star}^4}{g_{K^\star}^2} BW_{K^{+\star}}(p_1 + p_2)\right)\\
    &- \frac{\sqrt{m_D} f_D \lambda g_v m_{K^\star}^2 \left(p_1 \cdot k - (v \cdot k)(v \cdot p_1)\right)}{g_{K^\star} \sqrt{m_{D_s}}(v \cdot p_1 + v \cdot p_2 + \Delta)}BW_{K^{+\star}}(p_1 + p_2)
  \end{split}
\end{align}
\begin{align}
  \begin{split}
    c^\prime &= -\frac{\sqrt{m_D}f_D g}{\sqrt{2m_{D_s}^3}f_\pi f_K(v \cdot p_1 + \Delta)} \left[p_1 \cdot k + g\frac{(p_2 \cdot k)(v \cdot p_1) - (p_1 \cdot k)(v \cdot p_2)}{v \cdot p_1 + v \cdot p_2 + \Delta}\right] \\
    &+ \frac{\alpha_1 (p_1 \cdot k - p_2 \cdot k)}{\sqrt{2m_{D_s}^3}f_\pi f_K}\left(1 + \frac{f_\pi f_K m_{K^\star}^4}{g_{K^\star}^2} BW_{K^{+\star}}(p_1 + p_2)\right)\\
    &- \frac{\sqrt{m_D } f_D \lambda g_v m_{K^\star}^2 \left((p_2 \cdot k)(v \cdot p_1) - (p_1 \cdot k)(v \cdot p_2)\right)}{\sqrt{m_{D_s}^3}g_{K^\star}(v \cdot p_1 + v \cdot p_2 + \Delta)}BW_{K^{+\star}}(p_1 + p_2)
  \end{split}
\end{align}
\begin{align}
  \begin{split}
    h^\prime &=-\frac{\sqrt{m_D}f_D g }{2 \sqrt{2m_{D_s}^3} f_\pi f_K (v \cdot p_1 + \Delta)}\left(1 + g \frac{v \cdot k}{v \cdot p_1 + v \cdot p_2 + \Delta}\right)\\
    &- \frac{\alpha_1}{\sqrt{2m_{D_s}^3}f_\pi f_K}\left(1 + \frac{f_\pi f_K m_{K^\star}^4}{g_{K^\star}^2} BW_{K^{+\star}}(p_1 + p_2)\right)\\
    &- \frac{\sqrt{m_D } f_D \lambda g_v m_{K^\star}^2 \left(v \cdot k\right)}{2\sqrt{ m_{D_s}^3}g_{K^\star}(v \cdot p_1 + v \cdot p_2 + \Delta)}BW_{K^{+\star}}(p_1 + p_2)
  \end{split}
\end{align}

\noindent\underline{$D^+ \to K^+ \Kbar^0 \gamma$}

\begin{figure}
  \centering
  \subfigure[Contributions to the parity-even form factors $A$ and $E$. Note that the diagrams $A_1$ have two different factorizations. Additionally, for each of the diagrams $A_{1,2}$, $A_{1,3}$, $A_{2,3}$, $A_{3,3}$ and $A_{3,4}$ there is another one where the photon is coupled via a vector meson.]{\includegraphics[width=0.45\linewidth]{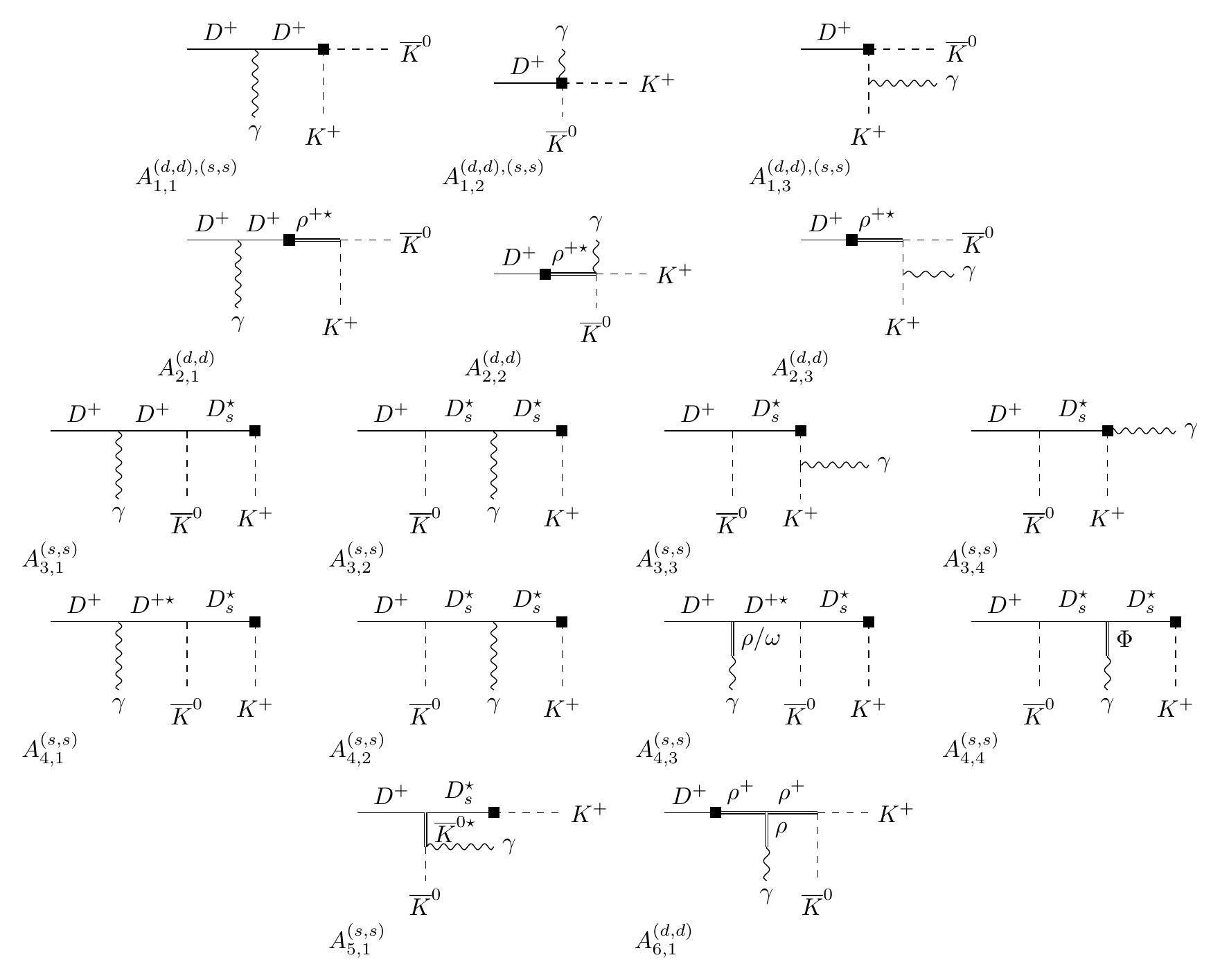}}
  \hfill
  \subfigure[Contributions to the parity-odd form factors $B$ and $D$. Note that the diagrams $B_{1,1/3}$ and $B_{3,2/3}$ have two different factorizations.]{\includegraphics[width=0.45\linewidth]{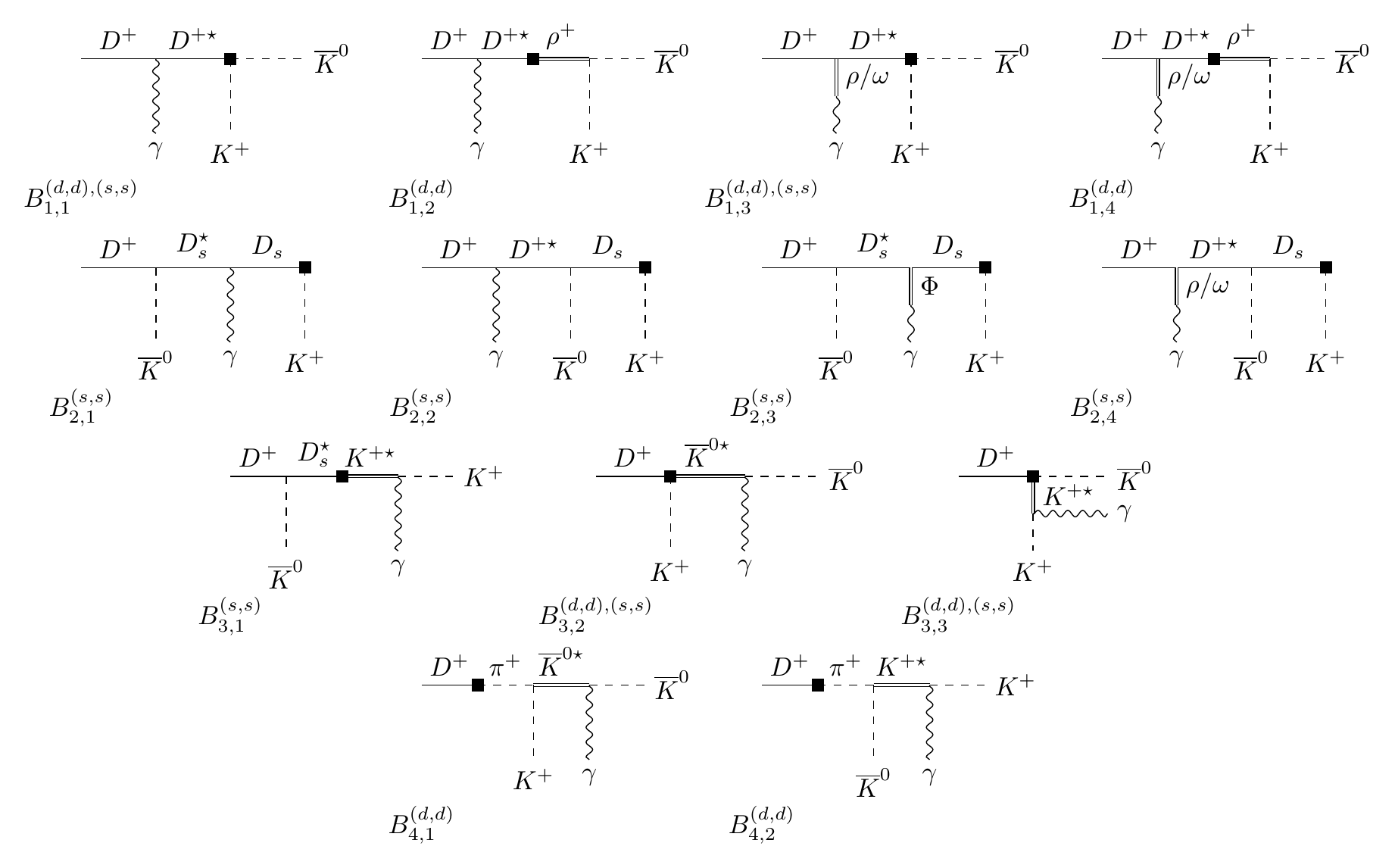}}
  \caption{Feynman diagrams contributing to the decay $D^+ \to K^+ \Kbar^0 \gamma$ within the SM.}
  \label{fig:Diagramme_Dplus_KplusKbar}
\end{figure}

\begin{align}
    A_1^{(s,s)} &= -i f_D\frac{v \cdot p_1 + v \cdot k}{(v \cdot k)(p_1 \cdot k)}\\
    A_{1+2}^{(d,d)} &= -i f_{D}\frac{v \cdot p_2 - v \cdot p_1 - v \cdot k}{(v \cdot k)(p_1 \cdot k)}\\
    A_3^{(s,s)} &= -i \sqrt{\frac{m_{D_s}}{m_D}}f_{D_s} g \frac{p_1 \cdot p_2 - (v \cdot p_1)(v \cdot p_2) + (v\cdot k)(M - v \cdot p_2)}{(v \cdot p_2 + \Delta)(v \cdot k)(p_1 \cdot k)}\\
    A_4^{(s,s)} &= -i \frac{\sqrt{m_{D_s}}f_{D_s} g (v \cdot k)}{\sqrt{m_D}(v \cdot k + v \cdot p_2 + \Delta)} \left[\frac{2\lambda^\prime + \frac{1}{\sqrt{2}}g_v\lambda\left(\frac{g_\omega}{3m_\omega^2} - \frac{g_\rho}{m_\rho^2}\right)}{v \cdot k + \Delta} - \frac{2\lambda^\prime - \frac{\sqrt{2}}{3}g_v\lambda\frac{g_\Phi}{m_\Phi^2}}{v \cdot p_2 + \Delta} \right]\\
    A_6^{(d,d)} &= -i 2 f_D \frac{p_1 \cdot p_2}{m_D(v \cdot k)} BW_{\rho^+}(p_1+p_2)
\end{align}

\begin{align}
  \begin{split}
    B_1^{(s,s)} &= \frac{f_{D}}{(v \cdot k + \Delta)} \left[2\lambda^\prime + \frac{1}{\sqrt{2}}g_v\lambda\left(\frac{g_\omega}{3m_\omega^2} - \frac{g_\rho}{m_\rho^2}\right)\right]\\
    B_1^{(d,d)} &= -\frac{f_{D}}{v \cdot k + \Delta} \left[1 - m_{\rho}^2 BW_{\rho^+}(p_1 +p_2)\right]\left[2\lambda^\prime + \frac{1}{\sqrt{2}}g_v\lambda\left(\frac{g_\omega}{3m_\omega^2} - \frac{g_\rho}{m_\rho^2}\right)\right]
  \end{split}\\
    B_2^{(s,s)} &= \frac{\sqrt{m_{D_s}}f_{D_s} g (v \cdot p_1)}{\sqrt{m_D}(v \cdot k + v \cdot p_2)}\left[\frac{2\lambda^\prime - \frac{\sqrt{2}}{3}g_v\lambda\frac{g_\Phi}{m_\Phi^2}}{v \cdot p_2 + \Delta} + \frac{2\lambda^\prime + \frac{1}{\sqrt{2}}g_v\lambda\left(\frac{g_\omega}{3m_\omega^2} - \frac{g_\rho}{m_\rho^2}\right)}{v \cdot k + \Delta}\right]\\
  \begin{split}
    B_3^{(s,s)} &= -\frac{g_{K^\star} g_{K^{\pm\star} K^\pm \gamma}}{f_K}\left(f_D + gf_{D_s}\sqrt{\frac{m_{D_s}}{m_D}}\frac{m_{D} - v \cdot p_2}{v \cdot p_2 + \Delta}\right)BW_{K^{+\star}}(p_1 + k) \\
    &+ \frac{2f_K (m_{D} \alpha_1 - \alpha_2 v \cdot p_1)}{\sqrt{m_D}}\frac{m_{K^\star}^2}{g_{K^\star}} g_{K^\star K \gamma} BW_{K^\star}(p_2 + k) )\\
    B_3^{(d,d)} &= \frac{f_{D} g_{K^\star}}{f_K}\left(g_{K^\star K \gamma}BW_{K^\star}(p_2 + k) + g_{K^{\pm\star} K^\pm \gamma} BW_{K^{+\star}}(p_1 + k)\right)
  \end{split}\\
  B_4^{(d,d)} &= -\frac{m_{D}^2 f_{D} f_\pi}{ (m_{D}^2 - m_\pi^2)}\frac{m_{K^\star}^2}{g_{K^\star}}\left(g_{K^\star K \gamma}BW_{K^\star}(p_2 + k) + g_{K^{\pm\star} K^\pm \gamma} BW_{K^{+\star}}(p_1 + k)\right)
\end{align}

\begin{align}
  \begin{split}
    a^\prime &= - \frac{f_D g^2 \left(p_2 \cdot k - (v \cdot k)(v \cdot p_2)\right)}{f_K^2 (v \cdot p_1 + v \cdot p_2 + \Delta)(v \cdot p_2 + \Delta)}\\
    &+ \frac{\alpha_1 (v \cdot k)}{f_K^2\sqrt{m_{D}}} \left(1 + \frac{f_K^2 m_{\rho}^4}{g_{\rho}^2} BW_{\rho^{+}}(p_1 + p_2)\right)\\
    &-\frac{\sqrt{2} f_D \lambda g_v m_{\rho}^2 \left(p_2 \cdot k - (v \cdot k)(v \cdot p_2)\right)}{g_{\rho}(v \cdot p_1 + v \cdot p_2 + \Delta)}BW_{\rho^+}(p_1 + p_2)
  \end{split}
\end{align}
\begin{align}
  \begin{split}
    b^\prime &= \frac{ g }{f_K^2 (v \cdot p_2 + \Delta)} \left[\sqrt{\frac{m_{D_s}}{m_D}} f_{D_s} (v \cdot k) + f_D g\frac{p_1 \cdot k - (v \cdot k)(v \cdot p_1)}{v \cdot p_1 + v \cdot p_2 + \Delta}\right] \\
    &- \frac{\alpha_1 (v \cdot k)}{f_K^2\sqrt{m_{D}}}\left(1 + \frac{f_K^2 m_{\rho}^4}{g_{\rho}^2} BW_{\rho^{+}}(p_1 + p_2)\right)\\
    &+ \frac{\sqrt{2}f_D \lambda g_v m_{\rho}^2 \left(p_1 \cdot k - (v \cdot k)(v \cdot p_1)\right)}{g_{\rho}(v \cdot p_1 + v \cdot p_2 + \Delta)}BW_{\rho^+}(p_1 + p_2)
  \end{split}
\end{align}
\begin{align}
  \begin{split}
    c^\prime &= -\frac{g}{f_K^2 m_D (v \cdot p_2 + \Delta)} \left[\sqrt{\frac{m_{D_s}}{m_D}}f_{D_s}(p_2 \cdot k) - f_D g\frac{(p_2 \cdot k)(v \cdot p_1) - (p_1 \cdot k)(v \cdot p_2)}{v \cdot p_1 + v \cdot p_2 + \Delta}\right] \\
    &- \frac{\alpha_1 (p_1 \cdot k - p_2 \cdot k)}{\sqrt{m_{D}^3}f_K^2}\left(1 + \frac{f_K^2 m_{\rho}^4}{g_{\rho}^2} BW_{\rho^{+}}(p_1 + p_2)\right)\\
    &+ \frac{\sqrt{2} f_D \lambda g_v m_{\rho}^2 \left((p_2 \cdot k)(v \cdot p_1) - (p_1 \cdot k)(v \cdot p_2)\right)}{g_{\rho} m_D (v \cdot p_1 + v \cdot p_2 + \Delta)}BW_{\rho^+}(p_1 + p_2)
  \end{split}
\end{align}
\begin{align}
  \begin{split}
    h^\prime &=\frac{g}{2 f_K^2 m_D (v \cdot p_2 + \Delta)}\left(\sqrt{\frac{m_{D_s}}{m_D}}f_{D_s} + f_D g \frac{v \cdot k}{v \cdot p_1 + v \cdot p_2 + \Delta}\right)\\
    &+ \frac{\alpha_1}{\sqrt{m_{D}^3}f_K^2}\left(1 + \frac{f_K^2 m_{\rho}^4}{g_{\rho}^2} BW_{\rho^{+}}(p_1 + p_2)\right)\\
    &+ \frac{f_D \lambda g_v m_{\rho}^2 \left(v \cdot k\right)}{\sqrt{2}g_{\rho} m_D (v \cdot p_1 + v \cdot p_2 + \Delta)}BW_{\rho^+}(p_1 + p_2)
  \end{split}
\end{align}

\subsection{Doubly Cabibbo-suppressed decay modes}

\noindent\underline{$D^+ \to \pi^+ K^0 \gamma$}

\begin{figure}
  \centering
  \subfigure[Contributions to the parity-even form factors $A$ and $E$. Additionally, for each of the diagrams $A_{1,2}$, $A_{1,3}$, $A_{2,3}$, $E_{1,2}$ und $E_{2,3}$ there is another one where the photon is coupled via a vector meson.]{\includegraphics[width=0.45\linewidth]{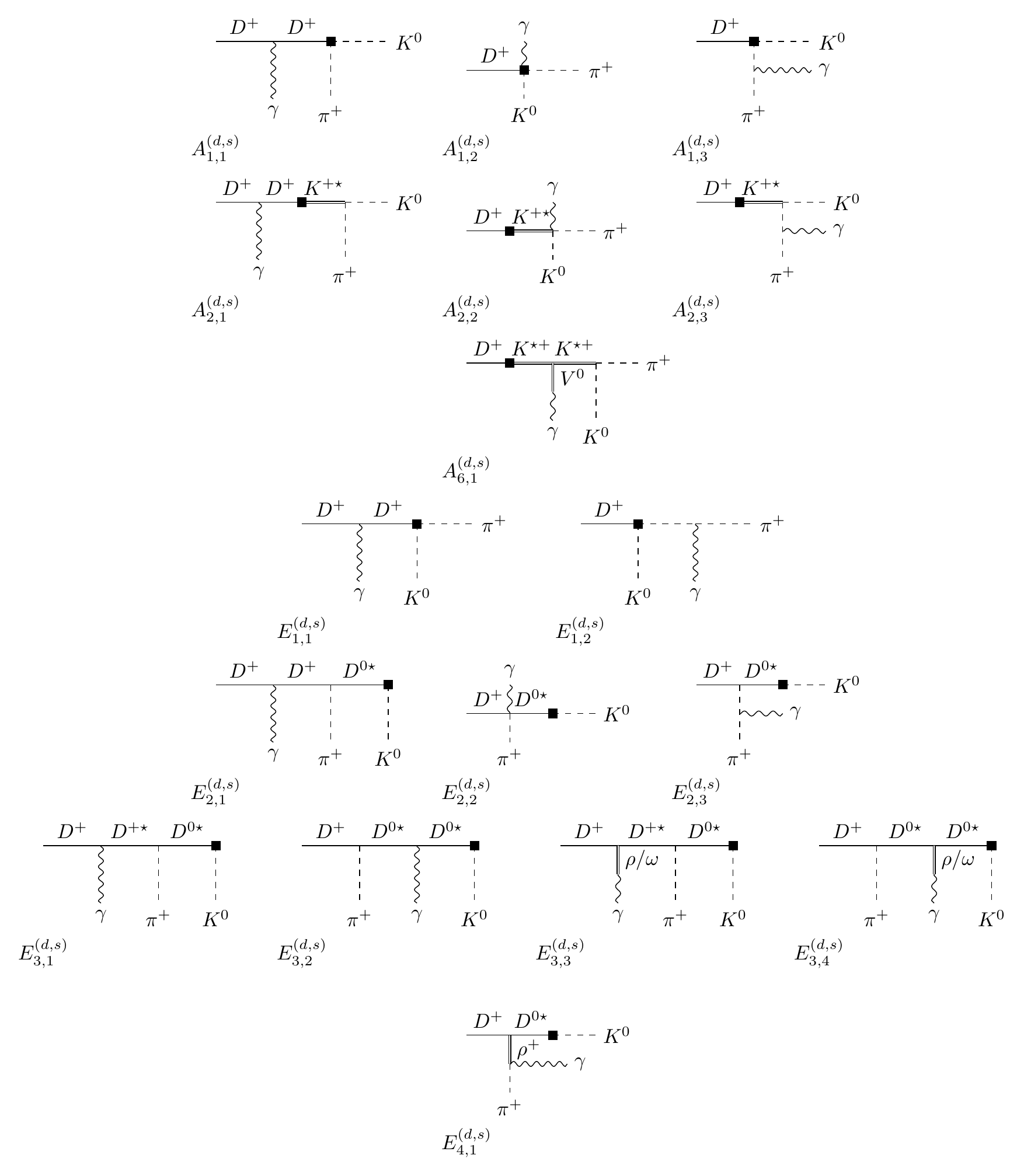}}
  \hfill
  \subfigure[Contributions to the parity-odd form factors $B$ and $D$.]{\includegraphics[width=0.45\linewidth]{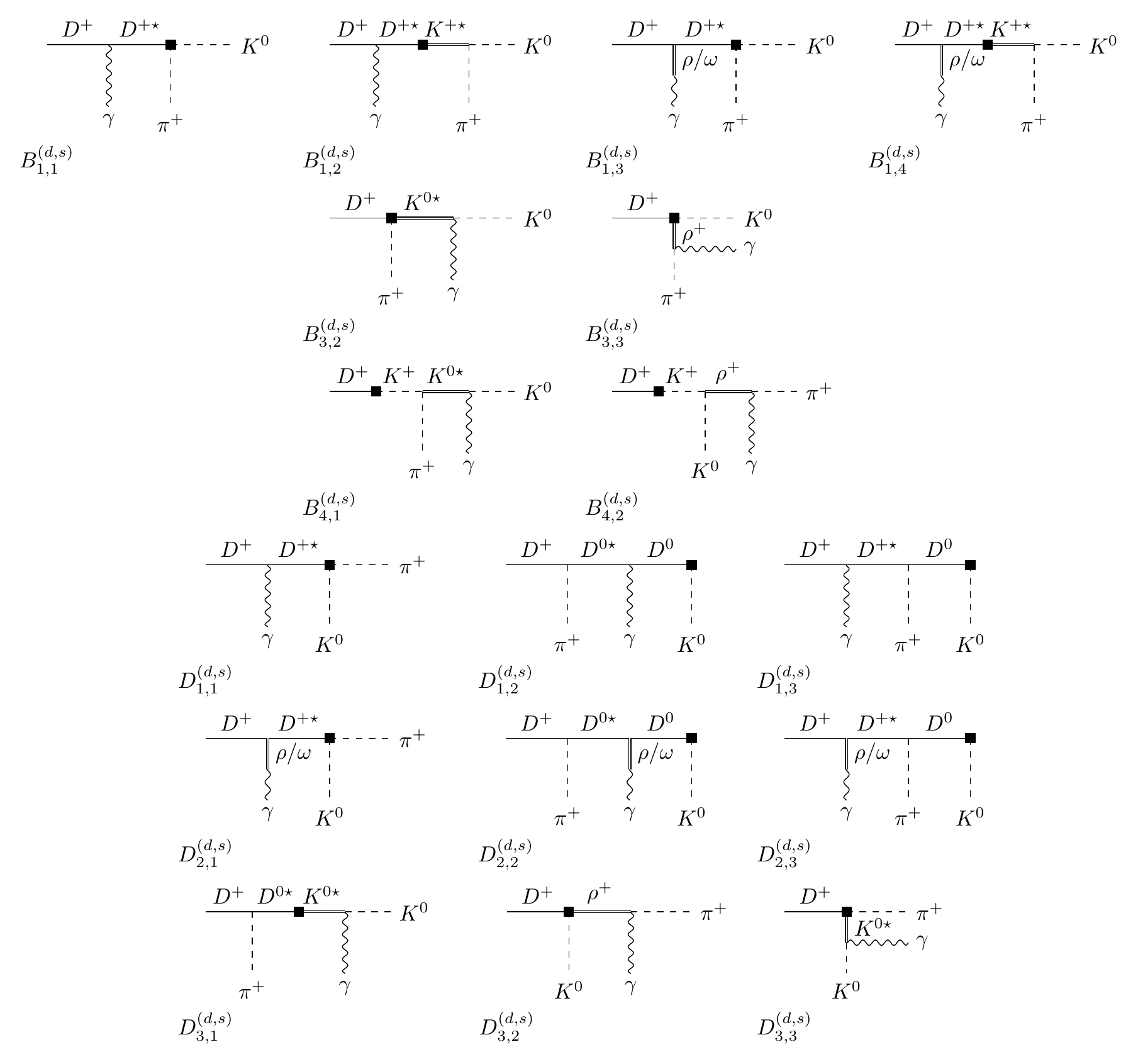}}
  \caption{Feynman diagrams contributing to the decay $D^+ \to \pi^+ K^0 \gamma$ within the SM.}
  \label{fig:Diagramme_Dplus_piplusK}
\end{figure}

\begin{align}
  \begin{split}
    A_{1+2}^{(d,s)} &= -i f_D\frac{v \cdot p_2 - v \cdot p_1 - v \cdot k}{(v \cdot k)(p_1 \cdot k)}\\
    A_6^{(d,s)} &= -i 2 f_D \frac{p_1 \cdot p_2}{m_D(v \cdot k)} BW_{K^{\star+}}(p_1+p_2)
  \end{split} 
\end{align}

\begin{align}
  E_1^{(d,s)} &= -i \frac{f_D f_K}{f_\pi}\frac{v \cdot p_2}{(v \cdot k)(p_1 \cdot k)}\\
  E_2^{(d,s)} &= -i \frac{f_D f_K g}{f_\pi} \frac{p_1 \cdot p_2 - (v \cdot p_1)(v \cdot p_2) + (v\cdot k)(m_D - v \cdot p_2)}{(v \cdot k + v \cdot p_1 + \Delta)(v \cdot k)(p_1 \cdot k)}\\
  E_3^{(d,s)} &= -i \frac{f_D f_K g (v \cdot k)}{f_\pi(v \cdot k + v \cdot p_1 + \Delta)} \left[\frac{2\lambda^\prime + \frac{1}{\sqrt{2}}g_v\lambda\left(\frac{g_\omega}{3m_\omega^2} + \frac{g_\rho}{m_\rho^2}\right)}{v \cdot p_1 + \Delta} - \frac{2\lambda^\prime + \frac{1}{\sqrt{2}}g_v\lambda\left(\frac{g_\omega}{3m_\omega^2} - \frac{g_\rho}{m_\rho^2}\right)}{v \cdot k + \Delta} \right]
\end{align}

\begin{align}
  \begin{split}
    B_1^{(d,s)} &= -\frac{f_D}{v \cdot k + \Delta} \left[1 - m_{K^\star}^2 BW_{K^{+\star}}(p_1 +p_2)\right]\left[2\lambda^\prime + \frac{g_v\lambda}{\sqrt{2}}\left(\frac{g_\omega}{3m_\omega^2} - \frac{g_\rho}{m_\rho^2}\right)\right]
  \end{split}\\
  \begin{split}
    B_3^{(d,s)} &= \frac{f_D g_{K^\star}}{f_\pi} g_{K^\star K \gamma}BW_{K^\star}(p_2 + k) + \frac{f_D g_\rho}{f_K} g_{\rho^\pm \pi^\pm \gamma} BW_{\rho^+}(p_1 + k)
  \end{split}\\
  B_4^{(d,s)} &= -\frac{m_D^2 f_D f_K}{(m_D^2 - m_K^2)}\left( \frac{m_{K^\star}^2}{g_{K^\star}}g_{K^\star K \gamma}BW_{K^\star}(p_2 + k) + \frac{m_{\rho}^2}{g_{\rho}}g_{\rho^\pm \pi^\pm \gamma} BW_{\rho^+}(p_1 + k)\right)
\end{align}
\begin{align}
  D_1^{(d,s)} &= -2\frac{f_D f_K}{f_\pi} \lambda^\prime \left[\frac{1}{v \cdot k + \Delta} + \frac{g (v \cdot p_2)}{v \cdot k + v \cdot p_1}\left(\frac{1}{v \cdot k + \Delta} + \frac{1}{v \cdot p_1 + \Delta}\right)\right]\\
  D_2^{(d,s)} &= -\frac{f_D f_K g_v \lambda}{\sqrt{2}f_\pi} \left[\frac{\frac{g_\omega}{3m_\omega^2} - \frac{g_\rho}{m_\rho^2}}{v \cdot k + \Delta} + \frac{g (v \cdot p_2)}{v \cdot k + v \cdot p_1}\left(\frac{\frac{g_\omega}{3m_\omega^2} - \frac{g_\rho}{m_\rho^2}}{v \cdot k + \Delta} + \frac{\frac{g_\omega}{3m_\omega^2} + \frac{g_\rho}{m_\rho^2}}{v \cdot p_1 + \Delta}\right)\right]\\
  \begin{split}
    D_3^{(d,s)} &= \frac{f_Dg_{K^\star} g_{{K^\star} K \gamma}}{f_\pi}  \left(1 + g \frac{m_D - v \cdot p_1}{v \cdot p_1 + \Delta}\right)BW_{K^\star}(p_2 + k)\\
    &- \frac{2 f_K \left(m_D \alpha_1 - \alpha_2 v \cdot p_2\right)}{\sqrt{m_D}}\frac{m_\rho^2}{g_\rho} g_{\rho^\pm \pi^\pm \gamma} BW_{\rho^+}(p_1 + k)
  \end{split}
\end{align}

\noindent\underline{$D^+ \to K^+ \pi^0 \gamma$}

\begin{figure}
  \centering
  \subfigure[Contributions to the parity-even form factors $A$ and $E$. Note that the diagrams $A_1$ have two different factorizations. Additionally, for each of the diagrams $A_{1,2}$, $A_{1,3}$, $A_{2,3}$, $A_{3,3}$ and $A_{3,4}$ there is another one where the photon is coupled via a vector meson.]{\includegraphics[width=0.45\linewidth]{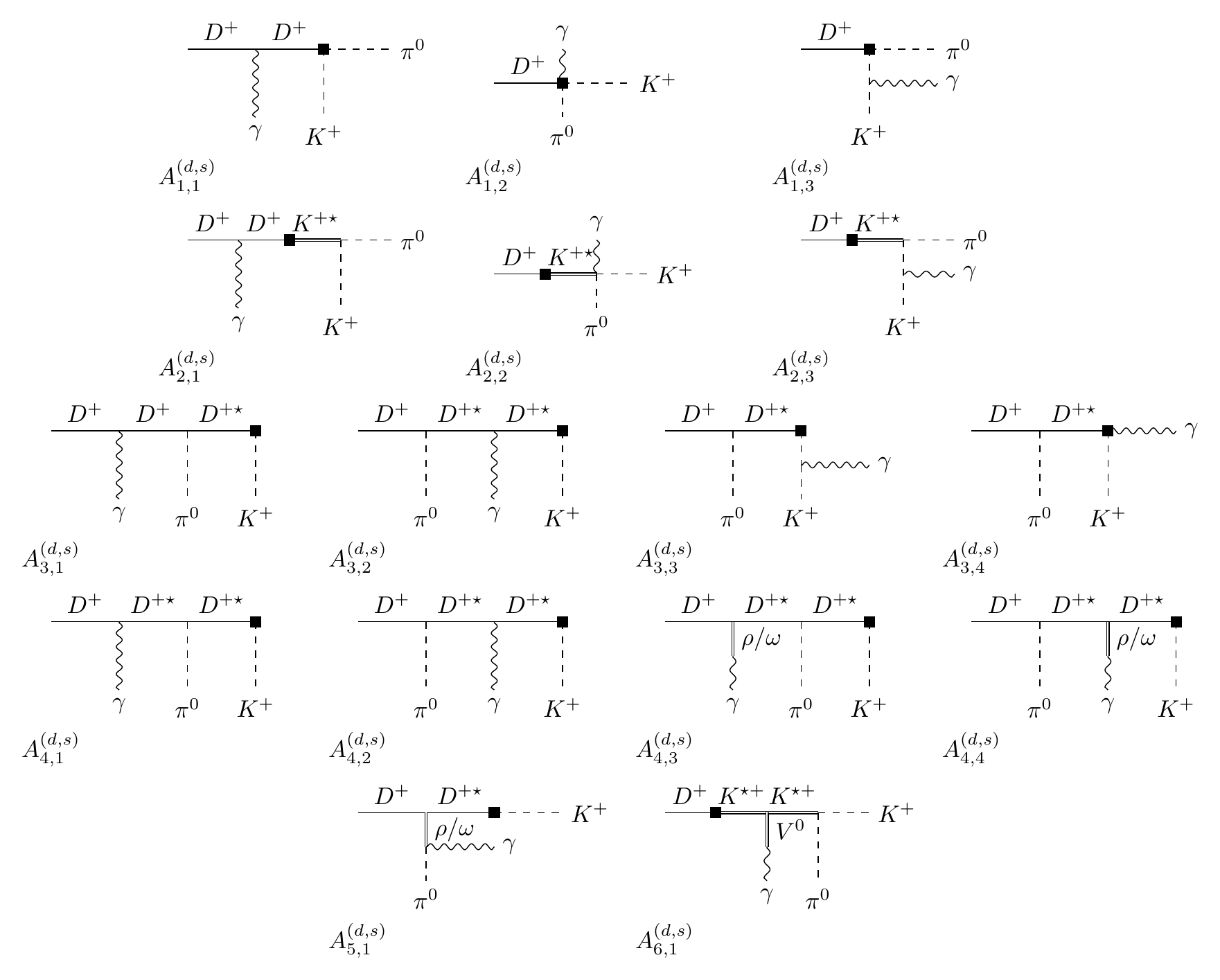}}
  \hfill
  \subfigure[Contributions to the parity-odd form factors $B$ and $D$. Note that the diagrams $B_{1,1/3}$ and $B_{3,2/3}$ have two different factorizations.]{\includegraphics[width=0.45\linewidth]{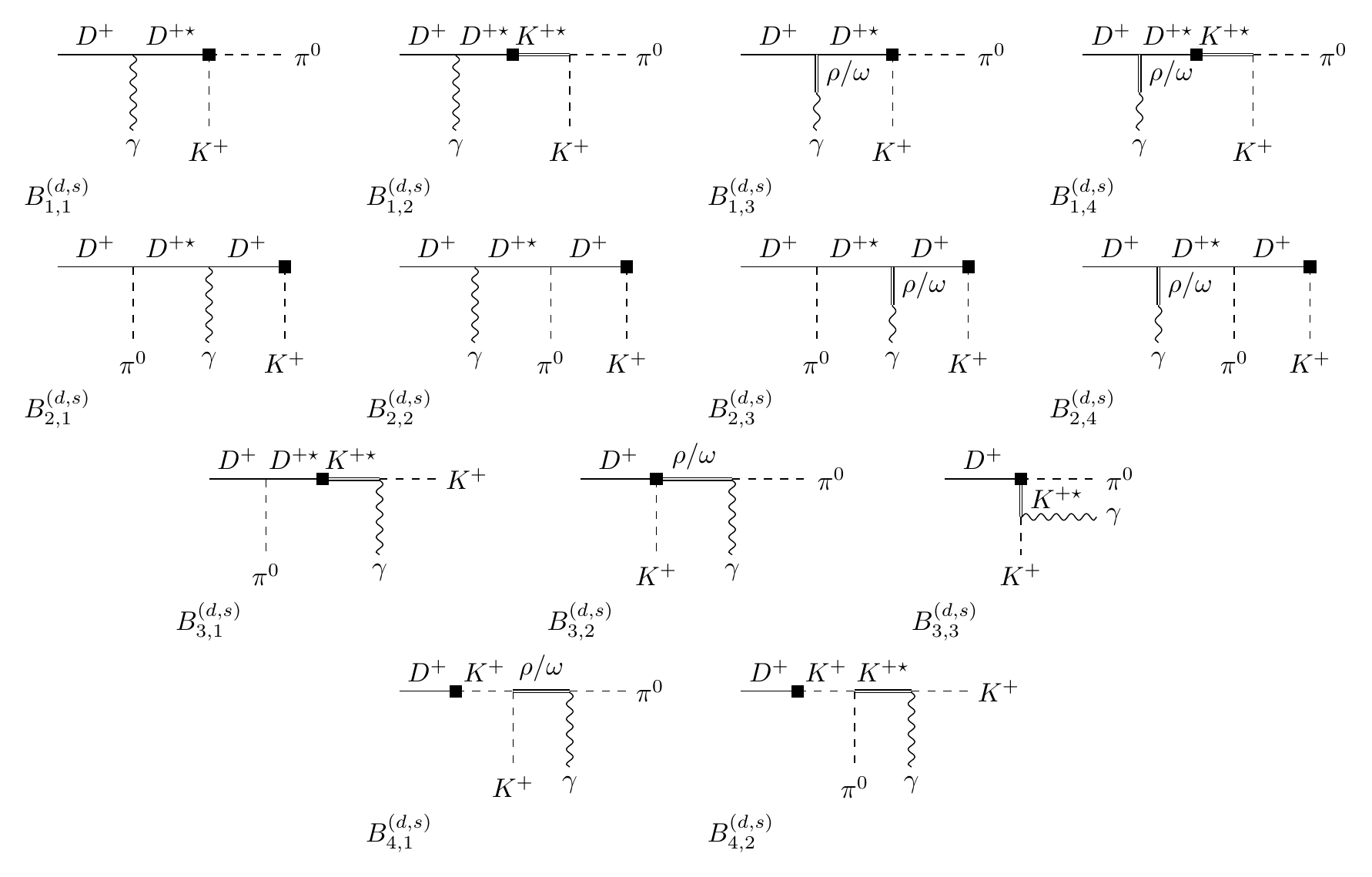}}
  \caption{Feynman diagrams contributing to the decay $D^+ \to K^+ \pi^0 \gamma$ within the SM.}
  \label{fig:Diagramme_Dplus_Kpluspi}
\end{figure}

\begin{align}
  \begin{split}
    A_{1+2}^{(d,s)} &= i \frac{f_D}{\sqrt{2}}\frac{v \cdot p_2 - v \cdot p_1 - v \cdot k}{(v \cdot k)(p_1 \cdot k)} +  i \frac{f_D f_K}{\sqrt{2} f_\pi}\frac{v \cdot p_1 + v \cdot k}{(v \cdot k)(p_1 \cdot k)}
  \end{split} \\
    A_3^{(d,s)} &= i \frac{f_D f_K g}{\sqrt{2} f_\pi} \frac{p_1 \cdot p_2 - (v \cdot p_1)(v \cdot p_2) + (v\cdot k)(M - v \cdot p_2)}{(v \cdot p_2 + \Delta)(v \cdot k)(p_1 \cdot k)}\\
    A_4^{(d,s)} &= i \frac{f_D f_K g (v \cdot k)}{f_\pi(v \cdot k + v \cdot p_2 + \Delta)} \left[\frac{\sqrt{2}\lambda^\prime + \frac{1}{2}g_v\lambda\left(\frac{g_\omega}{3m_\omega^2} - \frac{g_\rho}{m_\rho^2}\right)}{v \cdot k + \Delta} - \frac{\sqrt{2}\lambda^\prime + \frac{1}{2}g_v\lambda\left(\frac{g_\omega}{3m_\omega^2} - \frac{g_\rho}{m_\rho^2}\right)}{v \cdot p_2 + \Delta} \right]\\
    A_6^{(d,s)} &= i \sqrt{2} f_D \frac{p_1 \cdot p_2}{m_D(v \cdot k)} BW_{K^{\star +}}(p_1+p_2)
\end{align}

\begin{align}
    B_1^{(d,s)} &= \frac{f_D}{v \cdot k + \Delta} \left[1 - \frac{f_K}{f_\pi} - m_{K^\star}^2 BW_{K^{+\star}}(p_1 +p_2)\right]\left[\sqrt{2}\lambda^\prime + \frac{1}{2}g_v\lambda\left(\frac{g_\omega}{3m_\omega^2} - \frac{g_\rho}{m_\rho^2}\right)\right]\\
    B_2^{(d,s)} &= -\frac{f_D f_K g (v \cdot p_1)}{f_\pi(v \cdot k + v \cdot p_2)}\left[\frac{\sqrt{2}\lambda^\prime + \frac{1}{2}g_v\lambda\left(\frac{g_\omega}{3m_\omega^2} - \frac{g_\rho}{m_\rho^2}\right)}{v \cdot p_2 + \Delta} + \frac{\sqrt{2}\lambda^\prime + \frac{1}{2}g_v\lambda\left(\frac{g_\omega}{3m_\omega^2} - \frac{g_\rho}{m_\rho^2}\right)}{v \cdot k + \Delta}\right]\\
  \begin{split}
    B_3^{(d,s)} &=  \frac{f_D g_{K^\star} g_{K^{\pm \star} K^\pm \gamma}g(m_D - v \cdot p_2)}{\sqrt{2}f_\pi (v \cdot p_2 + \Delta)}BW_{K^{+\star}}(p_1 + k) \\
    &+ \frac{\sqrt{2}f_K (m_D \alpha_1 - \alpha_2 v \cdot p_1)}{\sqrt{m_D}}\left(\frac{m_\omega^2}{g_\omega}g_{\omega \pi \gamma} BW_\omega(p_2 + k) - \frac{m_\rho^2}{g_\rho}g_{\rho \pi \gamma} BW_\rho(p_2 + k)\right)\\
    & -\frac{f_D}{\sqrt{2}}\left(\frac{g_\rho}{f_K}g_{\rho \pi \gamma}BW_\rho(p_2 + k) + \frac{g_\omega}{f_K}g_{\omega \pi \gamma}BW_\omega(p_2 + k)\right)
  \end{split}\\
  B_4^{(d,s)} &= \frac{m_D^2 f_D f_K}{\sqrt{2} (m_D^2 - m_K^2)}\left(\frac{m_\rho^2}{g_\rho}g_{\rho \pi \gamma}BW_\rho(p_2 + k) + \frac{m_\omega^2}{g_\omega}g_{\omega \pi \gamma}BW_\omega(p_2 + k) + \frac{2m_{K^\star}^2}{g_{K^\star}}g_{K^{\pm \star} K^\pm \gamma} BW_{K^{+\star}}(p_1 + k)\right)
\end{align}

\noindent\underline{$D_s \to K^+ K^0 \gamma$}

\begin{figure}
  \centering
  \subfigure[Contributions to the parity-even form factors $A$ and $E$. For each of the diagrams $A_{1,2}$, $A_{1,3}$, $A_{3,3}$, $A_{3,4}$, $E_{1,2}$ und $E_{2,3}$ there is another one where the photon is coupled via a vector meson.]{\includegraphics[width=0.45\linewidth]{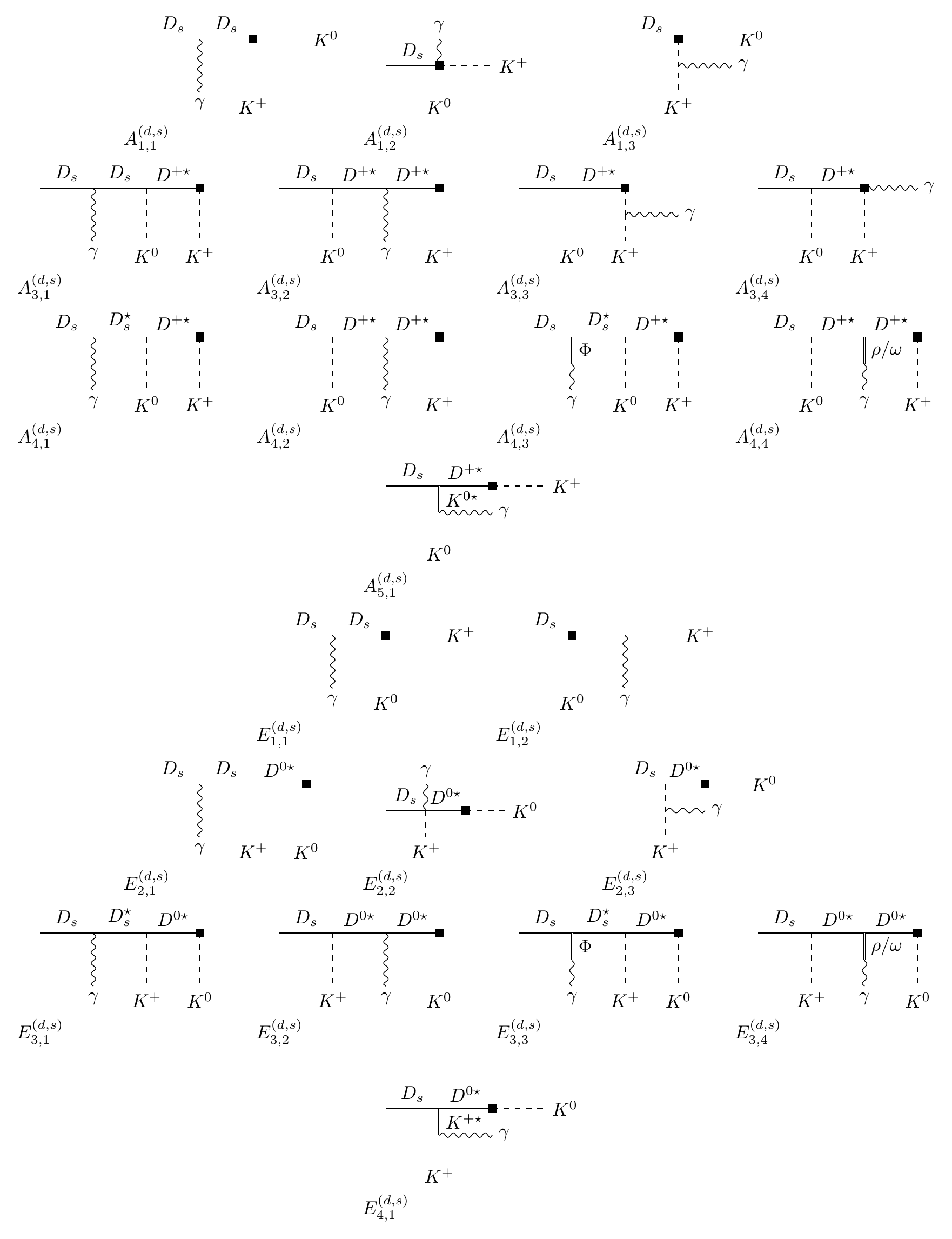}}
  \hfill
  \subfigure[Contributions to the parity-odd form factors $B$ and $D$.]{\includegraphics[width=0.45\linewidth]{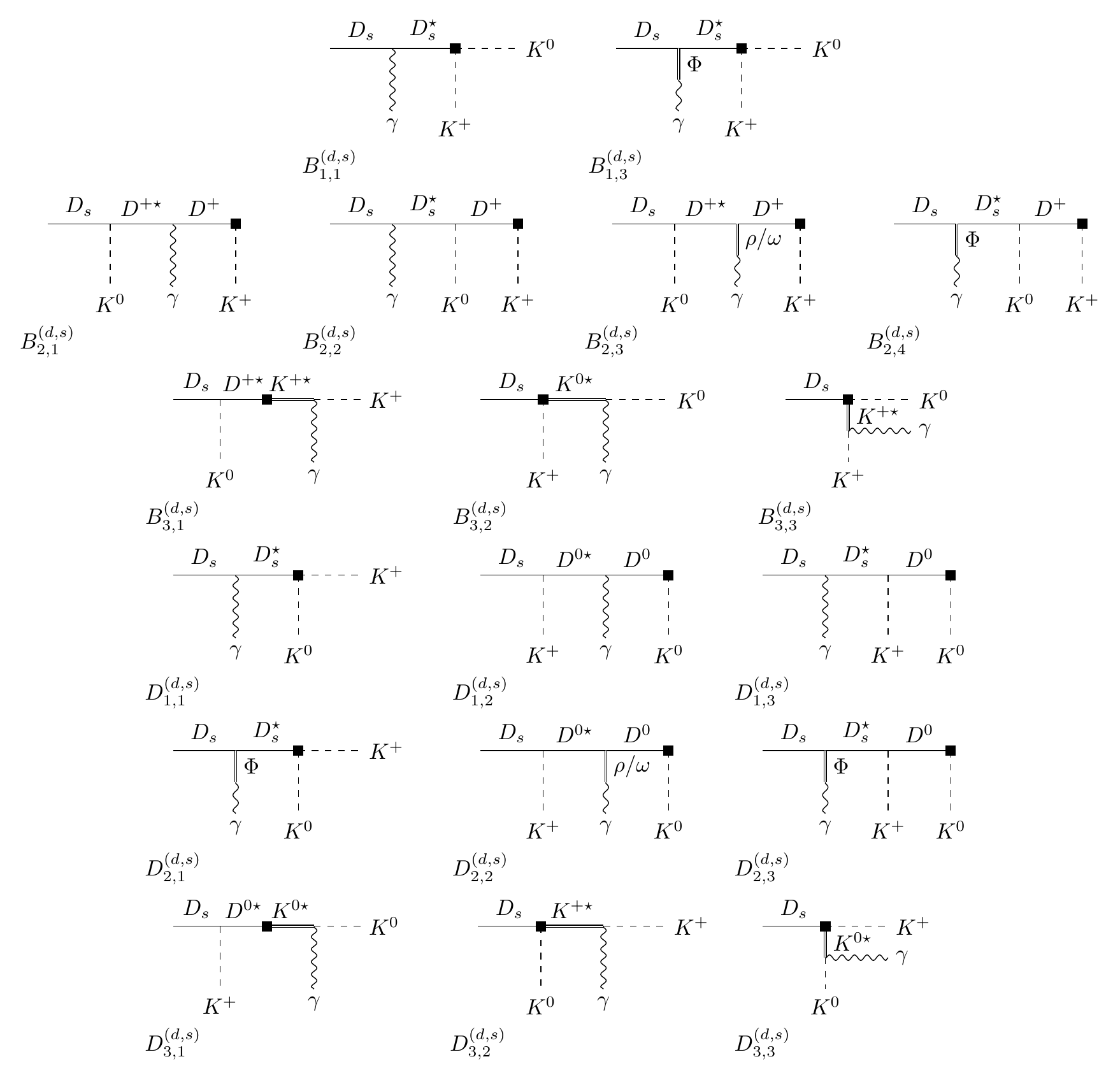}}
  \caption{Feynman diagrams contributing to the decay $D_s \to K^+ K^0 \gamma$ within the SM.}
  \label{fig:Diagramme_Ds_KplusK}
\end{figure}

\begin{align}
    A_1^{(d,s)} &= -i f_{D_s}\frac{v \cdot p_1 + v \cdot k}{(v \cdot k)(p_1 \cdot k)}\\
    A_3^{(d,s)} &= -i \sqrt{\frac{m_D}{m_{D_s}}} f_D  g \frac{p_1 \cdot p_2 - (v \cdot p_1)(v \cdot p_2) + (v\cdot k)(m_D - v \cdot p_2)}{(v \cdot p_2 + \Delta)(v \cdot k)(p_1 \cdot k)}\\
    A_4^{(d,s)} &= -i \sqrt{\frac{m_D}{m_{D_s}}} \frac{f_D g (v \cdot k)}{v \cdot k + v \cdot p_2 + \Delta} \left[\frac{2\lambda^\prime - \frac{\sqrt{2}}{3}g_v\lambda \frac{g_\Phi}{m_\Phi^2}}{v \cdot k + \Delta} - \frac{2\lambda^\prime  + \frac{1}{\sqrt{2}}g_v\lambda\left(\frac{g_\omega}{3m_\omega^2} - \frac{g_\rho}{m_\rho^2}\right)}{v \cdot p_2 + \Delta} \right]
\end{align}

\begin{align}
  E_1^{(d,s)} &= - i f_{D_s}\frac{v \cdot p_2}{(v \cdot k)(p_1 \cdot k)}\\
  E_2^{(d,s)} &= - i \sqrt{\frac{m_D}{m_{D_s}}} f_D g \frac{p_1 \cdot p_2 - (v \cdot p_1)(v \cdot p_2) + (v\cdot k)(m_D - v \cdot p_2)}{(v \cdot k + v \cdot p_1 + \Delta)(v \cdot k)(p_1 \cdot k)}\\
  E_3^{(d,s)} &= -i \sqrt{\frac{m_D}{m_{D_s}}}\frac{f_D g (v \cdot k)}{v \cdot k + v \cdot p_1 + \Delta} \left[\frac{2\lambda^\prime + \frac{1}{\sqrt{2}}g_v\lambda\left(\frac{g_\omega}{3m_\omega^2} + \frac{g_\rho}{m_\rho^2}\right)}{v \cdot p_1 + \Delta} - \frac{2\lambda^\prime - \frac{\sqrt{2}}{3}g_v\lambda \frac{g_\Phi}{m_\Phi^2}}{v \cdot k + \Delta} \right]
\end{align}

\begin{align}
  B_1^{(d,s)} &= \frac{f_{D_s}}{v \cdot k + \Delta} \left[2\lambda^\prime - \frac{\sqrt{2}}{3}g_v\lambda \frac{g_\Phi}{m_\Phi^2}\right]\\
  B_2^{(d,s)} &= \sqrt{\frac{m_D}{m_{D_s}}} \frac{f_D g (v \cdot p_1)}{v \cdot k + v \cdot p_2}\left[\frac{2\lambda^\prime+ \frac{1}{\sqrt{2}}g_v\lambda\left(\frac{g_\omega}{3m_\omega^2} - \frac{g_\rho}{m_\rho^2}\right) }{v \cdot p_2 + \Delta} + \frac{2\lambda^\prime - \frac{\sqrt{2}}{3}g_v\lambda \frac{g_\Phi}{m_\Phi^2}}{v \cdot k + \Delta}\right]\\
  \begin{split}
    B_3^{(d,s)} &= -\frac{g_{K^\star} g_{K^{\pm \star} K^\pm \gamma}}{f_K}\left(f_{D_s} + \sqrt{\frac{m_D}{m_{D_s}}}f_D g\frac{m_{D_s} - v \cdot p_2}{v \cdot p_2 + \Delta}\right)BW_{K^{+\star}}(p_1 + k) \\
    &+ \frac{2f_K (m_{D_s} \alpha_1 - \alpha_2 v \cdot p_1)}{\sqrt{m_{D_s}}}\frac{m_{K^\star}^2}{g_{K^\star}}g_{{K^\star} K \gamma} BW_{K^\star}(p_2 + k)
  \end{split}
\end{align}
\begin{align}
  D_1^{(d,s)} &= -2  \lambda^\prime \left[\frac{f_{D_s}}{v \cdot k + \Delta} + \sqrt{\frac{m_D}{m_{D_s}}}\frac{f_D g (v \cdot p_2)}{v \cdot k + v \cdot p_1}\left(\frac{1}{v \cdot k + \Delta} + \frac{1}{v \cdot p_1 + \Delta}\right)\right]\\
  D_2^{(d,s)} &= -g_v \lambda \left[\frac{- f_{D_s} \frac{\sqrt{2}}{3} \frac{g_\Phi}{m_\Phi^2}}{v \cdot k + \Delta} + \sqrt{\frac{m_D}{m_{D_s}}}\frac{f_D g (v \cdot p_2)}{v \cdot k + v \cdot p_1}\left(\frac{- \frac{\sqrt{2}}{3} \frac{g_\Phi}{m_\Phi^2}}{v \cdot k + \Delta} + \frac{\frac{1}{\sqrt{2}}\left(\frac{g_\omega}{3m_\omega^2} + \frac{g_\rho}{m_\rho^2}\right)}{v \cdot p_1 + \Delta}\right)\right]\\
  \begin{split}
    D_3^{(d,s)} &= \frac{g_{K^\star} g_{{K^\star} K \gamma}}{f_K}\left(f_{D_s} + \sqrt{\frac{m_D}{m_{D_s}}} f_D g \frac{m_{D_s} - v \cdot p_1}{v \cdot p_1 + \Delta}\right) BW_{K^\star}(p_2 + k) \\
    &- \frac{2 f_K  \left(m_{D_s} \alpha_1 - \alpha_2 v \cdot p_2\right)}{\sqrt{m_{D_s}}}\frac{m_{K^\star}^2 }{g_{K^\star}} g_{K^{\pm \star} K^\pm \gamma} BW_{K^{\pm \star}}(p_1 + k)
  \end{split}
\end{align}

\section{Differences with respect to \cite{Fajfer:2002bq} \label{sec:diff}}

We provide a list of some differences between our form factors and
those obtained in Ref.~\cite{Fajfer:2002bq}. More comments are given in \cite{Adolph:2020ema}.
\begin{enumerate}
	\item The contributions of the diagrams $A_{4,1}^+$ and $C_{4,1}^+$ vanish in our calculation.
	\item Missing global minus sign in $C_2^+$.
	\item We believe that there are diagrams that have not been shown in Ref.~\cite{Fajfer:2002bq}: For each of the diagrams $A_{1,2}^+$, $A_{1,3}^+$, $A_{3,3}^+$, $A_{3,4}^+$, $E_{1,2}^0$ and $E_{2,3}^0$ there is another one in which the photon couples via a vector meson. However, we obtain the same form factors for the bremsstrahlung contributions.
	\item We obtain a relative minus sign for each vector meson in a diagram; however, we get the same relative signs for $R_\gamma^{0/+}$ as given in Eqs.~(24) and~(25)~\cite{Bajc:1994ui}.
\end{enumerate}

\end{document}